\documentclass[aps,twocolumn,superscriptaddress,nofootinbib,prb,floatfix,citeautoscript,preprintnumbers]{revtex4-1}
\usepackage{graphicx,color,epsfig}
\usepackage{amssymb}
\usepackage{amsmath}
\usepackage{bm}
\usepackage{mathtools}
\usepackage{amsfonts}
\usepackage{dsfont}
\usepackage{epstopdf}
\usepackage[normalem]{ulem}
\usepackage{comment}
\usepackage{breakcites}
\usepackage[breaklinks=true]{hyperref}
\usepackage[caption=false]{subfig}
\usepackage[dvipsnames]{xcolor}

\newcommand{\beq}{\begin{equation}}
\newcommand{\eeq}{\end{equation}}
\newcommand{\bsf}{\begin{subfigure}}
\newcommand{\esf}{\end{subfigure}}
\newcommand{\bal}{\begin{align}}
\newcommand{\eal}{\end{align}}

\newcommand{\nn}{{\nonumber}}

\def\mya{a}
\def\myb{b}
\def\myc{c}

\newcommand{\fmnd}[1]{ {{c_#1'}^\dagger} }
\newcommand{\fmn}[1]{ {c_#1' } }

\begin{document}

\title{Markovianity of an emitter coupled to a structured spin chain bath}

\author{J. Roos}
\affiliation{Max-Planck-Institut f\"{u}r Quantenoptik, Hans-Kopfermann-Str. 1, D-85748 Garching, Germany}
\affiliation{Munich Center for Quantum Science and Technology (MCQST), Schellingstr. 4, D-80799 M\"unchen, Germany}
\author{J. I. Cirac}
\affiliation{Max-Planck-Institut f\"{u}r Quantenoptik, Hans-Kopfermann-Str. 1, D-85748 Garching, Germany}
\affiliation{Munich Center for Quantum Science and Technology (MCQST), Schellingstr. 4, D-80799 M\"unchen, Germany}
\author{M. C.  Ba\~nuls}
\affiliation{Max-Planck-Institut f\"{u}r Quantenoptik, Hans-Kopfermann-Str. 1, D-85748 Garching, Germany}
\affiliation{Munich Center for Quantum Science and Technology (MCQST), Schellingstr. 4, D-80799 M\"unchen, Germany}

\date{\today }

\begin{abstract}

We analyze the dynamics of a spin 1/2 subsystem coupled to a spin chain. We simulate numerically the full quantum many-body system for various sets of parameters and initial states of the chain, and characterize the divisibility of the subsystem dynamics, i.e. whether it is Markovian and can be described by a (time dependent) master equation. We identify regimes in which the subsystem admits such Markovian description, despite the many-body setting, and provide insight about why the same is not possible in other regimes. Interestingly, coupling the subsystem at the edge, instead of the center, of the chain gives rise to qualitatively distinct behavior.

\end{abstract}

\maketitle

\section{Introduction}
\label{sec:intro}

The subject of open quantum systems (OQS) focuses on the description of quantum systems coupled to a (typically much larger) environment. In general, solving the time evolution of the total system is out of reach due to the macroscopic number of environmental degrees of freedom and the exponentially large Hilbert space. Instead one tries to obtain an effective reduced description which involves the degrees of freedom of the OQS only. In this context, the distinction between Markovian and non-Markovian dynamics is a central theme.\cite{RevInes,RevRivas,RevBreuer} Originally, the former denoted situations that allowed for the derivation of a 'Markovian' master equation.\cite{history} This is a specific differential equation generating the dynamics of the OQS that has the property of being 'memoryless' (Markovian), in the sense that the evolution of the OQS at a given time depends only on its state at that time. 

The same OQS perspective can be applied to study the evolution of a subsystem in a closed many-body quantum system. For instance, although the full system is in a pure state, the computation of local observables requires only to know the state of a (small) subsystem, for which the rest of the system would play the role of environment. Again, solving the long time evolution of the full system is in general out of reach due to build-up of entanglement, such that it might be desirable to find an effective reduced description for the subsystem (OQS) only. However, the standard 'Markovian' master equation derivation is based on weak coupling and a separation of time scales between open system and environment;\cite{tannoudji} two conditions that are generally not fulfilled in the quantum many-body setup. It is thus interesting to analyze whether there are cases in which the dynamics still admits a reduced description in this setting, and, in case such a description exists, whether it is Markovian.

In recent years a variety of non-Markovianity measures were put forward  that go beyond the original Markovianity conditions mentioned above and attempt to quantify deviations from Markovian dynamics from a quantum information theoretical perspective.\cite{AssessingWolfCirac,blp1,rhp,nm1,nm2,nm3,nmdif1,nm4,nm5,nm6,nm7,nm8,nm9,nm9b,nm9c,nm9d,nm9e,nm10,nm11,nm12a,nm12b,nm12c,nm13,nm14,CPdivNotMeanMark} Two of the most widely used are the one introduced by Breuer, Laine, Piilo\cite{blp1,blp2} (BLP measure), which detects the non-monotonicity of the trace distance between pairs of states evolving in time, and the more stringent one introduced by Rivas, Huelga, Plenio\cite{rhp} (RHP measure), which detects non-divisibility of the quantum channel mapping initial OQS states to their time-evolved states. These measures are not equivalent, since there exist cases which are characterized as BLP-Markovian but RHP-non-Markovian.\cite{history,nmdif1,nmdif2,nmrev1,nmrev2} For cases with Markovian dynamics according to RHP, a Markovian reduced description in the above sense exists, i.e. the (time dependent) equation governing the reduced dynamics does not explicitly depend on past system states and is called 'time dependent Markovian' master equation.\cite{DivQuantChan,rhp} We quantify non-Markovianity by its robustness, as originally introduced for 'snapshots' of quantum evolution in reference\cite{AssessingWolfCirac} (and here generalized to continuous evolution), i.e. how much noise can be added to the dynamics before RHP Markovianity is recovered, thus providing a physical interpretation.

In this paper we explore the above questions in the particular case of a spin (the OQS) coupled to a XY spin chain, which plays the role of environment. In particular, we identify regimes that allow for a description via a 'time dependent Markovian' master equation and provide insight into what prevents such a description. We consider two scenarios in OBC: (i) spin coupled to the center; (ii) spin coupled to the first site; and we analyze different initial states of the chain.
In some particular cases (namely in scenario (ii)\cite{apollaro}, and for (i) in the thermodynamic limit~\cite{AlexPaper} when the chain is initially in the vacuum) an exact solution is possible. In more general cases, we use tensor network methods (MPS and MPO, as we are in 1D) to simulate the evolution of the full system. 

The initial state of the chain can be empty (i.e. in the fermionic vacuum of the XY chain) or contain 'excitations' (if some of such modes are occupied). We find that while in the first case, BLP and RHP Markovianity is equivalent, in the second case, only the divisibility (RHP) measure detects all the non-Markovianity appearing in regions of the parameter space. 
One possibility we explore to populate fermionic modes in the initial environment state is using thermal states. Whilst small temperature induces additional non-Markovianity, we find for scenario (i) that increasing the temperature gradually removes the non-Markovianity until at high temperature the dynamics is captured by a 'time-dependent Markovian' master equation. This applies even at the band edges where the spectral density diverges, a scenario that is often associated with strong non-Markovian behavior.\cite{KesslerSelfCons,ClosBreuer,znidaric,bandgap} In contrast, in scenario (ii) we find that any RHP non-Markovianity of the vacuum case survives at all temperatures. We show that this remarkable difference between the two cases can be anticipated from the different decay of the environment correlation functions at high temperature in both cases. 

The paper is structured as follows: we start in ~\ref{sec:measures} by introducing the non-Markovianity measures and the 'time dependent Markovian' master equation. In \ref{sec:model} we present the details of the model and the specific form the non-Markovianity measure adopts in this model. We also discuss the conditions that allow for the derivation of a standard 'Markovian' master equation. The section closes with a summary of our numerical methods. In \ref{sec:vacCase} we review the analytically solvable case in which the spin is coupled to the center of the chain initialized in the vacuum. In \ref{sec:TnonMark} we introduce initial environmental excitations in this setup which yields the main results of this paper. In \ref{sec:Apollaro} we explore the qualitatively different nature of the dynamics obtained by coupling the spin to the first site in the chain and provide insight into why the two scenarios differ so much with respect to non-Markovianity. Finally, in \ref{sec:conclu} we conclude and summarize the main results.

\section{Quantifying non-Markovianity}
\label{sec:measures}

Over the past decade multiple inequivalent characterisations of non-Markovianity based on quantum information theory have been introduced.\cite{AssessingWolfCirac,blp1,rhp,nm1,nm2,nm3,nmdif1,nm4,nm5,nm6,nm7,nm8,nm9,nm9b,nm9c,nm9d,nm9e,nm10,nm11,nm12a,nm12b,nm12c,nm13,nm14,CPdivNotMeanMark} In the next paragraphs we review the measures that are relevant for the rest of our paper. 

\subsection{Non-Markovianity robustness and 'time-dependent Markovian' master equation}

We consider the set $\mathfrak{T}$ of finite dimensional, completely positive (CP), trace-preserving, linear maps (quantum channels) $\mathcal{T}: \mathcal{M}_d\rightarrow \mathcal{M}_d$ from the space $\mathcal{M}_d$ of $d\times d$ matrices into itself. $\mathcal{T}$ is called \emph{divisible}~\cite{DivQuantChan} if there exists a decomposition $\mathcal{T}=\mathcal{T}_1 \mathcal{T}_2$ with $\mathcal{T}_i \in \mathfrak{T}$ such that none of the $\mathcal{T}_i$ is a unitary conjugation. $\mathcal{T}$ is called \emph{infinitesimal divisible} if for all $\epsilon>0$ there exists a finite set of channels $\mathcal{T}_i \in \mathfrak{T}$ such that (i) $||\mathcal{T}_i - \mathds{1} || \le \epsilon$ and (ii) $\mathcal{T}=\prod_i \mathcal{T}_i$.

The time evolution of finite dimensional quantum systems is given by a one parameter family of quantum channels, known as dynamical map, that maps the initial state of the system, described by the density matrix $\rho(0)$, to the time evolved state $\rho(t)=\mathcal{T}(t)[\rho(0)]$. Denoting by $\mathcal{T}(t_2,t_1)$ the map for the evolution from time $t_1$ to $t_2$, we have, by continuity, $\mathcal{T}(t+\Delta t)=\mathcal{T}(t+\Delta t,t)\mathcal{T}(t)$ and thus
\beq
\mathcal{T}(t+\Delta t,t)=\mathcal{T}(t+\Delta t)\mathcal{T}(t)^{-1}.
\eeq
If for all $\epsilon>0$ there is a finite $\Delta t<\epsilon$ such that these maps are CP for all $t$, 
$\mathcal{T}(t)$ is infinitesimal divisible ($\mathcal{T}(t+\Delta t, t)\to \mathds{1}$ as $\Delta t \to 0$). Note that this condition is more restrictive than mere infinitesimal divisibility of a 'snapshot' of $\mathcal{T}(t)$ at a given time, since our decomposition needs to follow the dynamics at all times. To be consistent with the recent literature we drop the word 'infinitesimal' and call the time evolution divisible or (RHP\cite{rhp}) Markovian if the above is true. Notice that for $\mathcal{T}(t+\Delta t,t)$ to be defined unambiguously, $\mathcal{T}(t)^{-1}$ needs to exist. Since $\mathcal{T}(t\to0)\to\mathds{1} $, for $t$ small enough $\mathcal{T}(t)$ will be invertible. For later times, $\mathcal{T}(t)^{-1}$ may not exist, in which case one lacks essential information as a consequence of being blind to the environment part of the whole system. In that case one can resort to pseudoinverse techniques.\cite{rhp,pseudoinv} For the remainder of this paper $\Delta t$ denotes a small (but finite) time step.

We can represent the map by a matrix~\cite{DivQuantChan}
\beq
T(t)_{\alpha,\beta}=\operatorname{tr}\Big(F_{\alpha}^\dagger \mathcal{T}(t)[F_{\beta}]\Big),
\eeq
where $\{ F_\alpha \}_{\alpha=1,\dots,d^2}$ is an orthonormal basis in $\mathcal{M}_d$. We will use the canonical basis $\{ |i\rangle \langle j| \}_{i,j=1,\dots,d}$. In the rest of the paper by $T$ and $dT$ we denote the matrix representations of $\mathcal{T}(t)$ and $\mathcal{T}(t+\Delta t,t)$ respectively and omit their time dependence for convenience. The density matrix can be written as a linear combination of this basis with components $\langle ij | \rho \rangle$, where $\langle A,B \rangle=\operatorname{tr}\left(A^\dagger B\right)$ is the Hilbert-Schmidt scalar product and we identify $|ij\rangle \leftrightarrow |i \rangle \langle j |$. Given a map $\mathcal{T}$ on a $d$-dimensional space, its Choi state\cite{choi1,choi2} is 
\beq
T^\Gamma=d\left( \mathcal{T} \otimes \mathds{1}\right)[\omega] \nn,
\eeq
where $\omega$ is a maximally entangled state $\omega=|\omega\rangle \langle \omega |$, $|\omega \rangle = \frac{1 }{\sqrt{d}}\sum_{i=1}^d |ii \rangle$. \cite{AssessingWolfCirac} One has $\langle ij| T^\Gamma | kl \rangle = \langle ik | T |jl \rangle $. The map is CP iff $T^\Gamma$ is positive semidefinite. 

Divisibility is equivalent to $\rho(t)$ being a solution of a time-dependent Lindblad ('time-dependent Markovian') master equation:~\cite{DivQuantChan,rhp}
\begin{align}
\frac{d\rho}{dt}&=\mathcal{L}(t)[\rho]  \nn\\ 
& = i[\rho,H(t)] \nn \\ &   \quad + \sum_{i=1}^{d^2-1} \gamma_i(t) \left( L_i(t) \rho L^\dagger_i(t)-\frac{ \left\{ L^\dagger_i(t) L_i(t),\rho \right\}}{2}  \right),
\label{eq:lind}
\end{align}
where $\gamma_i\ge0$ and $L_i$ are called rates and Lindblad operators of the (time-dependent) Lindbladian $\mathcal{L}(t)$, and $H(t)=H^\dagger(t)$. If the dynamics is described by a time-dependent Lindblad master equation, the dynamical map $T$ can be decomposed into infinitesimal 'pieces' $dT=e^{L dt}$, where $L$ is the matrix representation of $\mathcal{L}(t)$. Note that the corresponding Choi state $L^\Gamma$ is hermitian. We can now quantify non-divisibility of the reduced dynamics by calculating how much $\operatorname{log}dT$ deviates from a valid Lindbladian. Following reference\cite{AssessingWolfCirac} this amounts to checking if (a) $(\operatorname{log}dT)^\Gamma$ is hermitian and (b) if there exists a branch of the logarithm such that $\omega_\perp (\operatorname{log}dT )^\Gamma \omega_\perp \ge 0$, where $\omega_\perp= \mathds{1} -\omega$ is the projector onto the orthogonal complement of the maximally entangled state.

If $dT$ is hermiticity-preserving (a necessary condition for it to be CP), its eigenvalues are either real or come in complex conjugate pairs.
The set of hermitian logarithms of $dT$ is then parametrized by a set of integers $m_c'\in\mathbb{Z}$, $L_{m'}=L_0+2\pi i \sum_c m_c' ( P_{c} - P_{\overline{c}})$, where $L_0$ denotes the principal branch and $P_c$ and $P_{\overline{c}}$ are projectors onto the eigenspaces associated to a complex conjugate pair of eigenvalues $\lambda_c$ and $\lambda_{\overline{c}}$ of $dT$. We define $A_0=\omega_\perp L_0^\Gamma \omega_\perp$ and $A_c=2\pi i \omega_\perp \left( P_{c} - P_{\overline{c}} \right)^\Gamma \omega_\perp$.  If $L_0^\Gamma$ and thus $A_0$ is hermitian, the dynamics is Markovian iff for any time there exists $\{m_c\}$ such that
\beq
A_0+\sum_c m_c A_c \ge 0.
\label{eq:Am}
\eeq

If~\eqref{eq:Am} is not satisfied during a given time step, adding noise may remove the non-Markovianity. In\cite{AssessingWolfCirac} the non-Markovianity is measured by its robustness, i.e. by the minimum amount of isotropic noise $\mu$ that achieves this:
\beq
\mu=\operatorname{inf}\left\{ \mu' \ge 0 : \exists m \in \mathbb{Z}^C :  A_0+\sum_c m_c A_c + \frac{\mu'}{d} \mathds{1} \ge 0 \right\}
\label{eq:noise}
\eeq
$L_m-\mu \omega_\perp$ is then a valid Lindbladian. 
Note that if $dT$ has some real negative eigenvalue, $L_0^\Gamma$ is non-hermitian and $\operatorname{log} dT$ cannot be made a valid Lindbladian by adding a finite amount of noise. If $dT$ is hermiticity-preserving and does not have real negative eigenvalues, we assign robustness according to Eq.~\eqref{eq:noise}, otherwise $\mu=\infty$.  
$\mu>0$ at some time implies that the evolution cannot be described with a valid 'time-dependent Markovian' master equation, which is equivalent to non-divisibility. It is thus obvious that this provides a necessary and sufficient criterion to decide about Markovianity. 

In practice we compute $T$ at discrete times separated by time steps $\Delta t$ until a final time $t_{\rm fin}=K \Delta t$. We then compute the minimum noise $\mu^{(n)}$ required to make the $n-$th time step Markovian. In order to compute a non-Markovianity measure for the whole evolution interval, we choose to average: $\overline{\mu}=\frac{1}{K} \sum_{n=1}^K \mu^{(n)}$. We then use the normalized degree of non-Markovianity\cite{AssessingWolfCirac}
\beq
\mathcal{N}=1-\operatorname{exp}\left[\overline{\mu} \left( 1-d^2 \right) \right],
\label{eq:nMdegree}
\eeq
where we have $\mathcal{N} \in [0,1]$ and the dynamics is Markovian until time $t_{\rm fin}$ iff $\mathcal{N}=0$. We choose $\Delta t$ sufficiently small such that $\mathcal{N}$ is converged with respect to the time step. Notice that this non-Markovian measure depends on the final time $t_{\rm fin}$.

Note from Eq.~\eqref{eq:noise} that $\mu^{(n)}$ only depends on the most negative eigenvalue of $A^{(n)}_0+\sum_c m^{(n)}_c A^{(n)}_c$, where the $m_c^{(n)}$ are chosen in a way to minimize the magnitude of the most negative eigenvalue. We can gain further insight into the dynamics by looking at the full spectrum. We are interested only in the non-zero eigenvalues $\lambda^{(n)}_i \in \mathbb{R}$ and corresponding eigenvectors $v^{(n)}_i \in \mathbb{C}^{d^2}$ and define
\begin{align}
 \langle j | L^{(n)}_i | l \rangle &= (v^{(n)}_i)_{jl} \label{eq:lindops1}\\ 
  \gamma^{(n)}_i&=\frac{1}{\operatorname{tr} \left( {L^{(n)}_i}^\dagger L^{(n)}_i\right)}\frac{\lambda^{(n)}_i}{\Delta t}.\label{eq:lindops2}
\end{align}
If $\gamma_i^{(n)}$, $L_i^{(n)}$ exist, $\rho\Big((n+1)\Delta t\Big)$ can now be obtained by evolving $\rho(n\Delta t)$ for a time interval $\Delta t$ with a differential equation of the form of Eq.~\eqref{eq:lind} with time-independent rates $\gamma_i^{(n)}$ and operators $L_i^{(n)}$.\cite{GoToLindblad} In the limit $\Delta t \to0$ the time dependent equation can be recovered. This is a 'time-dependent Markovian' master equation if the rates are non-negative at all times. If we denote the minimum rate and its corresponding operator by $ \gamma_{\text{min}}$ and $L_{\text{min}}$ respectively (omitting their time dependence for convenience), we have 
\begin{equation}
\mu = \begin{cases}
0 & \gamma_{\text{min}} \ge 0\\
-d \cdot \gamma_{\text{min}} \cdot \operatorname{tr} \left( L_{\text{min}}^\dagger L_{\text{min}}\right) & \gamma_{\text{min}} <0 ,
\label{eq:noise2}
\end{cases}
\end{equation}
and thus the measure is non-zero if any rate ever becomes negative. The rates contain the required robustness information whilst providing additional insight into what kind of process is responsible for making a given time step non-Markovian.

\subsection{BLP measure}
Another broadly used measure of non-Markovianity is the BLP measure\cite{blp1,blp2}, which is based on the study of the time behavior of the trace distance $\mathcal{D}$ between pairs of density matrices evolving in time. For a pair $\rho^{1,2}$ it is defined as $\mathcal{D}\left[ \rho^1,\rho^2 \right]=\frac{1}{2}\operatorname{tr}\left| \rho^1-\rho^2 \right|$. The idea is that since the trace distance is contractive under CP maps, Markovian processes, described by 'Markovian' or 'time-dependent Markovian' master equations, cannot increase it during time-evolution. 

The rate of change of the trace distance is defined as $\sigma\left(t,\rho^{1,2}(0)\right)=\frac{d}{dt}\mathcal{D}\left[\rho^1(t),\rho^2(t) \right]$. The BLP measure is defined as
\begin{align}
\mathcal{N}_{\text{BLP}}=\max_{\rho^{1,2}(0)}\int_{\sigma>0} \sigma\left(t,\rho^{1,2}(0)\right) dt.
\end{align}
Notice that the definition involves a maximisation over initial pairs of states $\rho^{1,2}(0)$.  A discretized version, in line with our construction in the previous section, uses a summation over time steps $\Delta t$, in which the trace distance has increased: $\mathcal{D}\left[ \rho^1(t+\Delta t),\rho^2(t+\Delta t)\right]-\mathcal{D}\left[ \rho^1(t),\rho^2(t)\right]>0$. Whilst the BLP measure does not require knowledge of the dynamical map, the maximisation over initial states in general cannot be performed exactly.

The trace distance allows for an immediate information theoretic interpretation in the sense that $\mathcal{D}\left[ \rho^1,\rho^2 \right]=1$ if the states are perfectly distinguishable while it is zero if they are identical.\cite{nielsenchuang1,nielsenchuang2}
The BLP measure exploits this by interpreting an increase of the trace distance as information backflow from the environment into the system making the states more distinguishable. Such a backflow is then identified as non-Markovian as the time-evolution of the states at that time depends on information about the states that flowed into the environment at previous times ('environment keeps memory'). We call the dynamics BLP-Markovian if such a back flow never occurs $\mathcal{N}_{\text{BLP}}=0$. 

RHP (divisibility) implies BLP Markovianity, but not the other way round.\cite{history,nmdif1,nmdif2,nmrev1,nmrev2} In particular, the family of P-divisible dynamics, for which $\mathcal{T}(t+\Delta t,t)$ is positive, but not necessarily CP, is BLP-Markovian.\cite{blp1,blp2}

\section{Setup, Model and Measure}
\label{sec:model}

As open quantum system we consider a single spin 1/2 ($S$), coupled to a spin chain environment ($E$) of length $N$ governed by a Hamiltonian of XY type: 
\beq
H_E=\sum_{m=1}^{N-1}  \frac{J}{2}\left(\sigma_m^x  \sigma_{m+1}^x +\sigma_m^y  \sigma_{m+1}^y \right) +\sum_{m=1}^{N}h \sigma_m^z,
\label{eq:HE}
\eeq
where $\sigma_m^\mu$ ($\mu\in\{x,y,z\}$) are Pauli operators acting on site $m$ and $h$ is an external magnetic field in the $z$-direction. The system is coupled to the $m_0$-th spin of the chain via an exchange interaction of strength $\Omega$

\beq
H_{SE}=\frac{\Omega}{2} \left(\tau^x  \sigma_{m_0}^x +\tau^y  \sigma_{m_0}^y \right), 
\label{eq:HQE}
\eeq
where $\tau^\mu$ are simply the Pauli operators acting on the spin.  Finally, the system Hamiltonian is

\beq
H_{S}=\Delta \tau^+ \tau^-,
\label{eq:HQ}
\eeq
where $\tau^\pm=\frac{1}{2}(\tau^x\pm i \tau^y)$.

The environment is exactly solved in terms of diagonal fermionic modes $d_k$ with energies $E_k=2J\cos{\frac{\pi k}{N+1}}+2h$ ($k=1,\dots ,N$).\cite{wonmin} In the continuum limit this gives an energy band from $2h-2J$ to $2h+2J$ with diverging density of states at the edges. The field $h$ acts like a chemical potential that allows us to move the band up and down in energy, hence, the detuning is defined as $\Delta_h=\Delta-2h$. We consider different initial states for the environment, corresponding to either ground states at different magnetic fields, or to thermal equilibrium states.

\subsection{Spin dynamics}
\label{subsec:dynMap}

Under the assumption of no initial system-environment correlations, the dynamics of an OQS is given by the dynamical map
\beq
\mathcal{T}(t)\left[ \rho(0) \right] = \text{tr}_E \left( e^{-iHt}\left[ \rho(0) \otimes \rho_E \right] e^{iHt} \right),
\label{eq:dynMap}
\eeq
where $\rho(0)$ and $\rho_E$ are the initial states of system and environment respectively and $H$ is the total Hamiltonian. $\mathcal{T}(t)$ can be obtained using quantum process tomography,\cite{processtomography} which requires knowledge of $\rho(t)$ for a number of different initial (pure) states $\rho(0)$.

The total Hamiltonian $H=H_E+H_S+H_{SE}$ conserves the total spin along the $z$-direction or, in terms of the fermion operators of the chain, $N_{\text{exc}}=\tau^+ \tau^- + \sum_k d^\dagger_k d_k$. In a slight abuse of notation, we will refer to this conserved quantity as the number of 'excitations' in the full system. Furthermore, we consider initial states of the environment which commute with the total number of fermions. Thus, using Eq.~\eqref{eq:dynMap}, the channel takes the following form

\beq
T= \begin{bmatrix}
    \mya       & 0 & 0 & \myc \\
    0   & \overline{\myb} & 0 & 0  \\
    0 & 0 & \myb & 0 \\
    1-\mya & 0 & 0 & 1-\myc
\end{bmatrix},
\label{eq:chanT}
\eeq
where $a$ and $c$ are the excited state populations at time $t$ for the open system initialized in the excited and ground state respectively and $\myb$ evolves the coherences: $\langle e | \rho(t) | g\rangle =\myb\langle e | \rho(0) | g \rangle$. We omit the time dependence of the channel elements for notational convenience. 
Here we labeled the basis elements $i=e,g$, corresponding to excited and ground states of $H_S$. The block structure of $T$ translates in a straightforward way to $dT$.

Following the procedure explained in section~\ref{sec:measures} and exploiting the structure of $dT$, we find the following. First, $A_c=0$, which simplifies Eqs.~\eqref{eq:Am} and~\eqref{eq:noise}. Also, the eigenvectors $v_i$ are independent of time and the reduced system dynamics is described by the following time-dependent differential equation:
\begin{align}
\frac{d\rho}{dt}=&iE_{\text{LS}}(t)[\rho, \tau^+ \tau^-] \label{eq:mastereq}\\
&+\gamma_1(t) (\tau_z \rho \tau_z -\rho)  \nn \\
&+\gamma_2(t) (\tau^+ \rho \tau^- - \frac{1}{2}\{\tau^- \tau^+ , \rho \}) \nn \\
&+\gamma_3(t) (\tau^- \rho \tau^+ - \frac{1}{2}\{\tau^+ \tau^- , \rho \})  \nn,
\intertext{with} 
E_{\text{LS}}(t)=&\frac{d}{dt}\operatorname{Im}\big(\log{\myb}\big) \label{eq:E_LS}\\
\gamma_1(t)=&\frac{1}{4} \frac{d}{dt} \log{\frac{\mya -\myc }{\left|\myb\right|^2}} \label{eq:gamma1}\\
\gamma_2(t)=&\frac{\mya \myc }{\mya -\myc } \cdot \frac{d}{dt} \log{\frac{\myc }{\mya }} \label{eq:gamma2}\\
\gamma_3(t)=&\frac{(1-\myc)(1-\mya)}{\mya -\myc} \cdot \frac{d}{dt} \log{\frac{1-\mya}{1-\myc}} \label{eq:gamma3},
\end{align}
where $\gamma_i(t)$\footnote{By using $\tau_z$ in Eq.~\eqref{eq:mastereq}, we chose not to normalize the Lindblad operator. This affects the value of $\gamma_1$ (see Eqs.~\eqref{eq:lindops1} and~\eqref{eq:lindops2}).} is real ($a$ and $c$ take values between $0$ and $1$) and in Eq.~\ref{eq:E_LS} we take the principal branch. The coherent part of the equation with the Lamb shift energy $E_{\text{LS}}$ is obtained after subtracting the dissipative part from $\operatorname{log}dT$. 

Since we identify non-Markovianity with the occurrence of negative rates, in the following we focus on these expressions. In particular, for $\mya -\myc \ge0$, the dynamics is non-Markovian if the time-derivative of at least one of the fractions inside the logarithms is negative.

\subsection{Conditions for deriving a 'Markovian' master equation for the spin}
\label{subsec:kessler}

It is interesting to compare the previous constructions with the usual steps followed in the quantum optics formalism in order to derive a 'Markovian' master equation, which is a (time-independent) Lindblad equation. 

The standard derivation assumes a separation of time scales between system and environment. In the simplest case, where the system is coupled to a single environment operator $R$, the environment time scale is characterized by the correlation time $\tau_c$ after which the environment correlation function $\alpha(\tau)=\operatorname{tr} \Big( \rho_E \tilde{R}(\tau) \tilde{R}(0) \Big)$, with interaction picture operators $\tilde{R}(t)$, has decayed.\cite{tannoudji} 
As an aside, by system time scale it is typically meant the time scale on which the system changes due to its interaction with the environment. Thus, it depends on the intensity of the system-environment coupling $\Omega$. The standard derivation is valid if the following condition is satisfied:\cite{tannoudji}
\beq
\Omega \tau_c \ll 1
\label{eq:tannoudji}
\eeq
This means that the coupling has a weak effect during the correlation time of the environment fluctuations or, in other words, the reduced system dynamics at a given time only weakly depends on previous system states (Markovianity) since the environment memory of those persists only on the $\tau_c$ time scale.\cite{tannoudji}

We now discuss how the validity of the 'Markovian' master equation can be checked in our model. The typical starting point for its derivation is to express the reduced system dynamics by a specific integro-differential equation, which is valid to second order in the coupling (Born approximation).\cite{RevInes} For our spin it reads:\cite{KesslerSelfCons} 
\begin{align}
\frac{d\rho(t)}{dt}=&i\left[ \rho(t), H_S \right] - \int_0^t d s \operatorname{tr}_E\Big( \left[H_{SE},\left[\tilde{H}_{SE}( s), \right. \right. \nn \\
& \left. \left. e^{-iH_S s}\rho(t- s)e^{iH_S  s} \otimes \rho_E \vphantom{\tilde{H}_{SE}} \right] \right]\Big),
\label{eq:integrodiff}
\end{align}
where $\tilde{X}(t)=e^{-i(H_S+H_E)t}X e^{i(H_S+H_E) t}$. 

To simplify this equation we use the Jordan Wigner transformation to map the spin chain to free fermions: $\sigma_i^- = u_i c_i$, where $\sigma_i^\pm=\frac{1}{2}(\sigma_i^x\pm i \sigma_i^y)$, $u_i=e^{i\pi \sum_{j=1}^{i-1} c^\dagger_j c_j }$ and $c_i$ are fermionic operators. In this language the interaction Hamiltonian reads
\beq
H_{SE}=\Omega u_{m_0}\Big( \tau^+ c_{m_0} + \text{h.c.} \Big).
\label{eq:int2}
\eeq
Eq.~\eqref{eq:integrodiff} can then be rewritten as follows:\cite{RevInes,ines2}
\begin{align}
\frac{d\rho(t)}{dt}=&i\left[ \rho(t), H_S \right] \nn \\
&+\Omega^2 \int_0^t d s \Big( \alpha^+(s) e^{-i\Delta s}\left[ \tau^+ e^{-iH_S s}\rho(t- s)e^{iH_S  s},\tau^- \right]\nn \\
&  + \alpha^-(s) e^{i\Delta s}\left[ \tau^- e^{-iH_S s}\rho(t- s)e^{iH_S  s},\tau^+ \right] +  \text{h.c.} \Big),
\label{eq:integrodiff2}
\end{align}
where
\begin{align}
\alpha^+(t)=&\operatorname{tr}_E \Big( \rho_E c^\dagger_{m_0} u_{m_0} \tilde{u}_{m_0}(t)  \tilde{c}_{m_0}(t) \Big) \label{eq:alphap}\\
\alpha^-(t)=&\operatorname{tr}_E \Big( \rho_E c_{m_0} u_{m_0} \tilde{u}_{m_0}(t) \tilde{c}^\dagger_{m_0}(t) \Big) \label{eq:alpham}
\end{align}
are the environment correlation functions in our model.

From Eq.~\eqref{eq:integrodiff2} we see that if the kernels $\operatorname{Re}(\alpha^+(t)e^{-i\Delta t})$ and $\operatorname{Re}(\alpha^-(t)e^{i\Delta t})$ decay sufficiently fast, i.e. $\Omega \tau_c \ll 1$, the integrand has decayed before the coupling has had a significant effect on the evolution of the spin such that we can replace $\rho(t-s)$ with its unperturbed evolution. The equation is then time-local in the sense that it does not depend on the state of the spin at previous times ('memoryless') and can be further manipulated until the 'Markovian' master equation is obtained. 

In Appendix~\ref{app:Tao} we show how to compute the correlation functions for thermal and ground states of the chain. They can be written $\alpha^\pm(t)=\sum_{k=1}^N e^{\pm i E_k t}\alpha^\pm_k(t)$ such that the kernels of Eq.~\eqref{eq:integrodiff2} are:
\beq
\operatorname{Re}\Big( \alpha^\pm(t) e^{\mp i \Delta t} \Big)=\operatorname{Re}\Big(  \sum_{k=1}^N e^{\pm i (E_k-\Delta )  t}\alpha^\pm_k(t) \Big)
\label{eq:kernels},
\eeq
with $\alpha^\pm_k(t)$ given in Eqs.~\eqref{eq:alphakp} and~\eqref{eq:alphakm}. 

In certain regimes of our model, namely, when the environment is initially in the vacuum or when the spin is coupled to the first site $m_0=1$, the string operator is the identity $u_{m_0}=\mathds{1}$, and the coefficients adopt a time-independent form: $\alpha^{\rm{ns}+}_k=|W_{m_0,k}|^2 f_k$ and $\alpha^{\rm{ns}-}_k=|W_{m_0,k}|^2 (1-f_k)$, where the $N$ dimensional matrix $W$ transforms the fermion operators to diagonal modes $d_k=\sum_{i=1}^N W_{k,i} c_i$\cite{wonmin}, and $f_k=\frac{1}{1+e^{\beta E_k}}$ is the Fermi-Dirac distribution. We labeled such cases 'ns' for 'no string operator'. 

In such case, we can check condition~\eqref{eq:tannoudji} using a self-consistency argument. In the thermodynamic limit and going to energy space $W_{m_0,k}\to W_{m_0}(E)$,
we introduce the spectral density $D(E)=|W_{m_0}(E)|^2  n(E)$~\cite{RevInes, KesslerSelfCons}, where $n(E)=\frac{1}{\pi\sqrt{4J^2-(E-2h)^2}} \Theta\left( 2J-|E-2h|\right)$ is the density of states.~\cite{AlexPaper} We have $\alpha^{\rm{ns}-}(t)e^{i\Delta t}=\int dE  e^{- i(E-\Delta)t}\alpha^{\rm{ns}-}(E)$ with $\alpha^{\rm{ns}-}(E)=D(E)\big(1-f(E)\big) $, and an analogous expression for $\alpha^{\rm{ns}+}(t)$. Now the argument proceeds as follows. We assume that $\alpha^{\rm{ns}\pm}(E)$ is flat around $\Delta$ over an energy range set by the characteristic frequency of the spin $\Gamma$ (e.g. dissipation rate). Within the integral over past times $s$, Eq.~\eqref{eq:integrodiff2}, it can then be replaced by its value at $\Delta$, and, using the relation $\int dE e^{\pm i E t} = 2\pi \delta(t) - 2 i \mathds{P} \frac{1}{t}$, where $\mathds{P}$ denotes the Cauchy principal value, we can replace the respective kernel by $2\pi D(\Delta)\big(1 -f(\Delta) \big)\delta(t)$.\cite{KesslerSelfCons} Hence, under that assumption, the spin dynamics is captured by the 'Markovian' master equation with dissipation rates $\Gamma^{\rm{ns}+}=2\pi \Omega^2 D(\Delta)f(\Delta)$ and $\Gamma^{\rm{ns}-}=2\pi \Omega^2 D(\Delta)\big(1 -f(\Delta) \big)$. The self consistency argument succeeds if  
 \beq
\left| \left. \frac{\partial \alpha^{\rm{ns}\pm}(E) }{\partial E} \right|_{\Delta} \right| \Gamma \ll 1,
\label{eq:kessler}
\eeq
 where we set $\Gamma=\operatorname{max}(\Gamma^{\rm{ns}\pm})$.\cite{KesslerSelfCons}

\subsection{Numerical method}
A MPS for an open boundary system of $N$ sites with physical dimension $d$ and local basis $\{|i\rangle\}_{i=1}^d$ is a state of the form $|\Psi\rangle=\sum_{i_1,\dots,i_N=1}^d  A_1^{i_1}\dots A_N^{i_N} |i_1,\dots,i_N\rangle$, where $A_m^{i_m}$ are $D\times D$ matrices, except for the first and the last, which are $1\times D$ and $D \times 1$ vectors respectively. The bond dimension $D$ sets the number of free parameters in the ansatz\cite{mps1,mps3,mps4}. 
MPS yield good approximations to ground states of gapped, local Hamiltonians.\cite{mpsHastings,mps5} Efficient numerical algorithms exist to find MPS approximations to ground states of much more general situations, and also to simulate real time evolution.\cite{mpsSchollwock,mps2,mps6,mps7,karrasch} On the other hand, thermal states of local Hamiltonians can be efficiently approximated by an analogous ansatz in the space of operators\cite{mpo1,mpo2,mpo3}, referred to as matrix product operators (MPOs).\cite{mpoHastings,mpoAndras}

We write the state of the full system as a MPS (if the environment is initially in the ground state) or a MPO (in the thermal case), and apply standard MPS methods\cite{mpo1,mps6} to simulate real time evolution.
For convenience, we include the system in the $m_0$-th site of the chain (which then has physical dimension $d^2$).
The initial state is built as the tensor product of the desired initial spin state and the MPS (resp. MPO) approximation of the spin chain state found with standard MPS algorithms. 
In the thermal case, we evolve a purification MPS\cite{mpo1,gemma,karrasch}, which ensures positivity of the evolved state.

Bond dimension, system size and trotter step $\delta$ were chosen such that for the evolution times reported in the text the results are converged. In particular, we used $D\leq200$, $N=200$ and $\delta=0.01$.

\section{Vacuum initial state}
\label{sec:vacCase}

Let us consider that the spin is coupled to the center of the chain, $m_0=N/2$. The simplest scenario for this setup is when the chain is initialized in the fermionic vacuum $\rho_E=|0\rangle \langle 0 |$, where $d_k |0\rangle =0$ for all $k$. Note that $|0\rangle$ is equivalent to the ground state if $h>J$ ($h<-J$ yields the same physics due to particle-hole symmetry). Then the number of excitations in the total system $N_{\text{exc}}$ is set by the initial system state and the only sectors involved are those of zero and one excitations, which are not mixed under the dynamics. We can write the total system-environment state at any time as
\beq
|\Psi(t) \rangle = \left[ C_g(t)+C_e(t) \tau^+ + \sum_k C_k(t) d^\dagger_k \right] |g,0 \rangle,
\label{eq:wavefunction1exc}
\eeq
where $C_k(0)=0$ and $C_g(t)=C_g(0)$ does not evolve. The dynamical map Eq.~\eqref{eq:chanT} in this particular case has elements $\mya^{\text{vac}} =|C_e|^2$, $\myb^{\text{vac}} =C_e$, $\myc^{\text{vac}}=0$. The expressions in Eqs.~\eqref{eq:gamma1} to~\eqref{eq:gamma3} thus simplify to $\gamma^{\text{vac}}_1=\gamma^{\text{vac}}_2=0$ and $\gamma^{\text{vac}}_3=-\frac{d}{dt}\log{|C_e|^2}$. 

\begin{figure}
\begin{center}
\includegraphics[width=1\columnwidth]{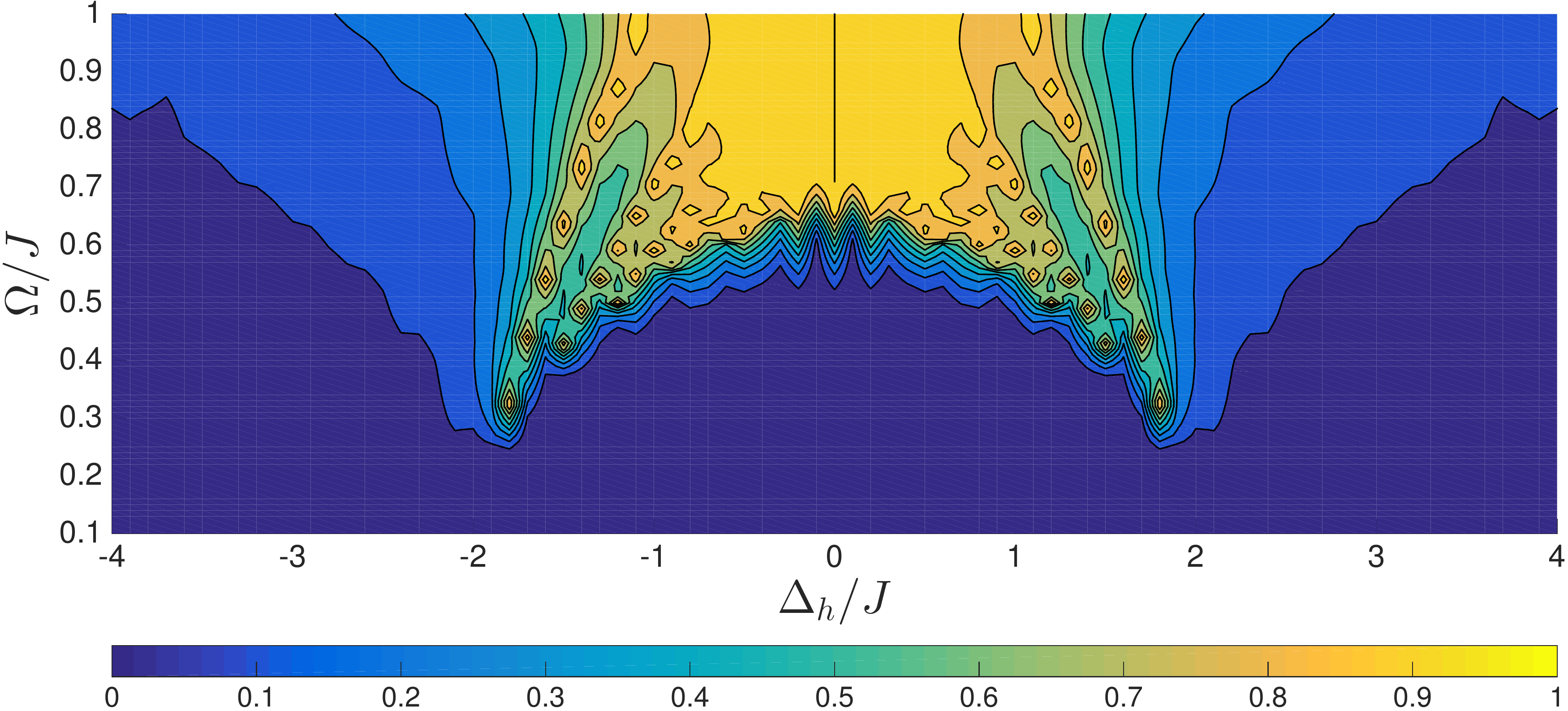}
\caption{non-Markovianity $\mathcal{N}$ for environment initially in the vacuum state, in the thermodynamic limit. The measure captures the evolution until time $ t_{\rm fin} J=20$.}
\label{fig:vacPhase}
\end{center}
\end{figure}

In this case, with at most one excitation present in the whole system, the fermionic chain we consider is completely equivalent to the bosonic one studied in reference\cite{AlexPaper}, and the analytical calculation of $C_e$ in the thermodynamic limit presented in that work is also applicable to our setup. We used this result to obtain the non-Markovianity degree $\mathcal{N}$, which we plot for a wide range of Hamiltonian parameters in Fig.~\ref{fig:vacPhase}. We find a non-Markovian region for detunings $\Delta_h$ around the band edges with a width that increases with the coupling strength. Note that for detunings in this region condition~\eqref{eq:kessler} for deriving a 'Markovian' master equation is obviously violated because the spectral density $D(E)$ diverges at the band edges. If we detune far outside the band $|\Delta_h\pm 2J |\gg \Omega$ the measure vanishes, which is what we expect because the system effectively decouples from the environment and we have a closed quantum system with coherent dynamics and thus $\gamma_i=0$ in Eq.~\eqref{eq:mastereq}. In contrast, strong coupling induces strong non-Markovian behavior because the model effectively reduces to the one of the system coupled to a single spin.~\footnote{If we would describe the system plus the spin it is coupled to, then the dynamics would be coherent and thus trivially Markovian, but this is not what we are looking at.}

\begin{figure}
\subfloat[\label{subfig:vacBothSidesOsc}]{\includegraphics[width=.24\textwidth]{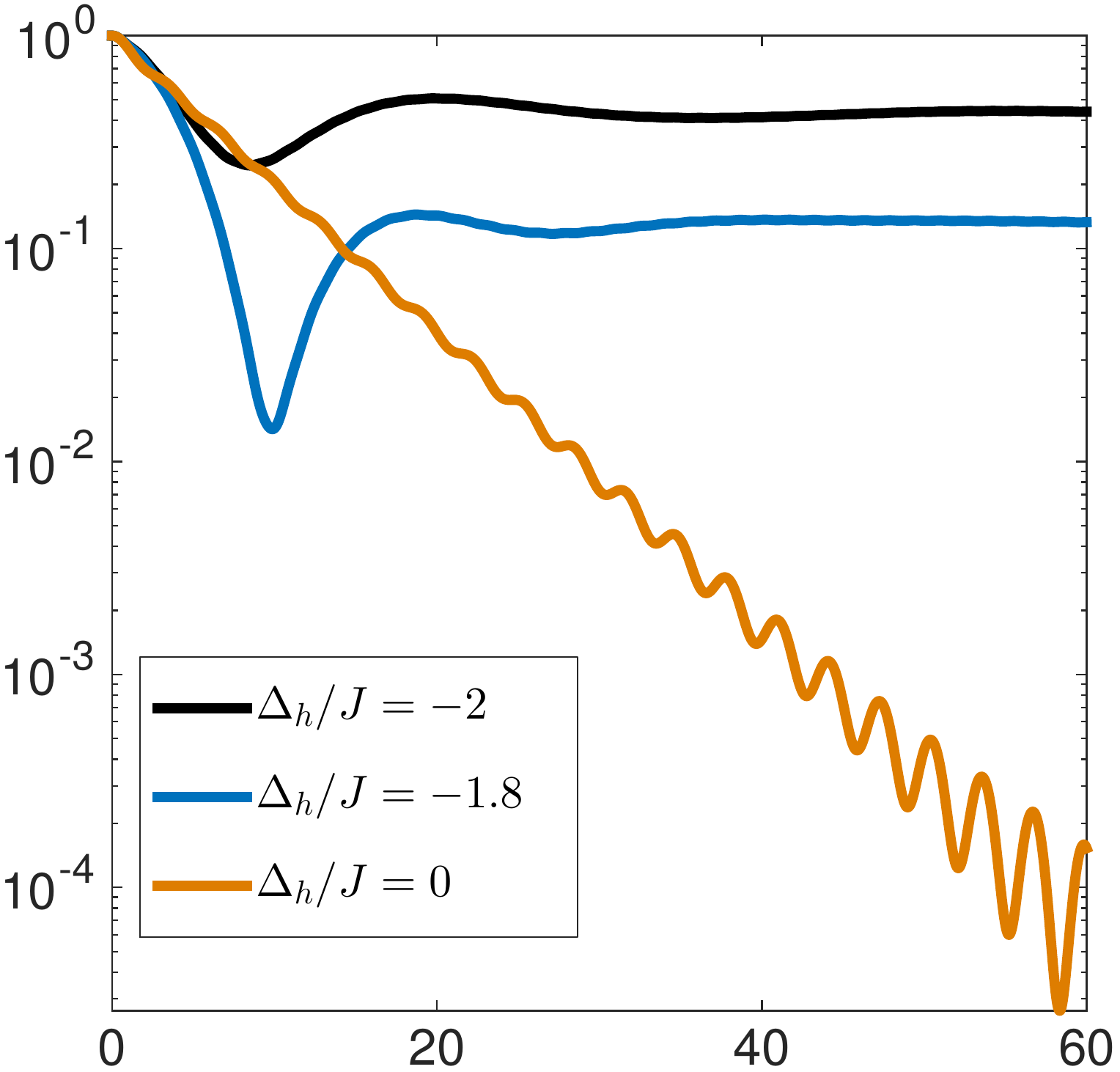}}
\hfill
\subfloat[\label{subfig:vacBothSidesRates}]{\includegraphics[width=.24\textwidth]{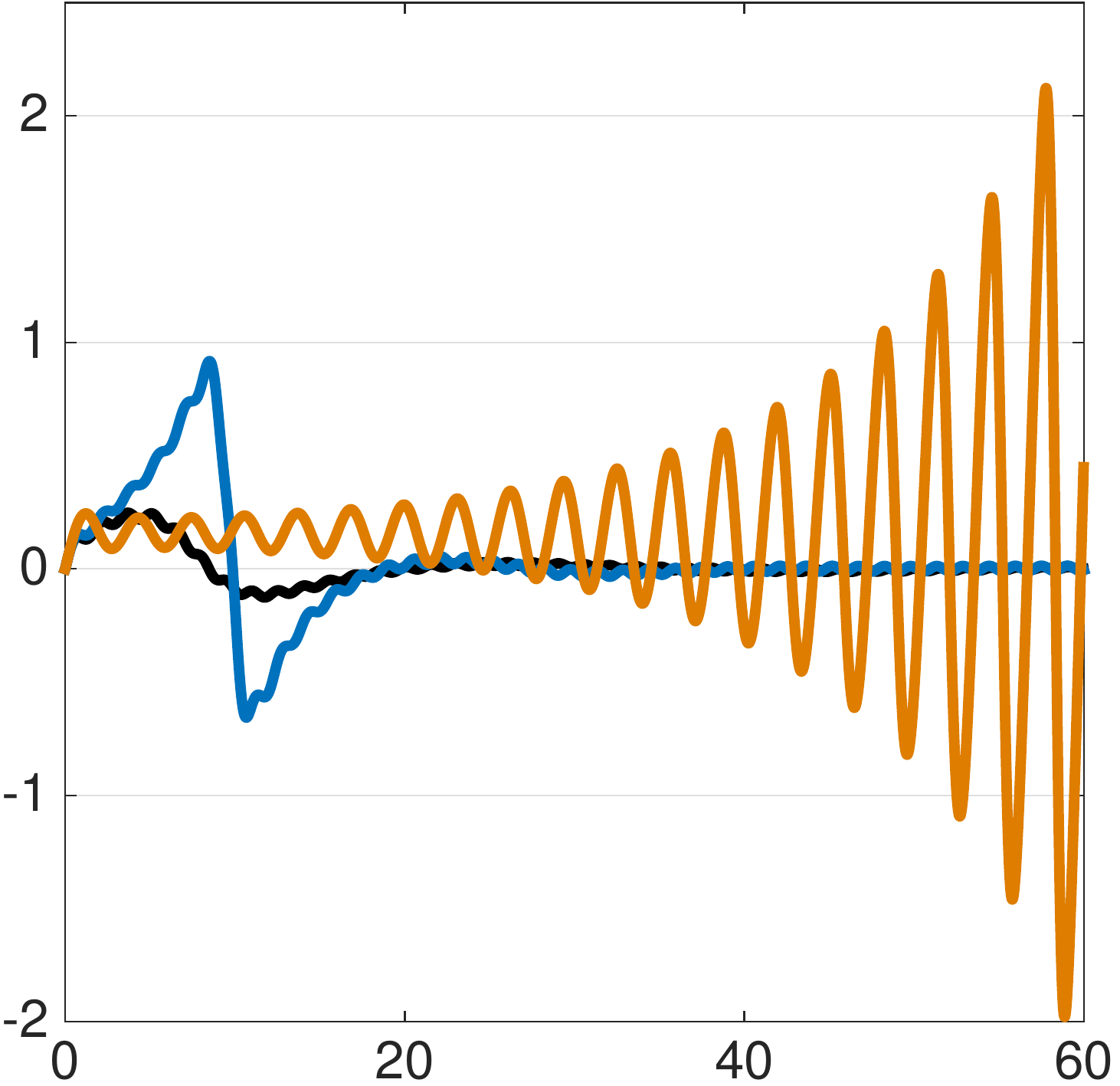}}
\caption{Time dependence of \protect\subref{subfig:vacBothSidesOsc} $|C_e|^2$ (logarithmic scale) and \protect\subref{subfig:vacBothSidesRates} $\gamma^{\text{vac}}_3$ for environment initially in the vacuum state, in the thermodynamic limit, for $\Omega/J=0.4$ at different $\Delta_h$. At the band edge (black), close to the band edge (blue) and at the band center (orange).}
\label{fig:vacBothSidesNonMark}
\end{figure}

We observe that the largest non-Markovianity is not at the band edge, but slightly shifted inside. This is due to the fact that at short times $|C_e|^2$ reaches values close to zero, smaller in this case than exactly at the band edge, leading to larger magnitude $\gamma^{\text{vac}}_3$, as shown in Fig.~\ref{fig:vacBothSidesNonMark} (black and blue lines). 

\begin{figure}
\subfloat[\label{subfig:vacoscCouplings}]{\includegraphics[width=.24\textwidth]{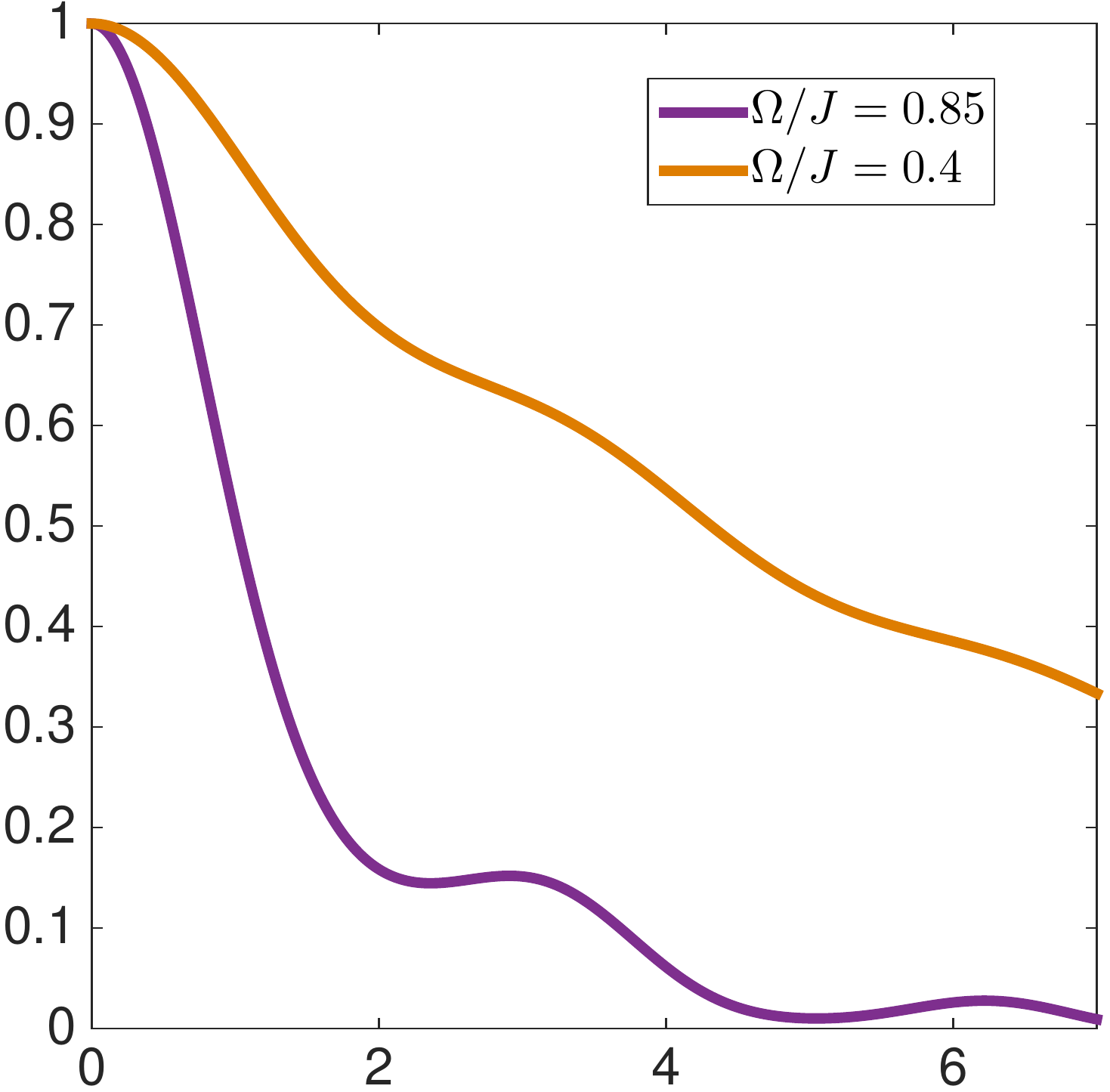}}
\hfill
\subfloat[\label{subfig:vacratesCouplings}]{\includegraphics[width=.24\textwidth]{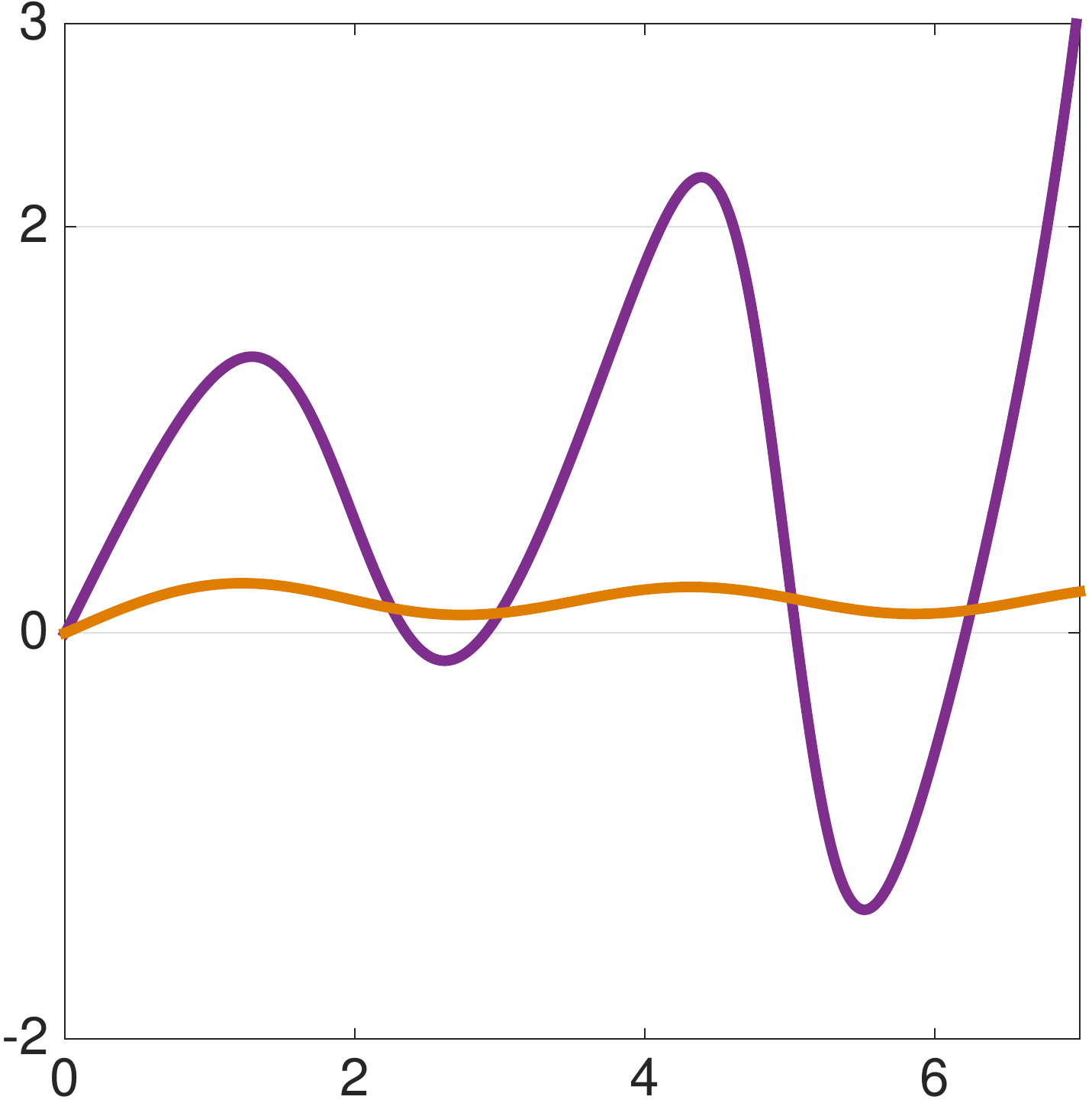}}
\caption{(Early) time dependence of \protect\subref{subfig:vacoscCouplings} $|C_e|^2$ and \protect\subref{subfig:vacratesCouplings} $\gamma^{\text{vac}}_3$ for environment initially in the vacuum state, in the thermodynamic limit, at the band center ($\Delta_h/J=0$) for different couplings.}
\label{fig:vacBandCenter}
\end{figure}

For $\Delta_h$ further inside the band, the system is more Markovian, in line with a classification based on condition~\eqref{eq:kessler}, which is closer to being satisfied due to smaller values of the spectral density and its derivative.~\footnote{For the present vacuum case ($m_0=\frac{N}{2}$) in the thermodynamic limit, the environment correlation function can be obtained in closed form using the (then equivalent) PB expression $W_{m_0,k}^{\text{PB}}=\frac{1}{\sqrt{N}} e^{-ikm_0}$. It reads $\alpha^-(t)=e^{- i2ht}J_0(2Jt)$, where $J_0(t)$ is a Bessel function of the first kind. An analogous argument as for case (ii) in Appendix~\ref{app:kernels} explains how such a slowly (power law) decaying correlation function can still induce Markovian dynamics captured by the 'Markovian' master equation at detunings away from the band edges $|\pm2-\frac{\Delta_h}{J}|\gg \frac{\Omega}{J}$. }
 In the middle of the band ($\Delta_h=0$), until intermediate times, the time dependent $|C_e|^2$ exhibits a monotonic decay, almost exponential, but modulated by oscillations (clearly appreciated in $\gamma^{\text{vac}}_3$) at a frequency approximately equal to $2J$ (Fig.~\ref{fig:vacBothSidesNonMark}, orange lines). The dynamics is captured by a 'time-dependent Markovian' master equation. Only when most of the population has decayed ($tJ \approx25$), monotonicity of $|C_e|^2$ is broken and the rate becomes negative. We focus on characterising the behavior at times before that happens and thus those late times do not enter our calculation of $\mathcal{N}$. When the coupling is increased while staying at the center of the band, the oscillations become stronger (see Fig.~\ref{fig:vacBandCenter}) until they break the monotonicity of $|C_e|^2$ and the description of the dynamics via a 'time dependent Markovian' master equation is no longer possible.

As shown in\cite{AlexPaper} (and reviewed in Appendix~\ref{app:Alex} for completeness), in the calculation of $C_e$ one identifies a number of terms which we call \emph{resonant}, \emph{edge} and \emph{bound state} contributions, since intuitively they can be connected to the overlap of the initial state $|e,0\rangle$ with continuum eigenstates of $H$ close to $\Delta$ and close to $2h\pm2J$ and with its two bound states\cite{AlexTaoBoundState} respectively. $C_e$ is the result of summing up these five contributions. The absolute square of each of these contributions is either a monotonically decaying or a constant function of time. Hence, the non-monotonicity of $|C_e|^2$ results from cross terms between pairs of those contributions, which oscillate with the difference of their corresponding frequencies $\nu$. The different non-Markovianity behavior obeys to which contributions are important for a given choice of parameters.

At short times, and for the detunings discussed above, $|C_e|^2$ is in general dominated by the exponentially decaying \emph{resonant} contribution.\cite{AlexPaper} If $\Delta_h$ is close to the band edge, also the corresponding \emph{edge} (power law decaying) and \emph{bound state} (constant magnitude) contributions become important\cite{AlexPaper}, and give rise to non-Markovianity via cross terms and to the incomplete decay of $|C_e|^2$ observed in Fig.~\ref{subfig:vacBothSidesOsc} (black and blue lines). 

If $\Delta_h$ is close to the middle of the band, the cross terms between the \emph{resonant} contribution ($\nu_{\mathrm{r}}\approx 2h$) and \emph{edge} and \emph{bound state} contributions ($\nu_{\mathrm{e}\pm}\approx \nu_{\mathrm{b}\pm}\approx 2h \pm2J$) oscillate with a frequency approximately equal to $2J$ ($|\nu_{\mathrm{r}}-\nu_{\mathrm{e}\pm}|$, $|\nu_{\mathrm{r}}-\nu_{\mathrm{b}\pm}|$). Their magnitude depends on the coupling strength, so that only if this is strong enough (see Fig.~\ref{fig:vacBandCenter}, pink lines) non-Markovianity appears at short times. However, at long times, non-Markovian behavior can appear even for weak coupling. That at long times the dynamics cannot be described by a 'Markovian' master equation, which is a stricter criterion, is well-known. There is always a transition from exponential to power law decay behavior.\cite{tannoudji}
In here we identify this non-Markovianity at long times with the relevant cross terms decaying exponentially only with half the rate with which the absolute square of the resonant contribution decays. The corresponding oscillations thus dominate at long times (Fig.~\ref{fig:vacBothSidesNonMark}, orange lines).
This is in fact a general feature of this model: no choice of parameters results in perfect Markovianity at all times, as for large enough times the constant contributions from bound states at both edges always give a non-monotonic behavior, after the other (non-constant) contributions have decayed. At sufficiently long times, when the cross terms involving the \emph{resonant} contribution have decayed, cross terms involving \emph{edge} and \emph{bound state} contributions from opposite sides that oscillate with a frequency approximately equal to $4J$ become visible ($|\nu_{\mathrm{e}+}-\nu_{\mathrm{e}-}|$, $|\nu_{\mathrm{e}+}-\nu_{\mathrm{b}-}|$, $\dots$).

We may ask how much of the non-Markovian behavior described here is detected by the BLP measure. It is in fact easy to see (see Appendix~\ref{app:blp}) that in the vacuum case, since we have $\mya -\myc =|\myb|^2=|C_e|^2$, information flows back in the sense of BLP iff $\frac{d}{dt}|C_e|^2>0$, and thus all (RHP) non-Markovian behavior is detected by the BLP measure.

\section{Excitations in the environment}
\label{sec:TnonMark}
In order to investigate how the presence of excitations in the environment affects the non-Markovianity analysis of the previous section, we study two scenarios. On the one hand, we consider the environment in a thermal state, for a chain that has the fermionic vacuum as ground state, $h=J$. This allows us to recover the previous case in the limit of low temperature $\beta J\to\infty$. On the other hand, by tuning the parameter $h$, the ground state of the chain can be chosen to contain the desired number of occupied modes. 
For these two types of thermal states, the dynamical map is still of the form of Eq.~\eqref{eq:chanT}, but it is no longer determined by a single parameter. Instead, $\gamma_1$ and $\gamma_2$, given by the general expressions in Eqs.~\eqref{eq:gamma1} and~\eqref{eq:gamma2} do not vanish, and can now become negative and give rise to non-Markovianity. 
In this case, no analytical (exact) solution is available, and the dynamical map is computed using tensor network techniques. 

\subsection{few excitations induce (new) non-Markovianity}
\label{subsec:newRates}

Thermal and ground states with few excitations correspond to populating the lower edge of the band, which can be achieved, respectively, by a low temperature ($\beta J \gg 1$) or suitable chemical potential ($0<1-h/J\ll 1$). In the rest of this section, we consider the cases $h=J$ at $\beta J=10$ and $h=0.95J$ at $\beta J\to\infty$ (ground state).

\begin{figure}
\subfloat[\label{subfig:beta10osc040}]{\includegraphics[width=.24\textwidth]{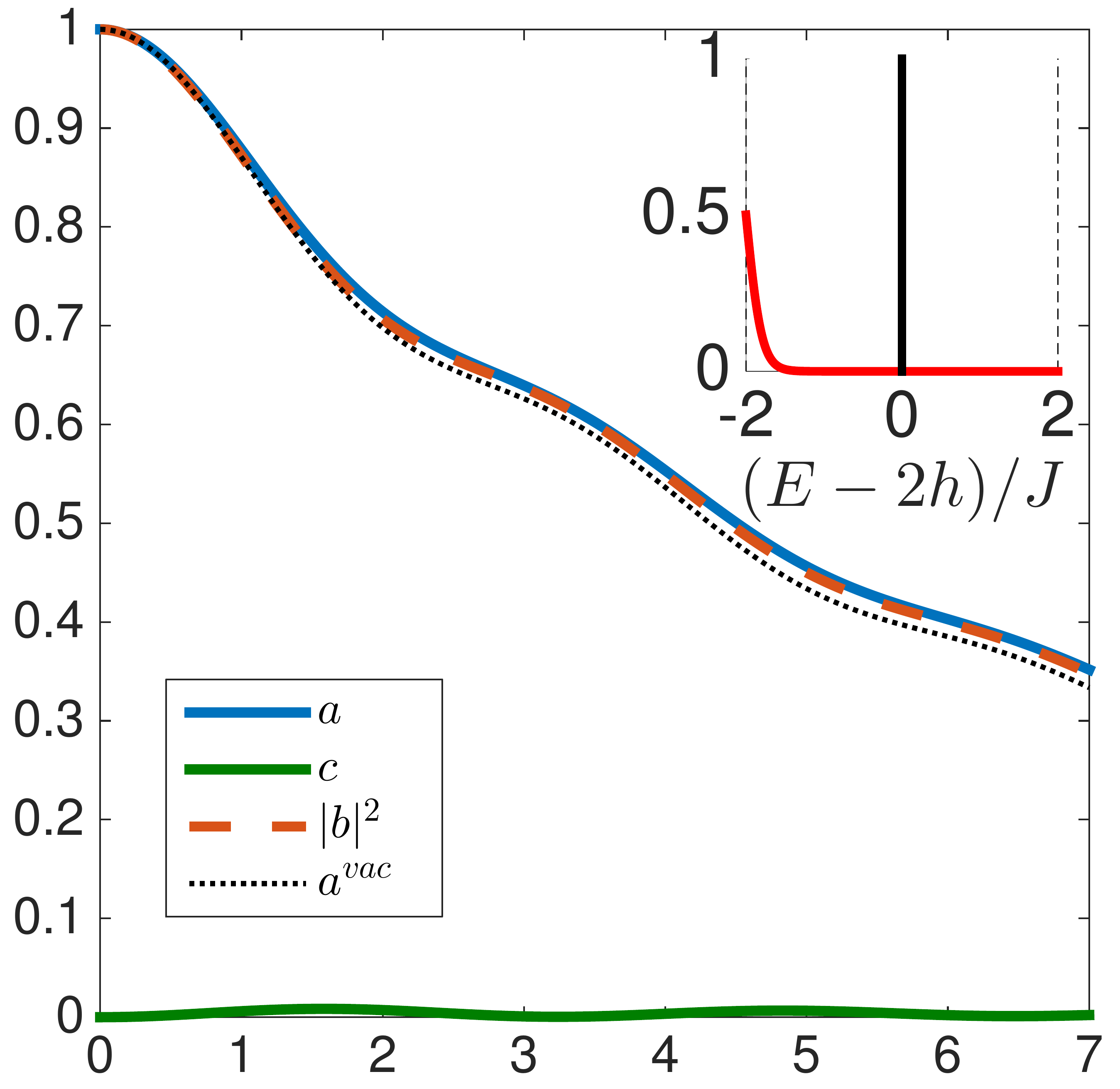}}
\hfill
\subfloat[\label{subfig:GSosc040}]{\includegraphics[width=.24\textwidth]{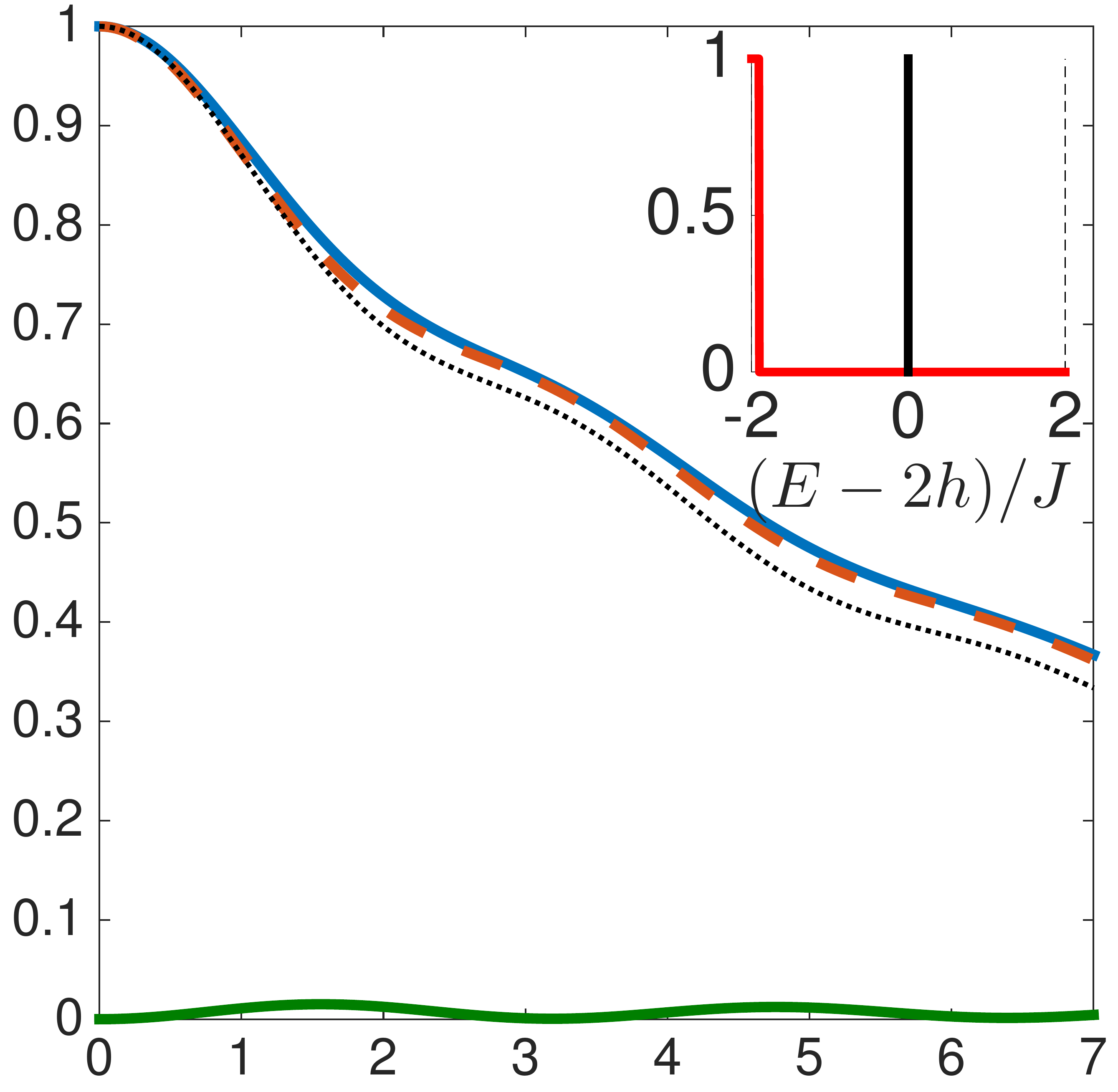}}\\
\subfloat[\label{subfig:beta10rates040}]{\includegraphics[width=.24\textwidth]{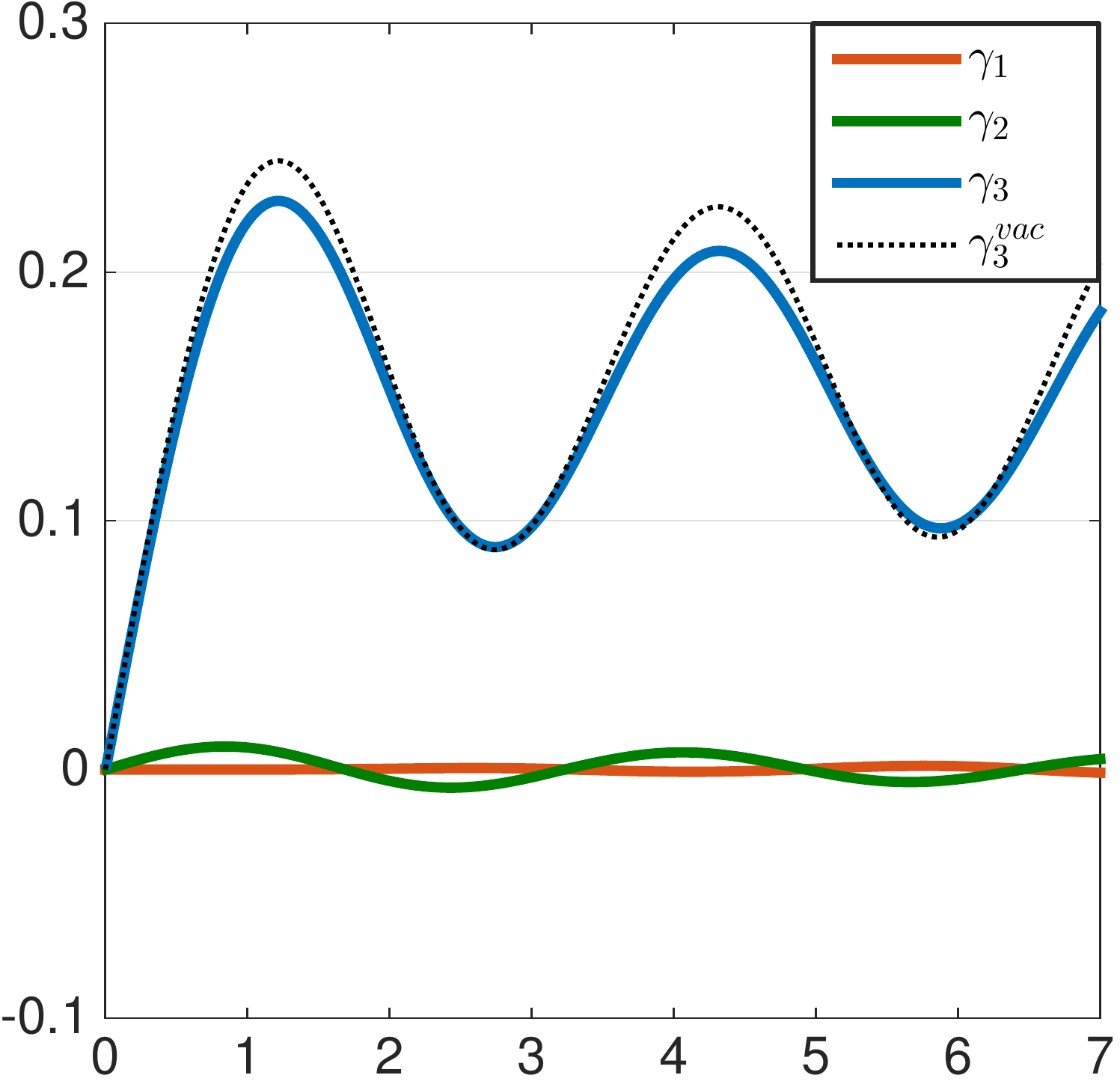}}
\hfill
\subfloat[\label{subfig:GSrates040}]{\includegraphics[width=.24\textwidth]{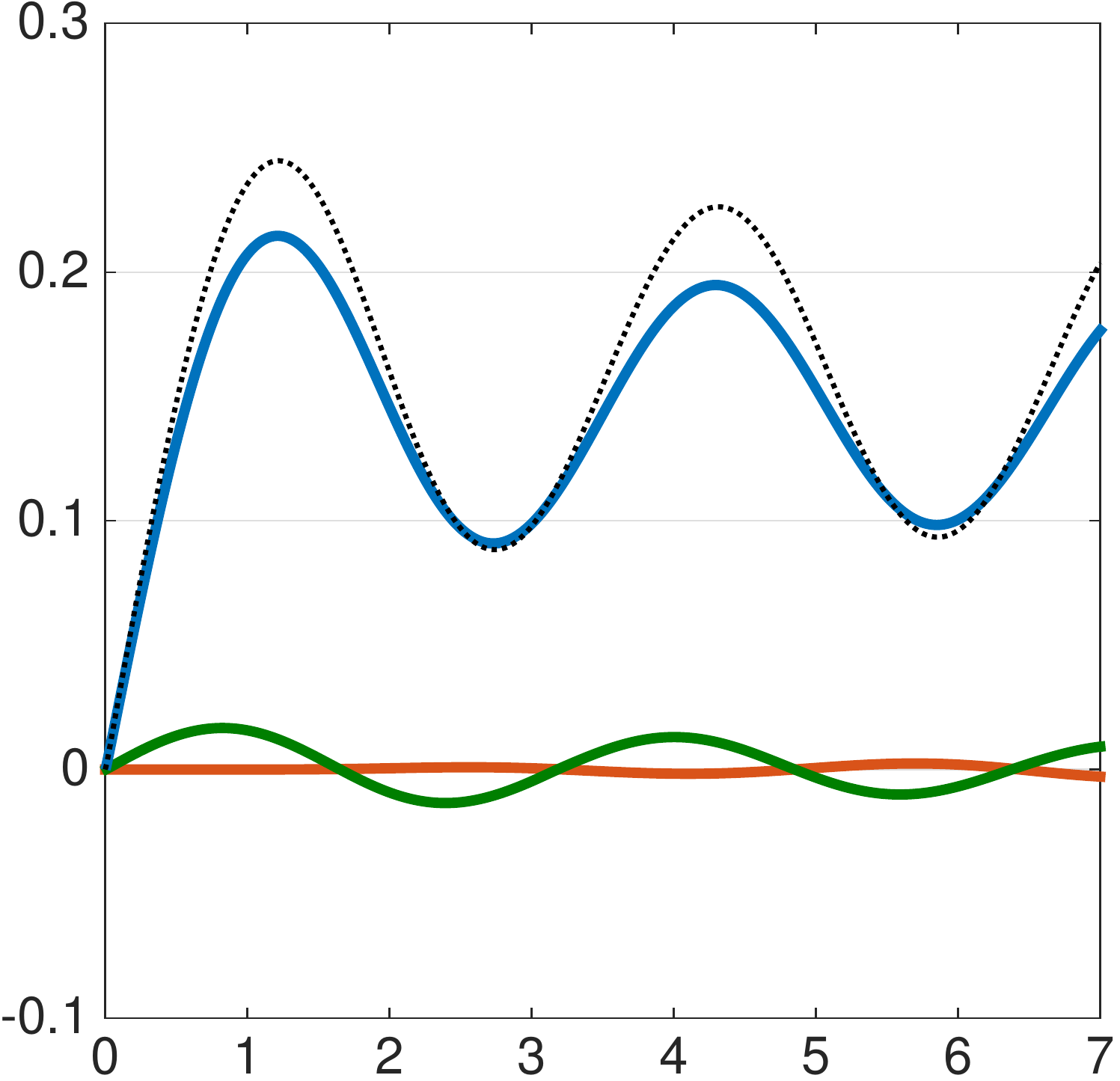}}
\caption{Time dependence of the channel elements (above) and rates (below) at the band center ($\Delta_h/J=0$) for $\Omega/J=0.4$: (\protect\subref*{subfig:beta10osc040}, \protect\subref*{subfig:beta10rates040}) Environment initially in thermal state with $\beta J=10$ ($h=J$). (\protect\subref*{subfig:GSosc040}, \protect\subref*{subfig:GSrates040}) Environment initially in the ground state for $h=0.95J$. For reference we plot the vacuum values as dotted lines (see also Fig.~\eqref{fig:vacBandCenter}, orange lines). Insets: Fermi-Dirac distribution $f(E)$ (red) from lower to upper band edge. The value of $\Delta_h/J$ is indicated as a vertical black line.}
\label{fig:excBandCenter}
\end{figure}

For $\Delta_h=0$, i.e. in the middle of the band, we observe that the map parameter $\mya $ varies very little with respect to the vacuum case (see Fig.~\ref{fig:excBandCenter}), which is not surprising, as the only excitations in the system are far off-resonant, close to the lower band edge. However, we obtain a contribution to non-Markovianity at short times from $\gamma_2$. This is originated from the monotonicity breaking oscillations of $\myc $, clearly observed in Fig.~\ref{subfig:beta10osc040}/\subref*{subfig:GSosc040}, which exhibit approximately the same frequency as the ones of $\mya $. As we argue in Appendix~\ref{app:Alex}, for the case of a single initial excitation in the environment, we may expect that the time dependence of the $c$ component of the dynamical map is determined by the same frequencies that appear in the vacuum case~\footnote{There is further contributions corresponding to the free propagation of the environment excitations, but these modify the observed frequencies in a significant way only in the case where the energy of the excitations are not close to the band edge.}, and that its oscillations may break monotonicity already at early times since the {resonance} contribution doesn't have the same dominating effect it has on $a^\text{vac}$.
Numerically, we find that this signature of the cross terms between \emph{resonant}, \emph{edge} and \emph{bound state} contributions seems to explain qualitatively also the case of few excitations.
 
\begin{figure}
\subfloat[\label{subfig:beta10crossing}]{\includegraphics[width=.24\textwidth]{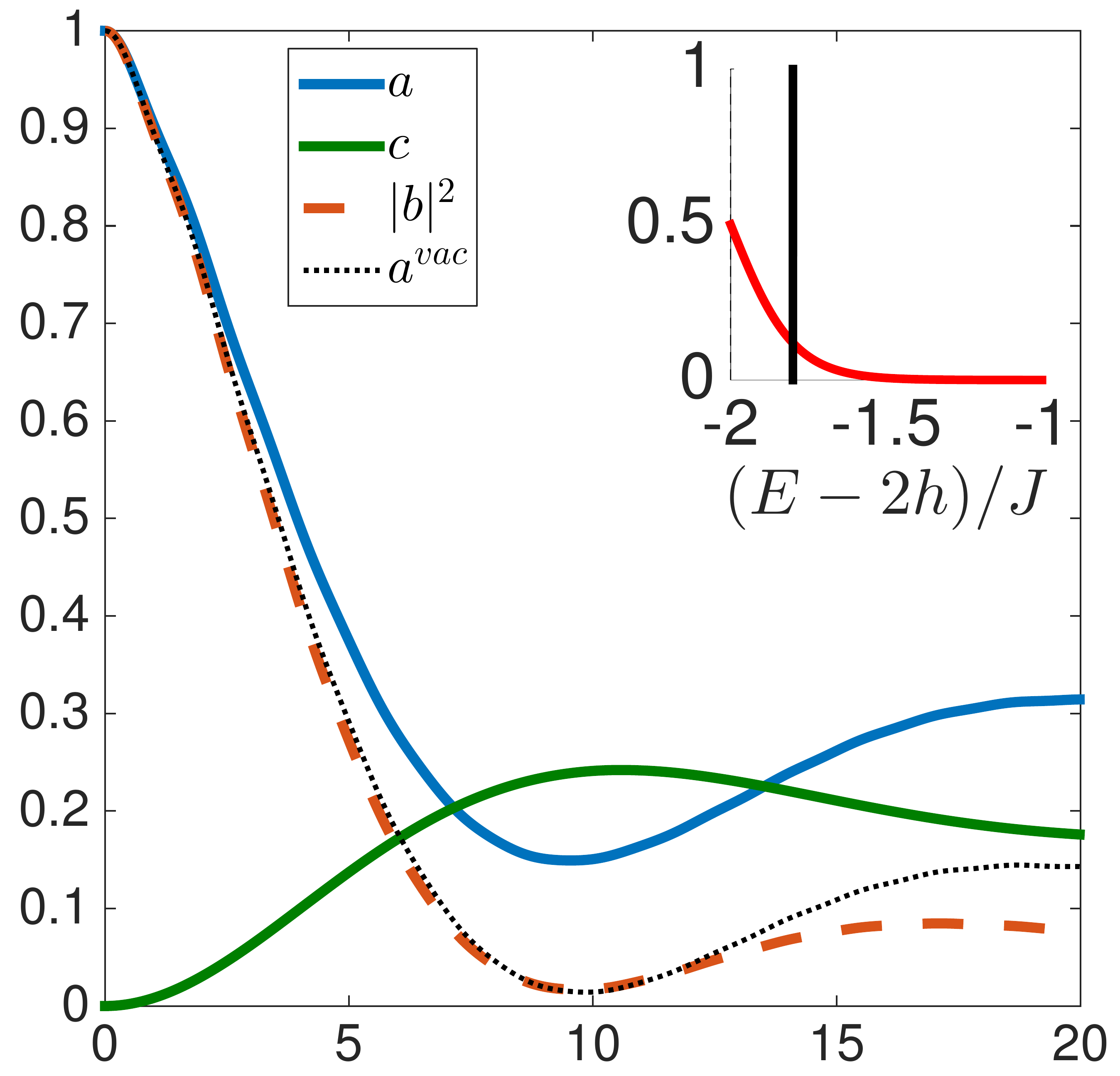}}
\hfill
\subfloat[\label{subfig:GScrossing}]{\includegraphics[width=.24\textwidth]{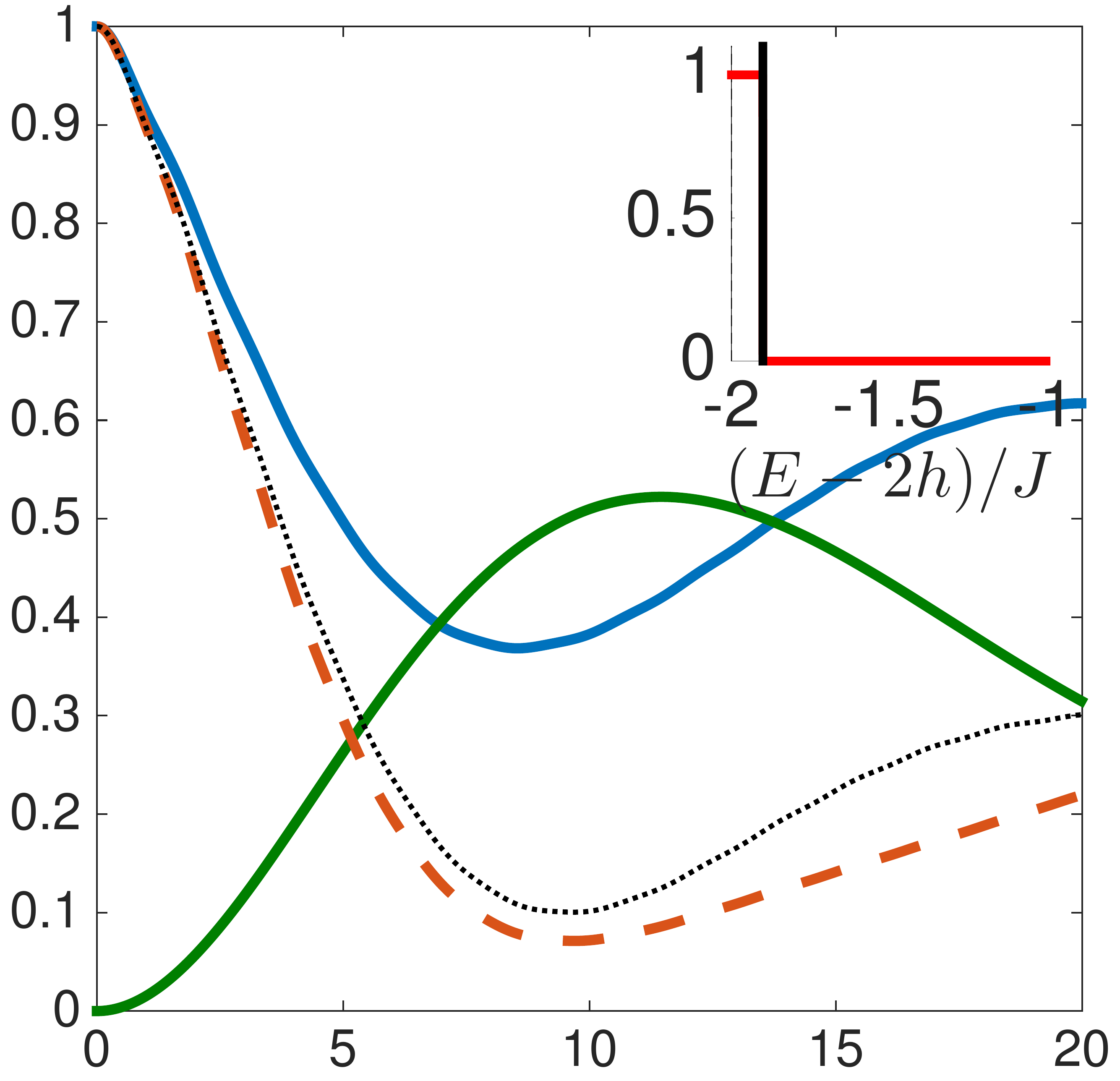}}\\
\subfloat[\label{subfig:beta10crossingrates}]{\includegraphics[width=.24\textwidth]{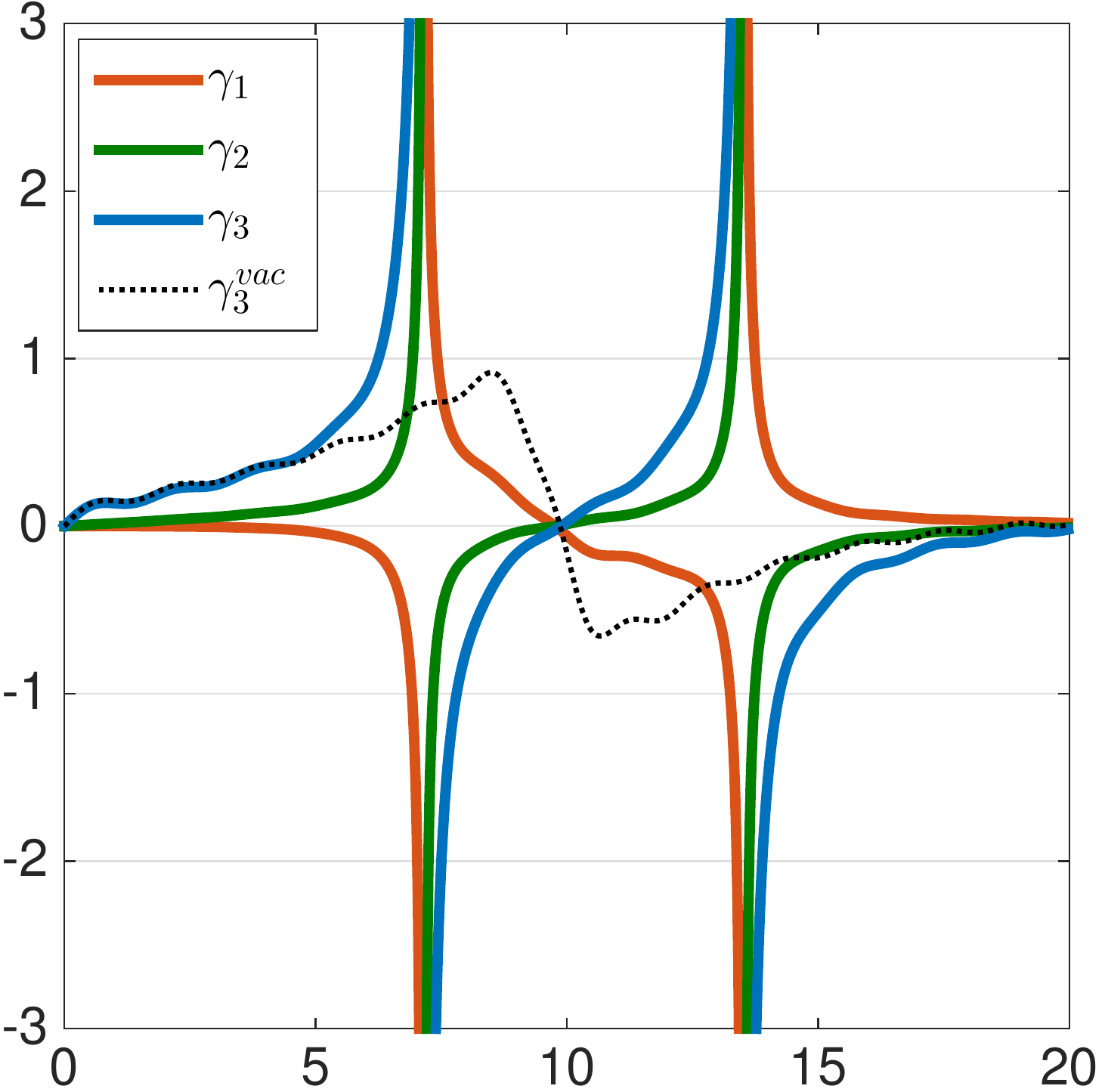}}
\hfill
\subfloat[\label{subfig:GScrossingrates}]{\includegraphics[width=.24\textwidth]{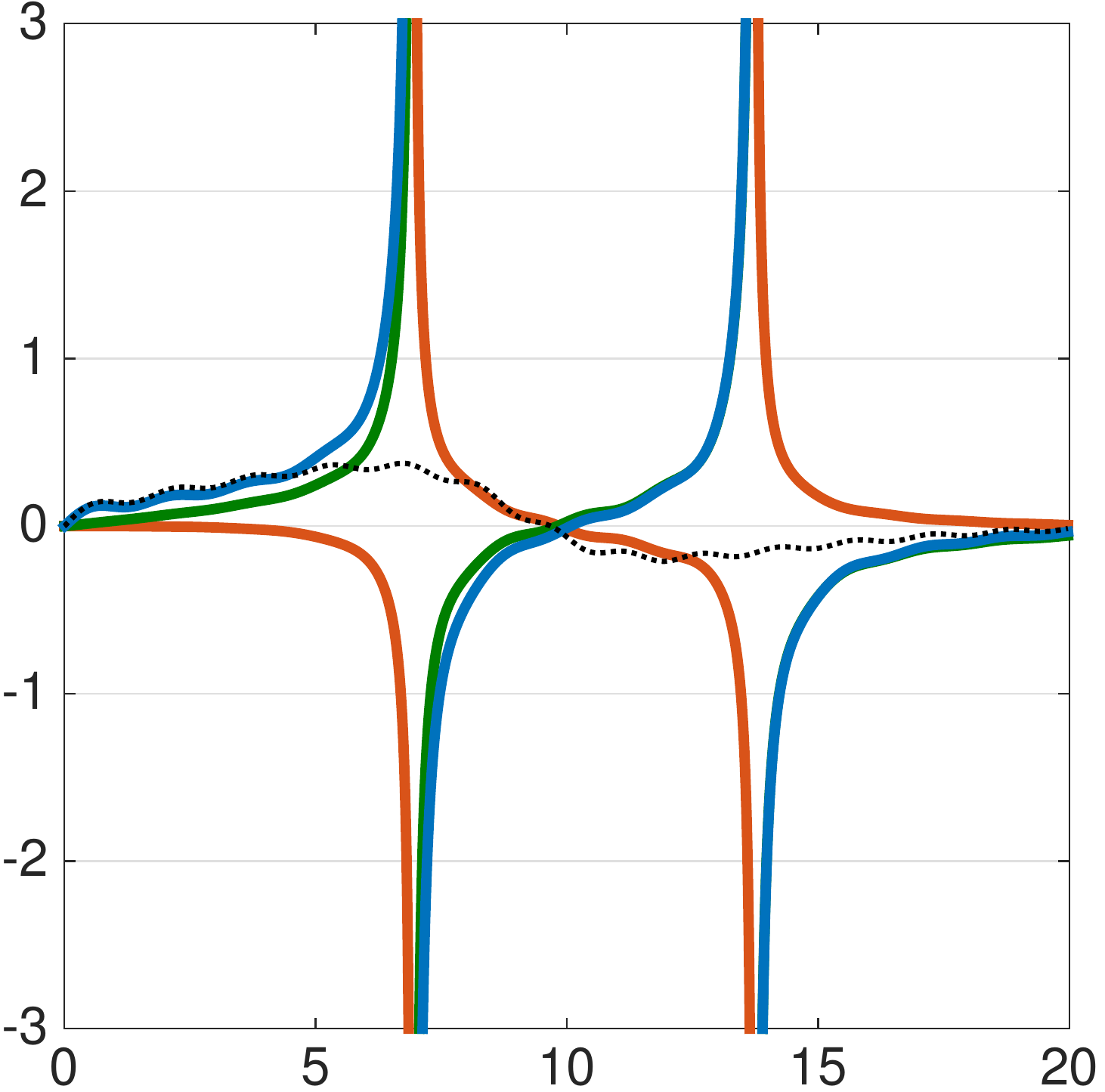}}
\caption{
Time dependence of the channel elements (above) and rates (below) close to the lower band edge for $\Omega/J=0.4$: (\protect\subref*{subfig:beta10crossing}, \protect\subref*{subfig:beta10crossingrates}) $\Delta_h/J=-1.8$ for environment initially in thermal state with $\beta J=10$ ($h=J$). (\protect\subref*{subfig:GScrossing}, \protect\subref*{subfig:GScrossingrates}) $\Delta_h/J=-1.9$ for environment initially in the ground state for $h=0.95J$. For reference we plot the vacuum values as dotted lines. Insets: Fermi-Dirac distribution $f(E)$ (red) across lower quarter of the band. The value of $\Delta_h/J$ is indicated as a vertical black line.
}
\label{fig:crossing}
\end{figure}

Next we consider setups, in which the detuning is chosen close to the lower band edge, for which the non-Markovianity was large in the vacuum case. We observe that in this case, the crossing of $\mya $ and $\myc $ results in vanishing denominators in Eqs.~\eqref{eq:gamma1} to~\eqref{eq:gamma3}, such that the rates diverge and change sign (see Fig.~\ref{fig:crossing}). The non-Markovianity in this case is thus more dramatic. Looking at Eq.~\eqref{eq:gamma1}, we notice that the early time non-Markovianity ($\gamma_1<0$) is due to $|\myb|^2$ decaying similarly whilst $\mya -\myc $ decaying faster than their respective values in the vacuum case.

It is interesting to notice that this non-divisibility at early times, corresponding to the divergence of $\gamma_1$, is not witnessed by the BLP measure. As shown in Appendix~\ref{app:blp}, there is no information backflow in the sense of BLP when $\mya -\myc \ge0$, $\frac{d}{dt}(\mya -\myc )\le0$ and $\frac{d}{dt}|\myb|^2\le0$. These conditions are in fact satisfied in this setup, until the time when $\mya $ and $\myc $ cross. They are also satisfied in the setup discussed in the previous paragraph, for $\Delta_h=0$, so that most of the new non-Markovian phenomena induced by a few excitations in the environment are not detected by BLP.

\subsection{high temperature leads to Markovian dynamics}
\label{subsec:highT}

\begin{figure}
\subfloat[\label{subfig:beta3crossing}]{\includegraphics[width=.24\textwidth]{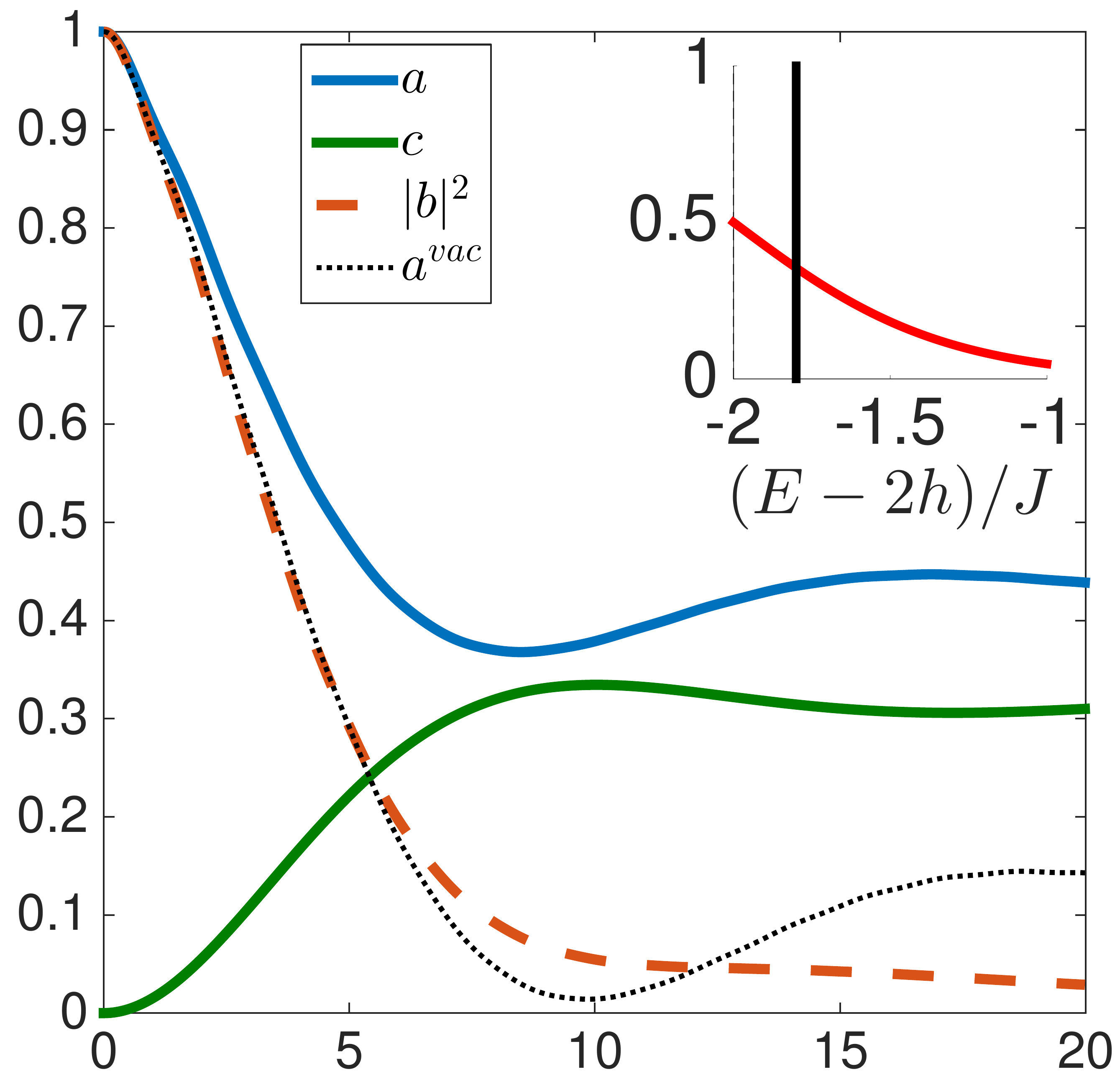}}
\hfill
\subfloat[\label{subfig:GSavoidedcrossing}]{\includegraphics[width=.24\textwidth]{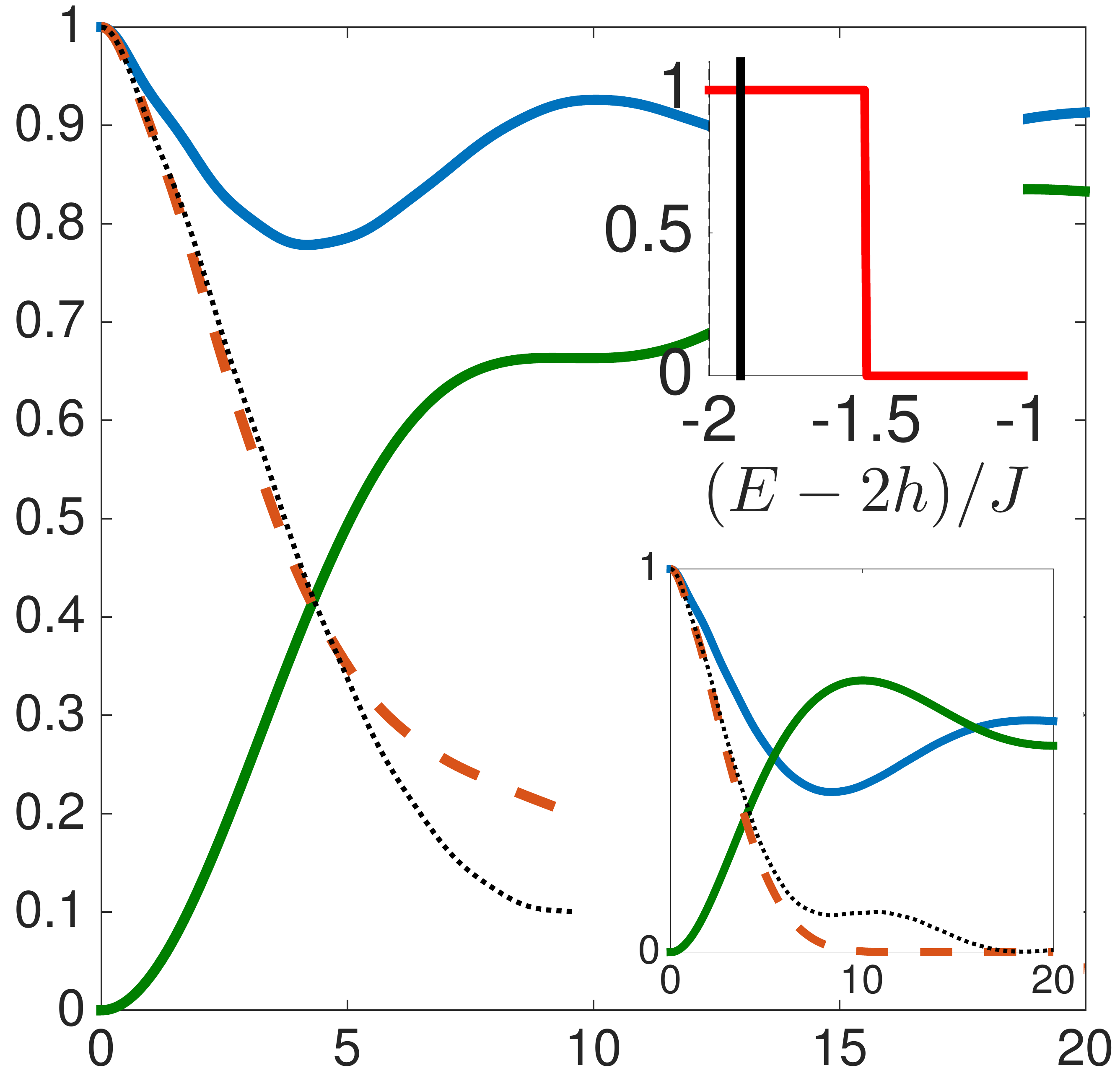}}\\
\subfloat[\label{subfig:beta3crossingrates}]{\includegraphics[width=.24\textwidth]{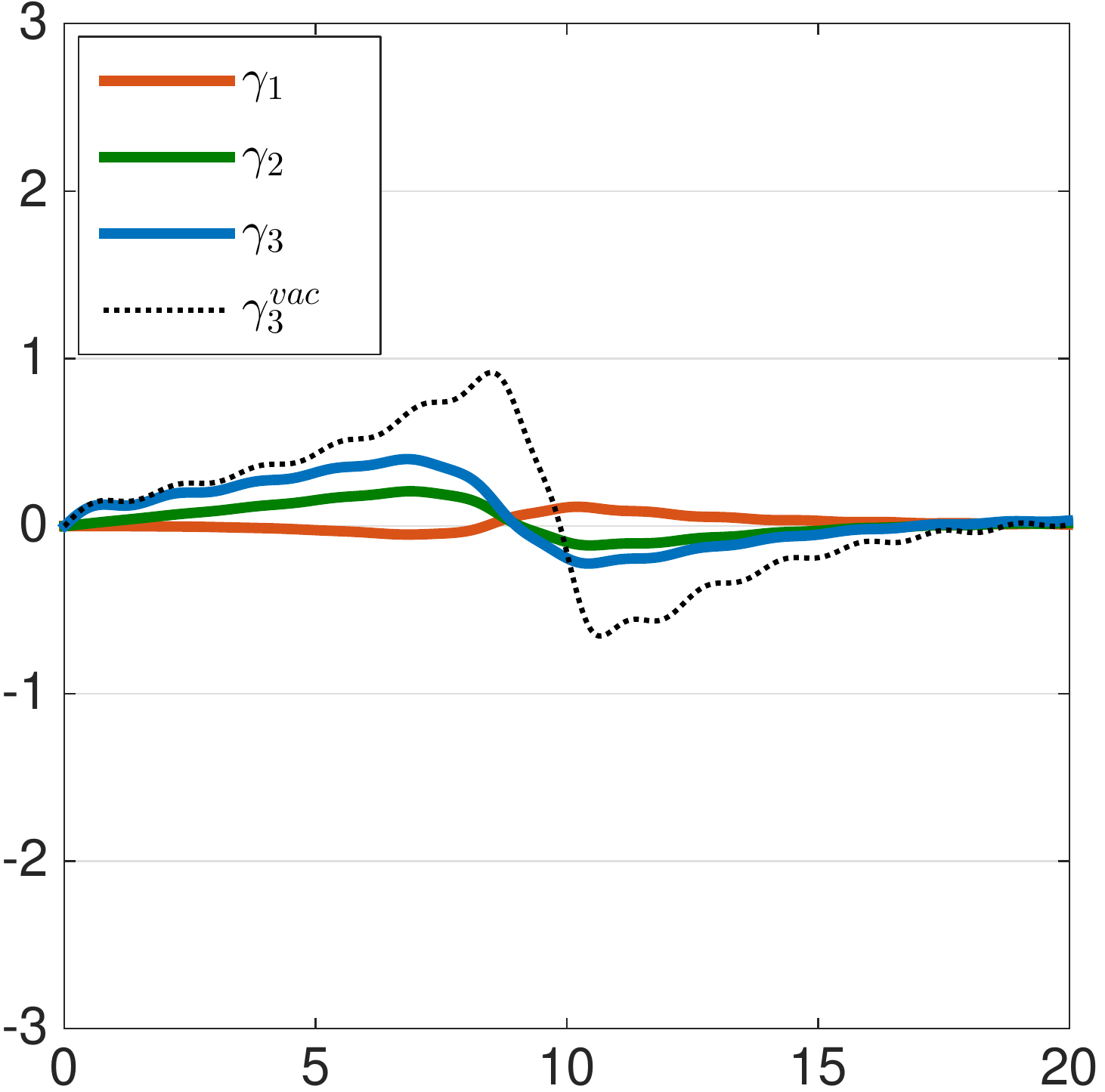}}
\hfill
\subfloat[\label{subfig:GSavoidedcrossingrates}]{\includegraphics[width=.24\textwidth]{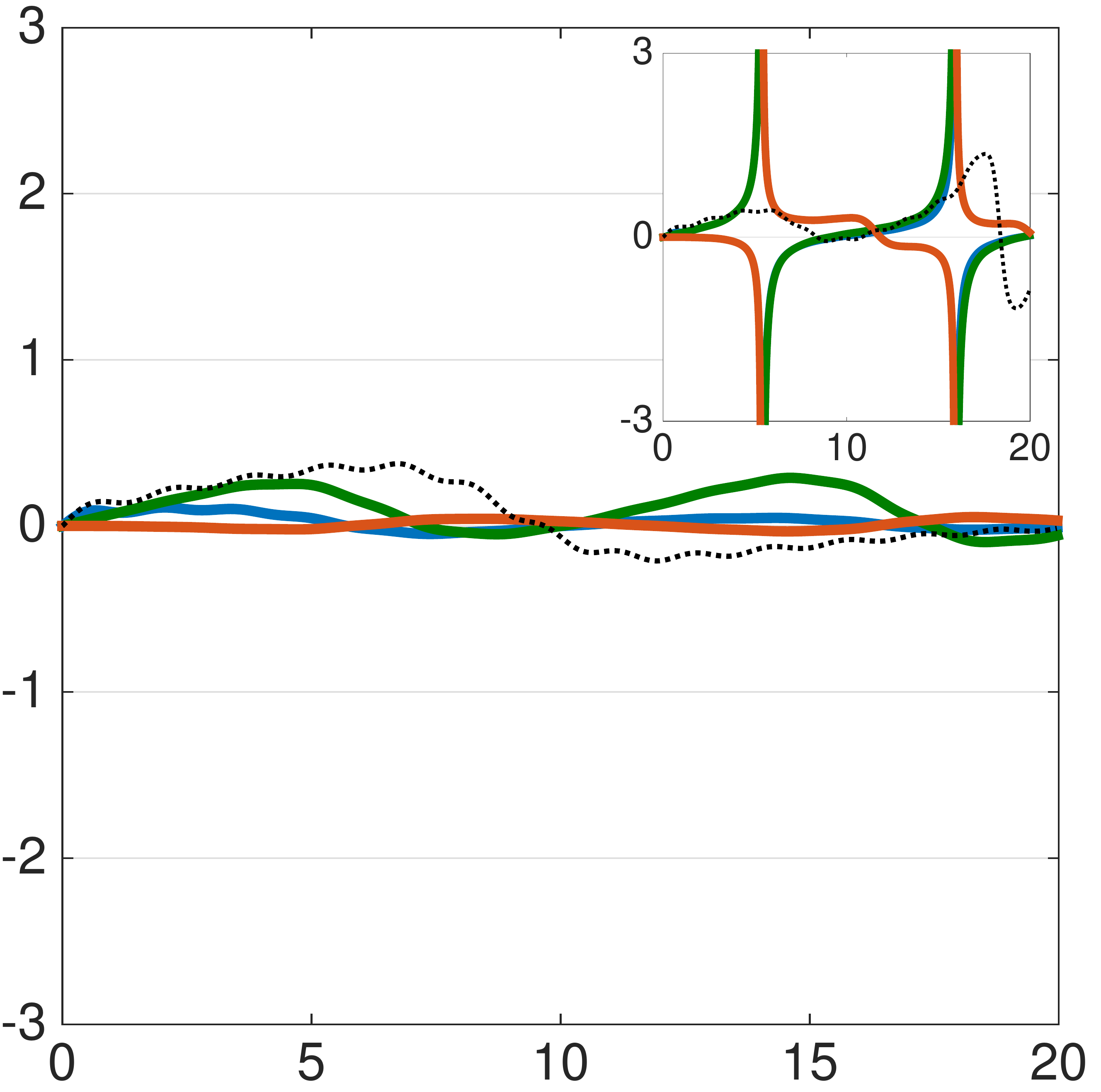}}
\caption{
Time dependence of the channel elements (above) and rates (below) close to the Fermi level for $\Omega/J=0.4$: (\protect\subref*{subfig:beta3crossing}, \protect\subref*{subfig:beta3crossingrates}) $\Delta_h/J=-1.8$ for environment initially in thermal state with $\beta J=3$ ($h=J$). (\protect\subref*{subfig:GSavoidedcrossing}, \protect\subref*{subfig:GSavoidedcrossingrates}) $\Delta_h/J=-1.9$ for environment initially in the ground state for $h=0.75 J$. For reference we plot the vacuum values as dotted lines. Insets: Fermi-Dirac distribution $f(E)$ (red) across lower quarter of the band. The value of $\Delta_h/J$ is indicated as a vertical black line. In (\protect\subref*{subfig:GSavoidedcrossing}, \protect\subref*{subfig:GSavoidedcrossingrates}) we show an additional pair of insets illustrating the case $\Delta_h/J=-1.5$ (at the Fermi level).
}
\label{fig:MoreExccrossing}
\end{figure}

We may ask how the picture changes with an increasing number of excitations in the environment, either due to a higher temperature, or to a lower chemical potential. First of all, we observe that for thermal states at a higher temperature (see Fig.~\ref{subfig:beta3crossing}/\subref*{subfig:beta3crossingrates}) and for ground states at a lower chemical potential (see Fig.~\ref{subfig:GSavoidedcrossing}/\subref*{subfig:GSavoidedcrossingrates}) the crossing of $\mya $ and $\myc $ described above does not take place. On the other hand, when, for the ground state case, we set the detuning at the Fermi level, we get a crossing (see insets in Fig.~\ref{subfig:GSavoidedcrossing}/\subref*{subfig:GSavoidedcrossingrates}).  
Together, these observations suggest that the occurrence of the crossing phenomenon, which is always accompanied by diverging non-Markovianity, is linked to the Fermi-Dirac distribution $f(E)$ changing sharply across $\Delta_h$ (compare insets in Fig.~\ref{fig:crossing} and Fig.~\ref{fig:MoreExccrossing}). 

\begin{figure}
\subfloat[\label{subfig:highTchanneldelta02}]{\includegraphics[width=.24\textwidth]{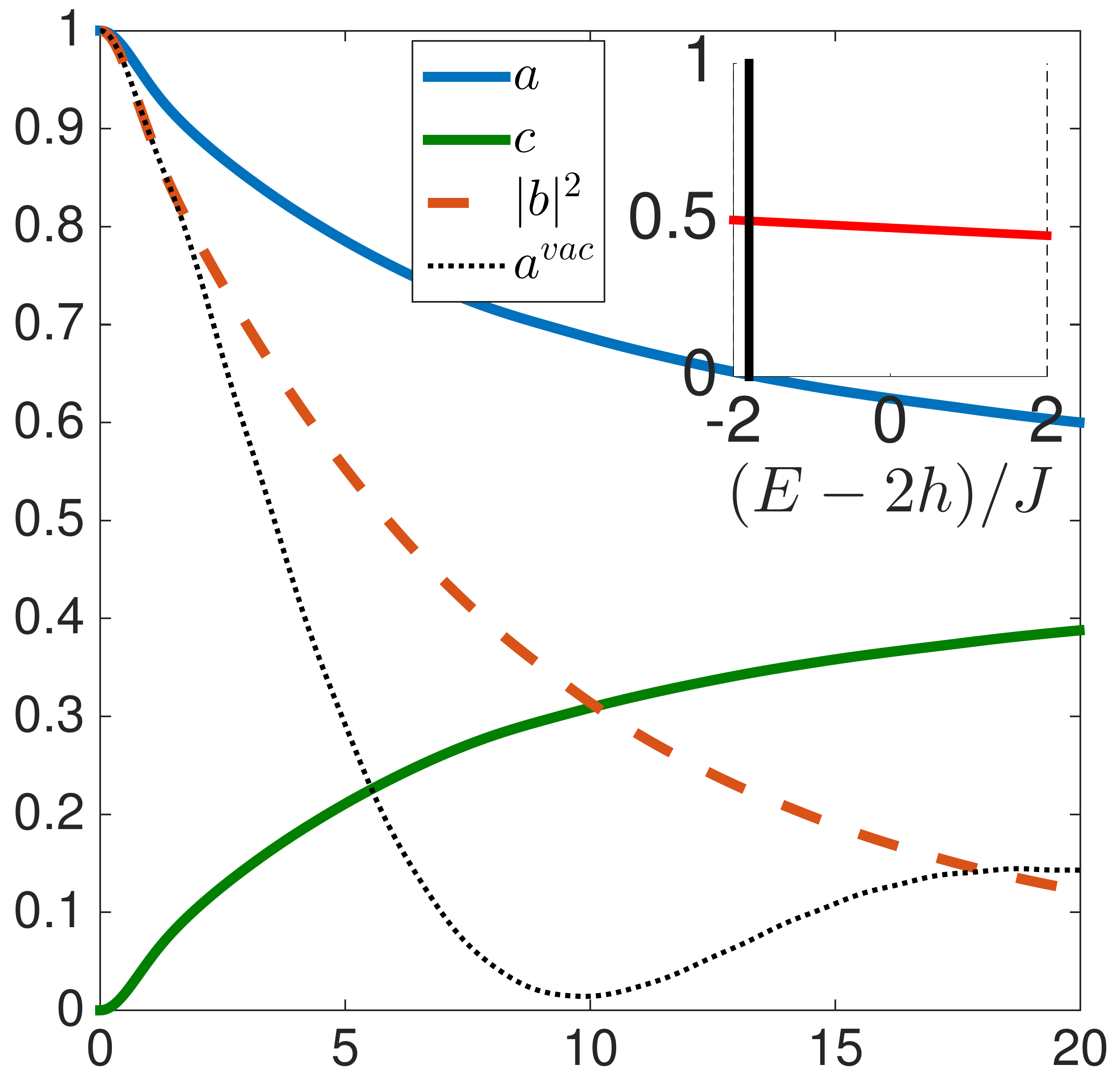}}
\hfill
\subfloat[\label{subfig:highTchanneldelta2}]{\includegraphics[width=.24\textwidth]{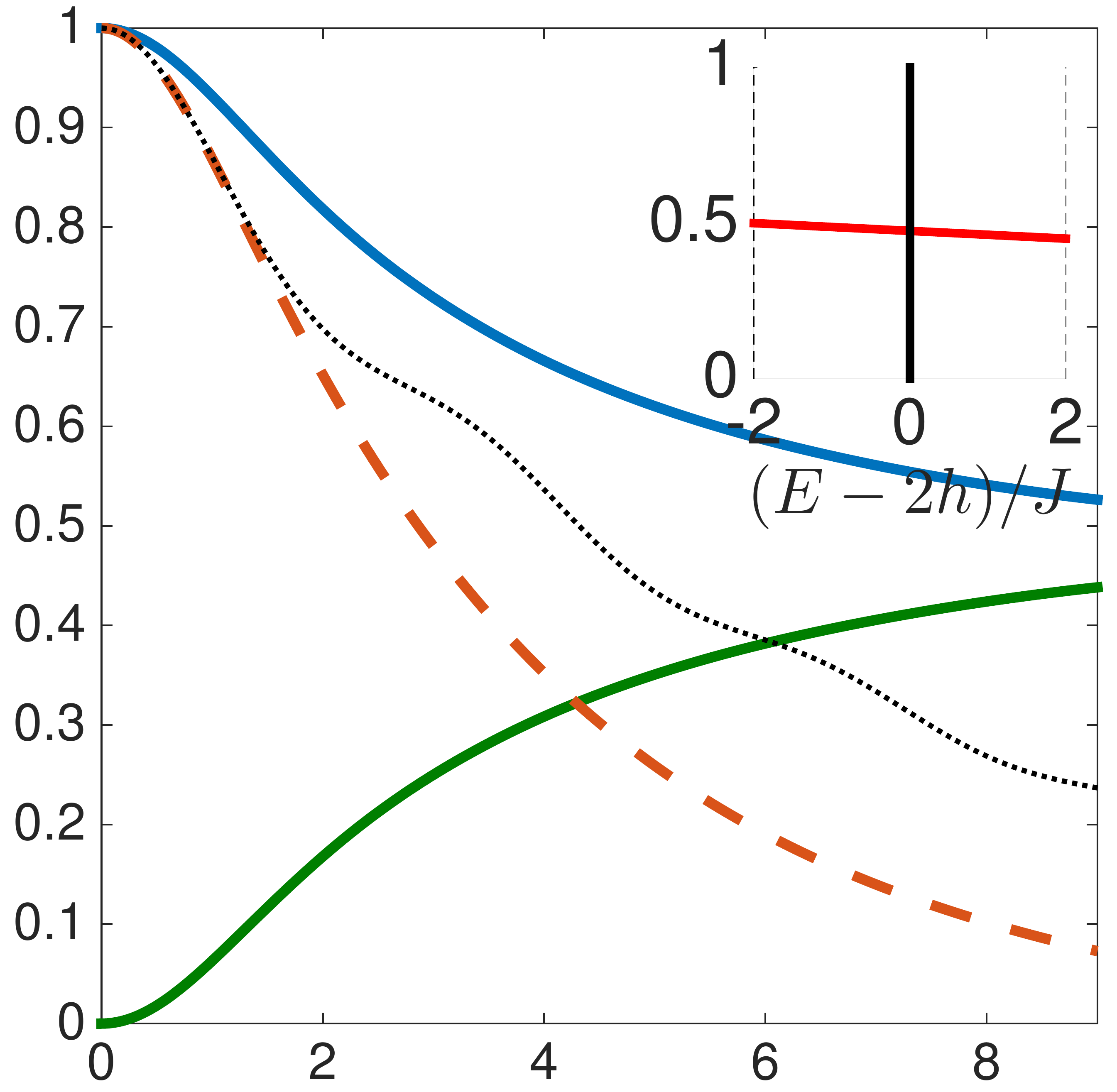}}\\
\subfloat[\label{subfig:highTratesdelta02}]{\includegraphics[width=.24\textwidth]{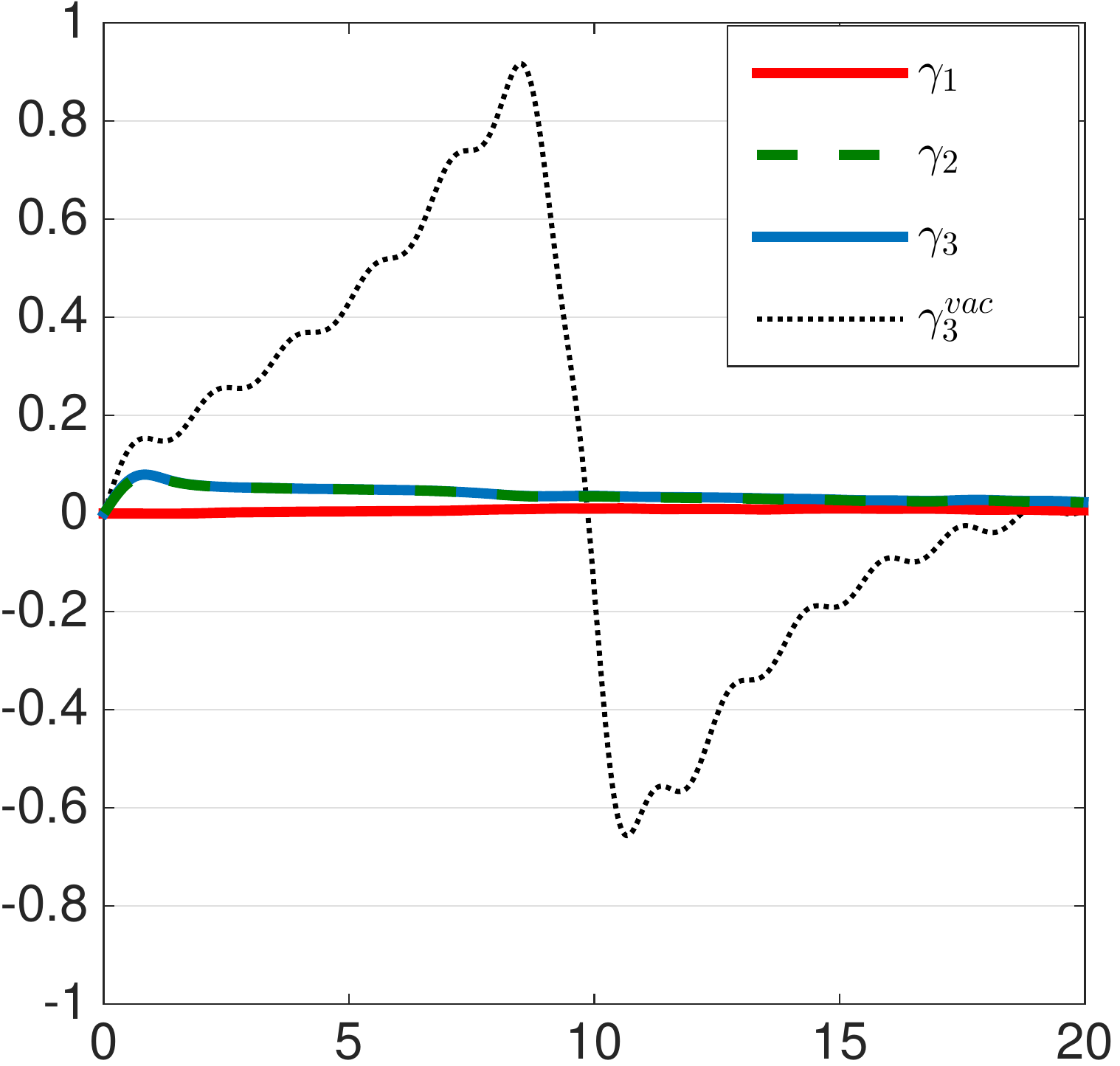}}
\hfill
\subfloat[\label{subfig:highTratesdelta2}]{\includegraphics[width=.24\textwidth]{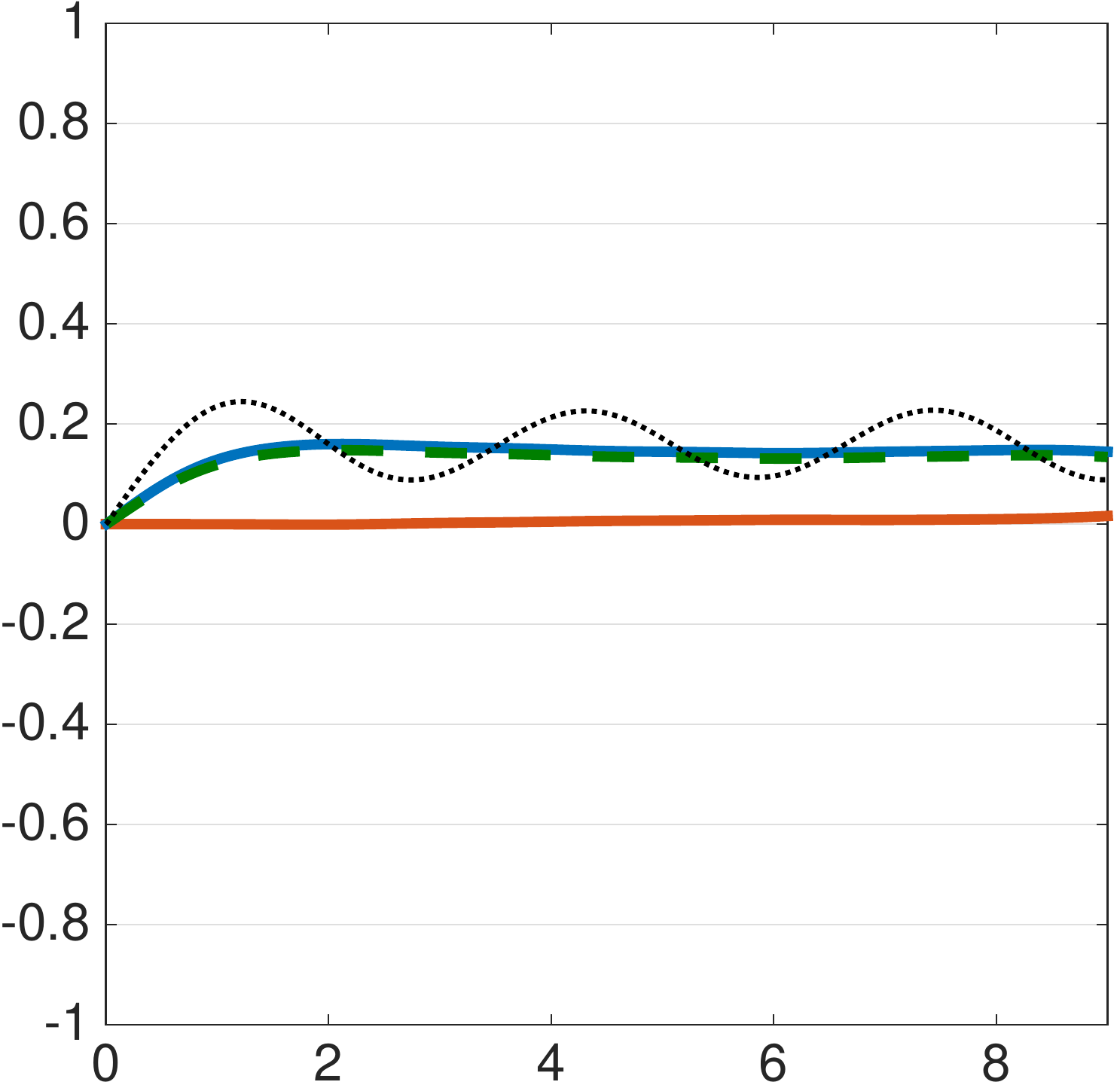}}
\caption{
Time dependence of the channel elements (above) and rates (below) for environment initially in a high temperature state ($\beta J=0.05$, $h=J$) for $\Omega/J=0.4$: (\protect\subref*{subfig:highTchanneldelta02}, \protect\subref*{subfig:highTratesdelta02}) close to the lower band edge ($\Delta_h/J=-1.8$). (\protect\subref*{subfig:highTchanneldelta2}, \protect\subref*{subfig:highTratesdelta2}) at the band center ($\Delta_h/J=0$). For reference we plot the vacuum values as dotted lines. Insets: Fermi-Dirac distribution $f(E)$ (red) from lower to upper band edge. The value of $\Delta_h/J$ is indicated as a vertical black line.
Note that in the band center case we show the results only for shorter times. This is due to the sensibility of the rates to truncation errors as $a-c$ gets small. 
}
\label{fig:highT}
\end{figure}

In Fig.~\ref{subfig:beta3crossingrates} we can also observe how the increased temperature smoothes out non-Markovian effects either introduced by few excitations (Fig.~\ref{fig:crossing}) or already present in the vacuum scenario (dotted line). It turns out that it is possible to obtain entirely Markovian dynamics if one chooses a high enough temperature. This is illustrated in Fig.~\ref{fig:highT} where we plot the transients of the channel elements and rates for system energies close to the band edge ($\Delta_h=-1.8J$, left panels) and at the band center ($\Delta_h=0$, right panels) at high temperature. We observe that the time dependence of the channel elements is monotonic and that the rates are positive for the times we can access with our simulations. 

\begin{figure}
\subfloat[\label{subfig:corrwithstring}]{\includegraphics[width=.24\textwidth]{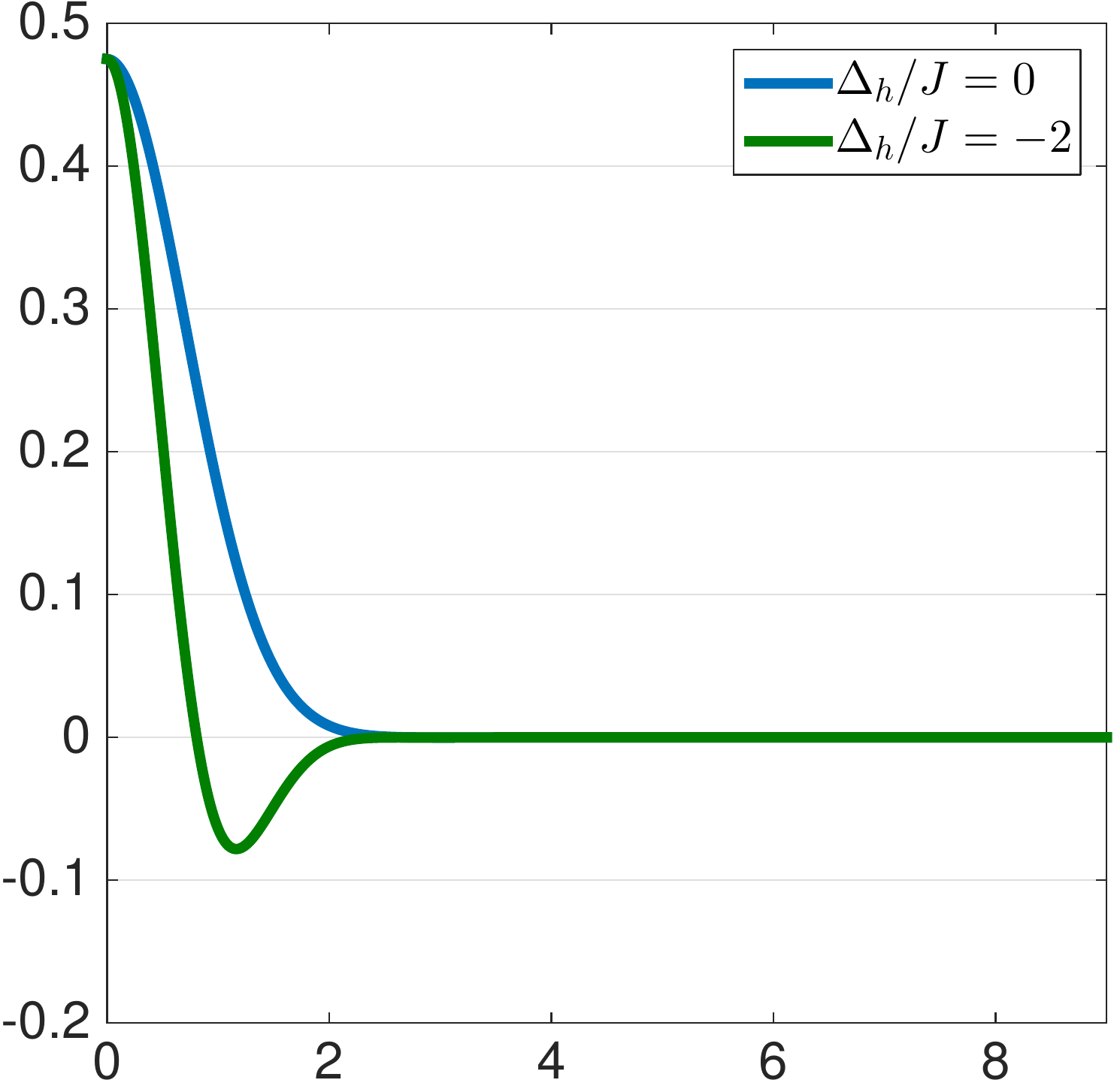}}
\caption{$\operatorname{Re}\big(\alpha^+(t) e^{-i\Delta t} \big)$ for environment initially in a high temperature state ($\beta J=0.05$, $h=J$). We plot the band center ($\Delta_h/J=0$) and lower band edge ($\Delta_h/J=-2$) cases.  
}
\label{fig:corr}
\end{figure}

One might ask if this Markovian behavior may be anticipated in the sense that condition~\eqref{eq:tannoudji} for deriving the 'Markovian' master equation is satisfied at high temperature. Using the explicit expression Eq.~\eqref{eq:alphakp}, we compute 
one of the kernels Eq.~\eqref{eq:kernels} for $\beta J=0.05$ (see Fig.~\ref{fig:corr}), and find that it decays rapidly to zero (at such high temperature the other one behaves qualitatively the same). Remarkably, the environment correlation time $\tau_c$ is essentially the same at the center and the edge of the band, although the corresponding spectral densities are very different (diverges for the latter), and, in the literature, the flatness of the spectral density is often associated to short correlation times.\cite{KesslerSelfCons,ClosBreuer,znidaric,bandgap}  At infinite temperature we can write a closed form for the correlation functions: $\alpha^\pm(t)=\frac{1}{2}e^{\pm i2ht}e^{-J^2t^2}$.\cite{spinspin} 
Their superexponential decay confirms that at high temperatures and small enough (but still intermediate) coupling strength (which sets the system time scale), the 'Markovian' master equation becomes a valid description at all detunings.~\footnote{It is interesting to contrast this with a model of a fermionic state coupled to a chain of free fermions, i.e. in the interaction term Eq.~\eqref{eq:int2} we exchange $\tau^-$ by a fermionic annihilation operator and remove the string operator $u_{m0}$. The infinite temperature environment correlation functions are then proportional to the one of the vacuum case of our model (Bessel function), and hence, as illustrated in Appendix~\ref{app:kernels}, the 'Markovian' master equation is only valid at detunings away from the band edges $|\pm 2 - \frac{\Delta_h}{J}|\gg \frac{\Omega}{J}$. Note that from the perspective of environment correlation functions the only difference to our spin model (if expressed in terms of fermionic modes Eq.~\eqref{eq:alphap}) is the absence of the string operator.}

In summary we found in this section that while few initial excitations, e.g. the ones present at small temperature, introduce a number of new non-Markovian features (early time negative $\gamma_2$ for $\Delta_h$ deep within the band and early time diverging negative $\gamma_1$ for $\Delta_h$ close to the lower band edge), these phenomena get smoothed out together with any non-Markovian features already present in the vacuum case as one increases the temperature to large values such that, at high temperature, we obtain divisible dynamics. We found that this high temperature Markovianity could already be anticipated from the observation that the conditions for deriving the 'Markovian' master equation are satisfied at intermediate couplings. The Markovianity at high temperatures is nevertheless not a completely general effect, as we discuss in the next section.

\section{Exactly solvable setup}
\label{sec:Apollaro}

If the system is coupled to the first site of the chain ($m_0=1$), the full model can be solved analytically, as it can be mapped to a quadratic fermionic Hamiltonian.
Using the exact solution, we analyze here the divisibility properties of this setup, and compare them with the case
discussed in the previous sections. 
Notice that the non-Markovianity in this scenario was already studied in reference\cite{apollaro}, with the chain initially in a ground state,
but with a focus on the (less strict) BLP measure, which in some cases gives a qualitatively different picture. 

The dynamical map is still of the form of Eq.~\eqref{eq:chanT} and we compute explicit expressions for its elements using Gaussian methods for free fermions (see Appendix~\ref{app:ApollaroFreeFermions}). We find that in this case  $\mya -\myc $ is the same for any environment initial state that is a ground state with different $h$ or any other thermal state, i.e., $\mya -\myc =\mya^{\text{vac}} $. Thus, different from the setup considered in the previous section, introducing a few excitations cannot affect the non-Markovianity dramatically (see Eqs.~\eqref{eq:gamma1} to~\eqref{eq:gamma3}), in particular, the crossing between $a$ and $c$ cannot happen.
Also, $\mya -\myc =|\myb|^2$ for the environment initial states we consider, which implies $\gamma_1=0$ Eq.~\eqref{eq:gamma1}.

Under these conditions, information backflow in the sense of BLP is equivalent to having $\frac{d}{dt}(\mya -\myc )>0$ (see Appendix~\ref{app:blp}). Thus, BLP-Markovianity is independent of such environment states and it is enough to check the simplest one, i.e. the vacuum (ground state for $h>J$). For this particular case BLP-Markovianity and divisibility turn out to be equivalent, and reduce to the condition $\frac{d}{dt}\mya^{\text{vac}}\leq0$ $\forall t$, in accordance with what we found in section~\ref{sec:vacCase}. Thus, for fixed parameters $\Delta_h/J$ and $\Omega/J$, non-divisibility in the vacuum case implies BLP-non-Markovianity (and hence non-divisibility) independent of the considered environment states. This is in contrast to the setup discussed in the previous section, in particular, it is not possible to obtain Markovian dynamics at high temperatures if the corresponding vacuum case is non-divisible. 

It is interesting to compare the predictions based on the behavior of the high temperature environment correlation functions for both setups. In Appendix~\ref{app:kernels} we show that (at infinite temperature, where they can be obtained in closed form) whilst in the previous case, with its superexponentially decaying correlation functions, the derivation of the 'Markovian' master equation is valid at all detunings if $\frac{\Omega}{J}\ll1$, in the present case ($\sim t^{-\frac{3}{2}}$) this is only true at detunings away from the band edges $|\pm2-\frac{\Delta_h}{J}|\gg \frac{\Omega}{J}$.  
If we move the coupling site into the chain, our numerics show the power law decay getting steeper as $\sim t^{m_0^2+\frac{1}{2}}$ (Appendix~\ref{app:Tao}).

The simple expression for the BLP measure in this setup allows us to reproduce one of the main results of\cite{apollaro}: for each value of the coupling $\Omega/J\leq1$ there is a specific detuning $\Delta_h=2J-\frac{\Omega^2}{J}$ for which the BLP measure is identically zero for any of the environment initial states we consider. We refer to this situation as the BLP-Markovian point. For any other set of parameters the measure is different from zero, with the largest non-Markovianity occurring at the center of the band ($\Delta_h=0$).\cite{apollaro}

\begin{figure}
\subfloat[\label{subfig:ApoDivChannel}]{\includegraphics[width=.24\textwidth]{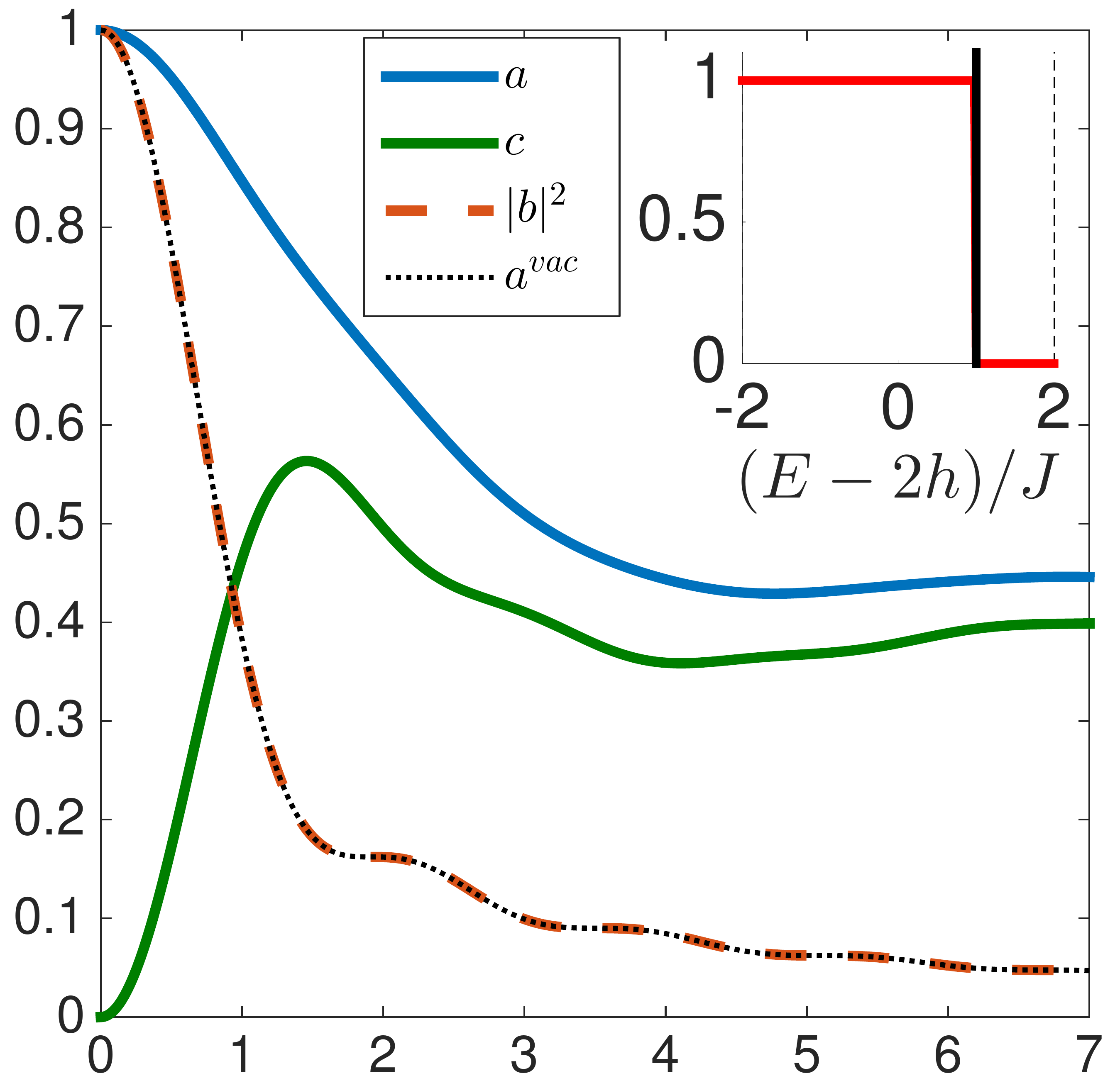}}
\hfill
\subfloat[\label{subfig:ApoDivRates}]{\includegraphics[width=.24\textwidth]{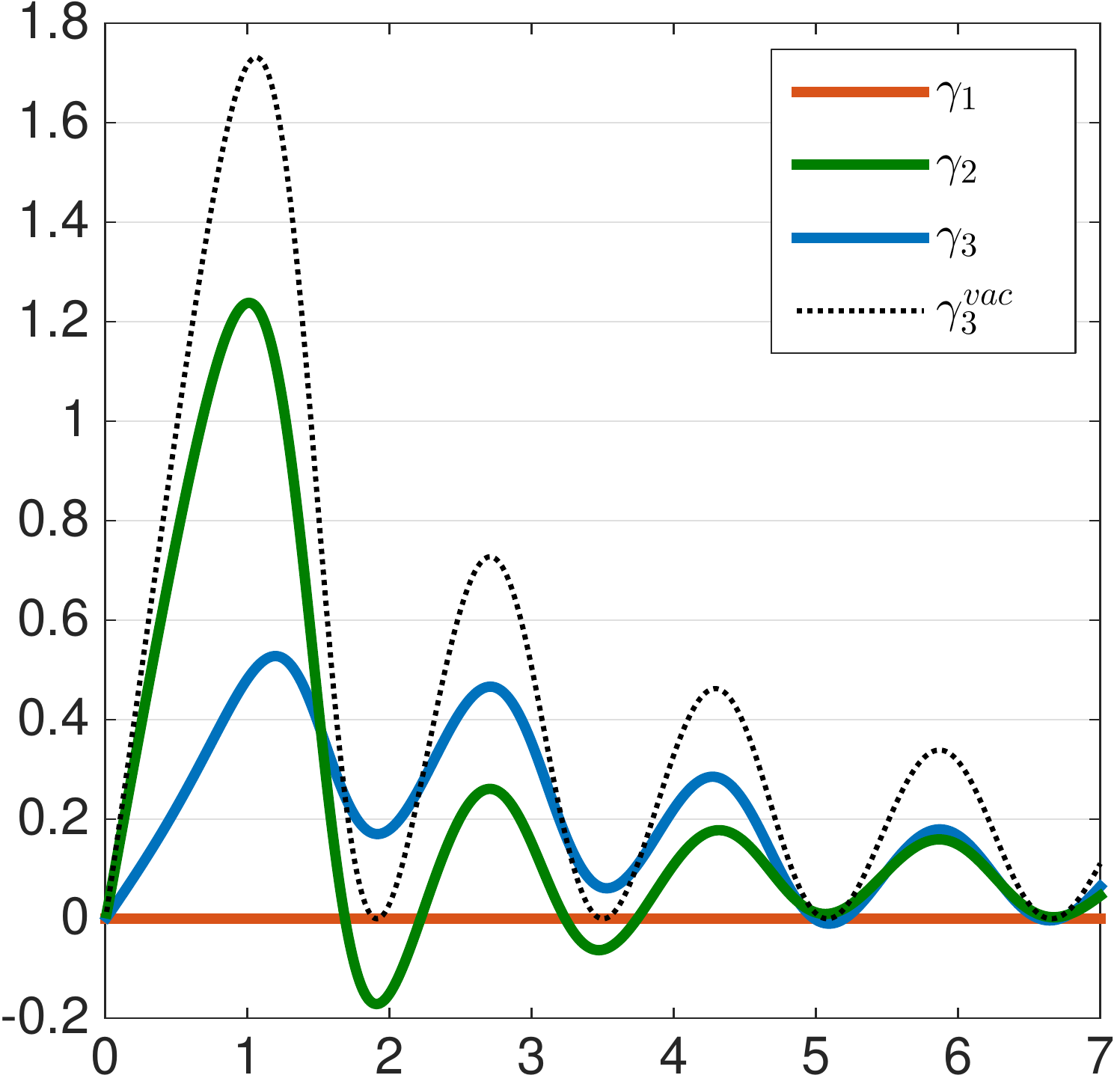}}
\caption{Time dependence of \protect\subref{subfig:ApoDivChannel} channel elements and \protect\subref{subfig:ApoDivRates} rates at a BLP-Markovian point ($\Delta_h/J=1$) in the $m_0=1$ model and for environment initially in the ground state for $h=-\frac{J}{2}$ ($\Omega/J=1$). For reference we plot the vacuum values as dotted lines. Inset: Fermi-Dirac distribution $f(E)$ (red) from lower to upper band edge. The value of $\Delta_h/J$ is indicated as a vertical black line.
}
\label{fig:ApoDiv}
\end{figure}

The divisibility of the channel, on the contrary, depends on the initial state of the environment. In general, $c$ does not vanish, and the evolution may be non-divisible, even for the detuning set at the BLP-Markovian point. This is shown explicitly in Fig.~\ref{fig:ApoDiv}: for $\Delta_h/J=1$, the BLP-Markovian point at coupling strength $\Omega/J=1$, and an initial state different from the vacuum, the rate $\gamma_2$ becomes negative at some times.

\begin{figure}
\subfloat[\label{subfig:ApoOm040FreqDoubling}]{\includegraphics[width=.24\textwidth]{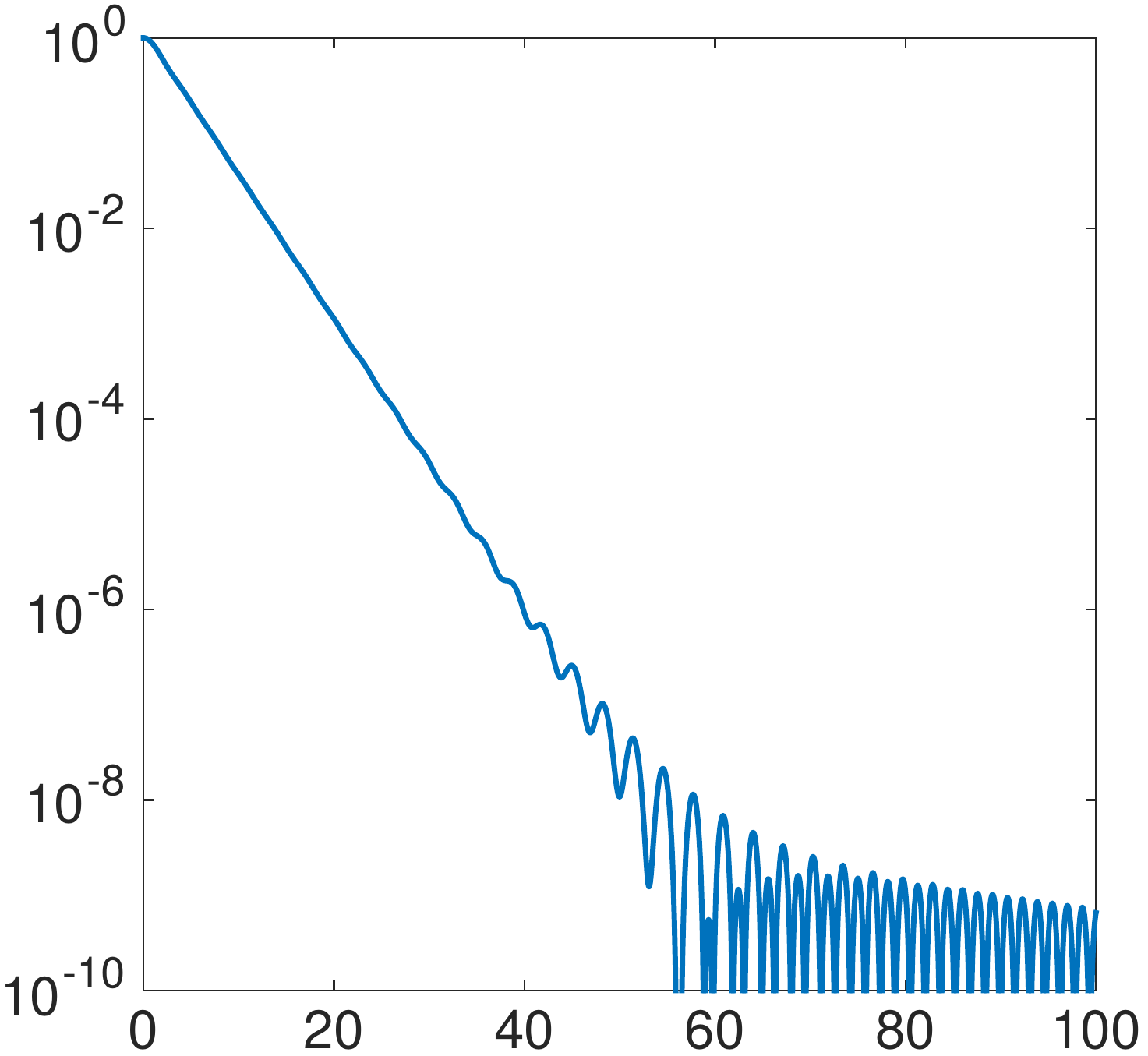}}
\hfill
\subfloat[\label{subfig:ApoOm040FreqDoublingRates}]{\includegraphics[width=.24\textwidth]{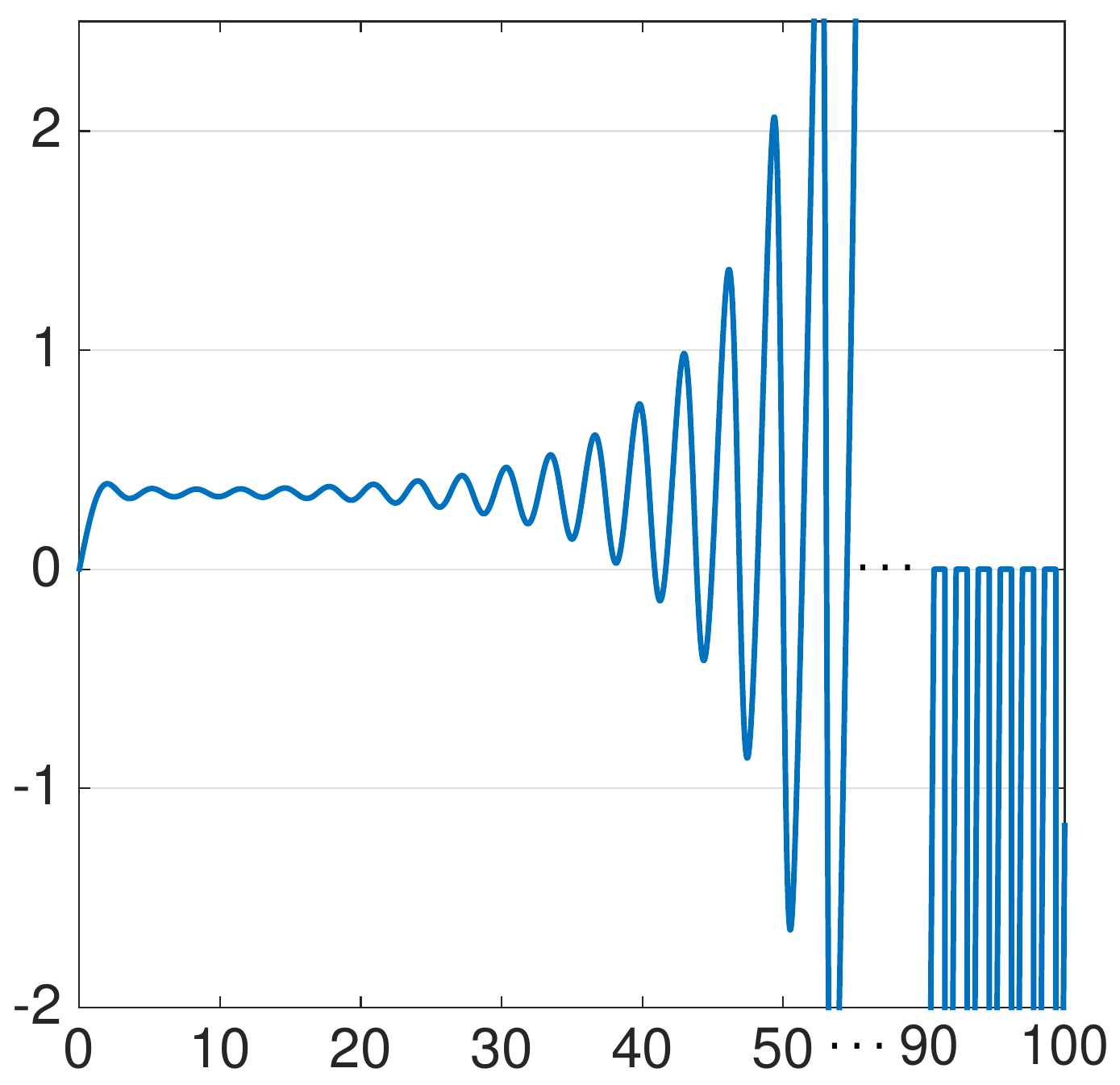}}
\caption{Time dependence of channel element and rate at the center of the band ($\Delta_h/J=0$) in the $m_0=1$ model and for environment initially in the vacuum state ($\Omega/J=0.4$). \protect\subref{subfig:ApoOm040FreqDoubling} $\mya^{\text{vac}} $ in logarithmic scale. \protect\subref{subfig:ApoOm040FreqDoublingRates} $\gamma^{\text{vac}}_3$, where we skipped a time interval in the figure (denoted by dots) and in the late times we projected positive values to zero for clarity. 
}
\label{fig:ApoOm040FreqDoubling}
\end{figure}

On the other hand, for any other parameters and any of the considered initial states, since the BLP measure is always non-zero, the channel is not divisible. However, it is interesting to analyze the time dependence of this  non-Markovianity. For the most non-Markovian setup ($\Delta_h=0$), shown in Fig.~\ref{fig:ApoOm040FreqDoubling}, we find that the dynamics (in the vacuum case) is indeed divisible according to our measure until intermediate times. The BLP-non-Markovianity arises only from the late times ($tJ \gtrsim40$), when the population has decayed so much that the oscillations break monotonicity of $\mya^{\text{vac}} $. This is analogous to our observation in section~\ref{sec:vacCase} (see Fig.~\ref{fig:vacBothSidesNonMark}, orange lines), in which we obtain large non-Markovianity contributions from late times because a vanishing population leads to diverging $\gamma^{\text{vac}}_3$.

A significant difference between this setup and the one discussed in the previous sections is that, as shown in reference\cite{apollaro}, in the case of the system coupled at the beginning of the chain, there is a single bound state of the interacting Hamiltonian for $\Omega/J<1.5$, and only if $|\Delta_h|\ge 2J-\frac{\Omega^2}{J}$.
Although the analytical calculation of section~\ref{sec:vacCase} (see also Appendix~ \ref{app:Alex}) is not directly applicable to this setup, we observe that the oscillations of $\mya^{\text{vac}} $ at early times (during the exponential transient) and at late times still have a frequency approximately equal to $2J$ and $4J$ respectively (see Fig.~\ref{fig:ApoOm040FreqDoubling}), suggesting that they still originate in cross terms involving \emph{resonant} and \emph{edge} contributions.

\begin{figure}
\subfloat[\label{subfig:MPexplanation}]{\includegraphics[width=.24\textwidth]{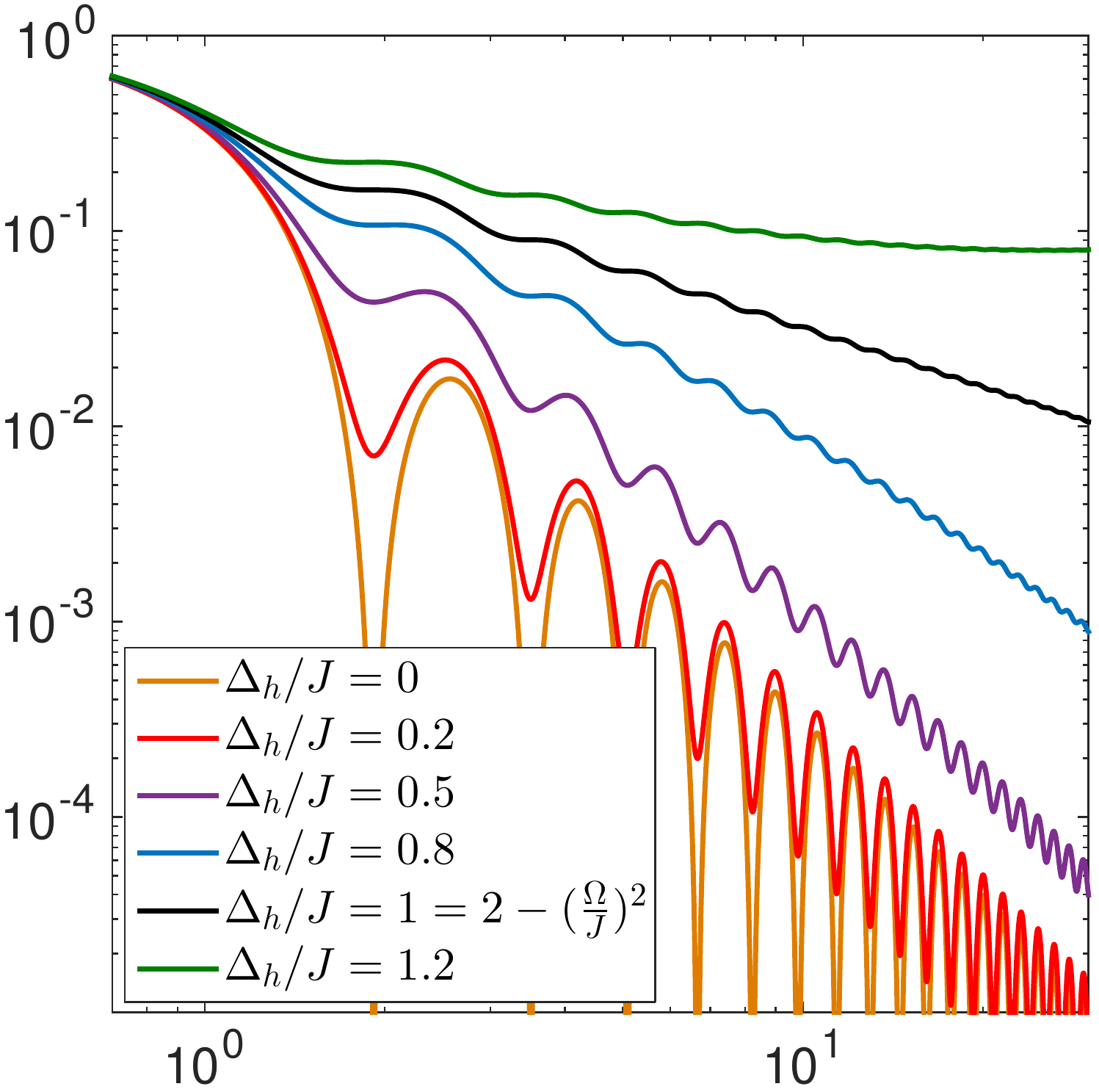}}
\hfill
\subfloat[\label{subfig:MPexplanationRates}]{\includegraphics[width=.24\textwidth]{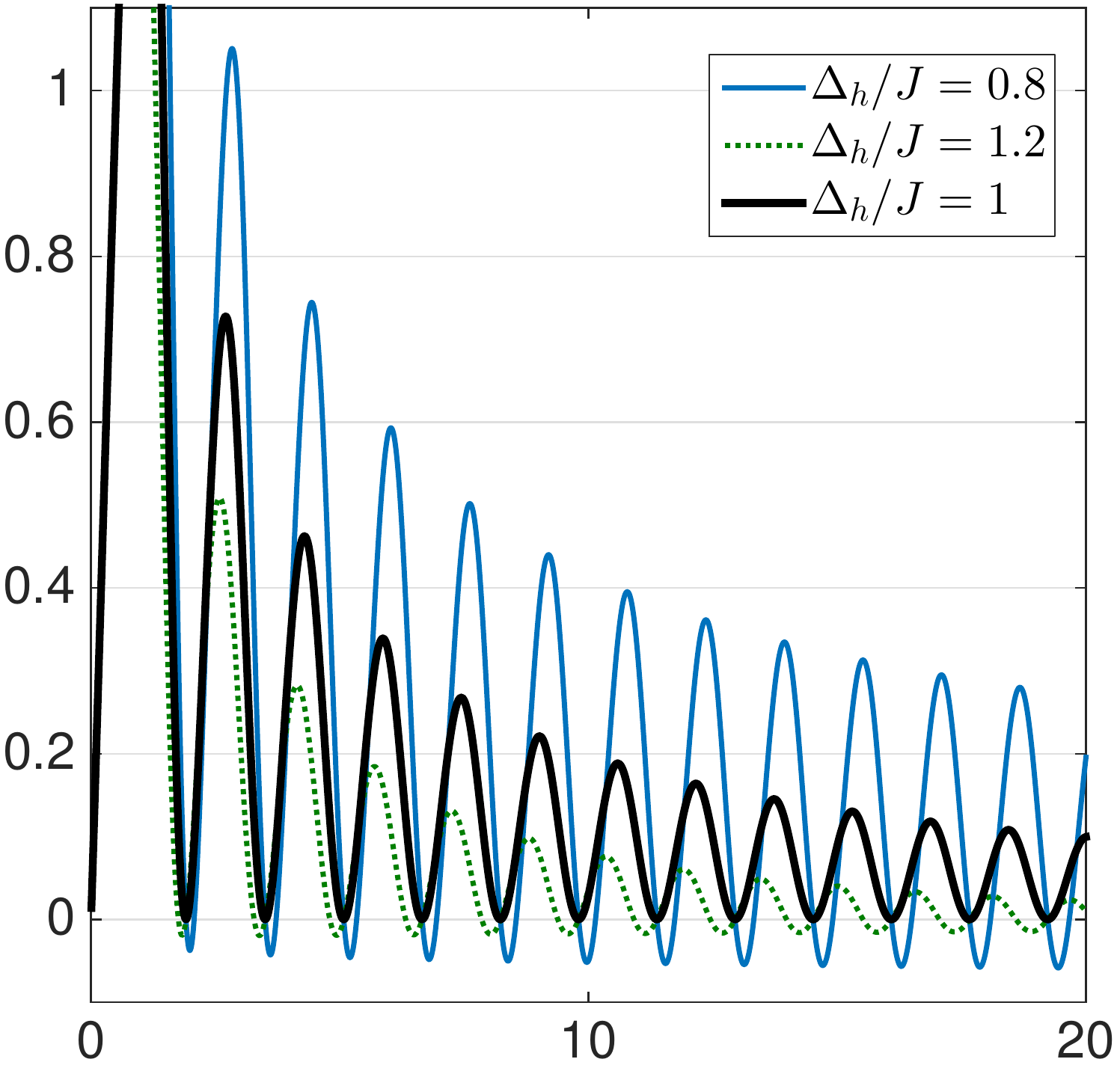}}
\caption{Time dependence of channel element and rate at different $\Delta_h$ in the $m_0=1$ model and for environment initially in the vacuum state ( $\Omega/J=1$). \protect\subref{subfig:MPexplanation} $\mya^{\text{vac}} $ in log-log scale \protect\subref{subfig:MPexplanationRates} $\gamma^{\text{vac}}_3$ 
}
\label{fig:MPexplanation}
\end{figure}

\begin{figure}
\subfloat[\label{subfig:MPexplanation2}]{\includegraphics[width=.24\textwidth]{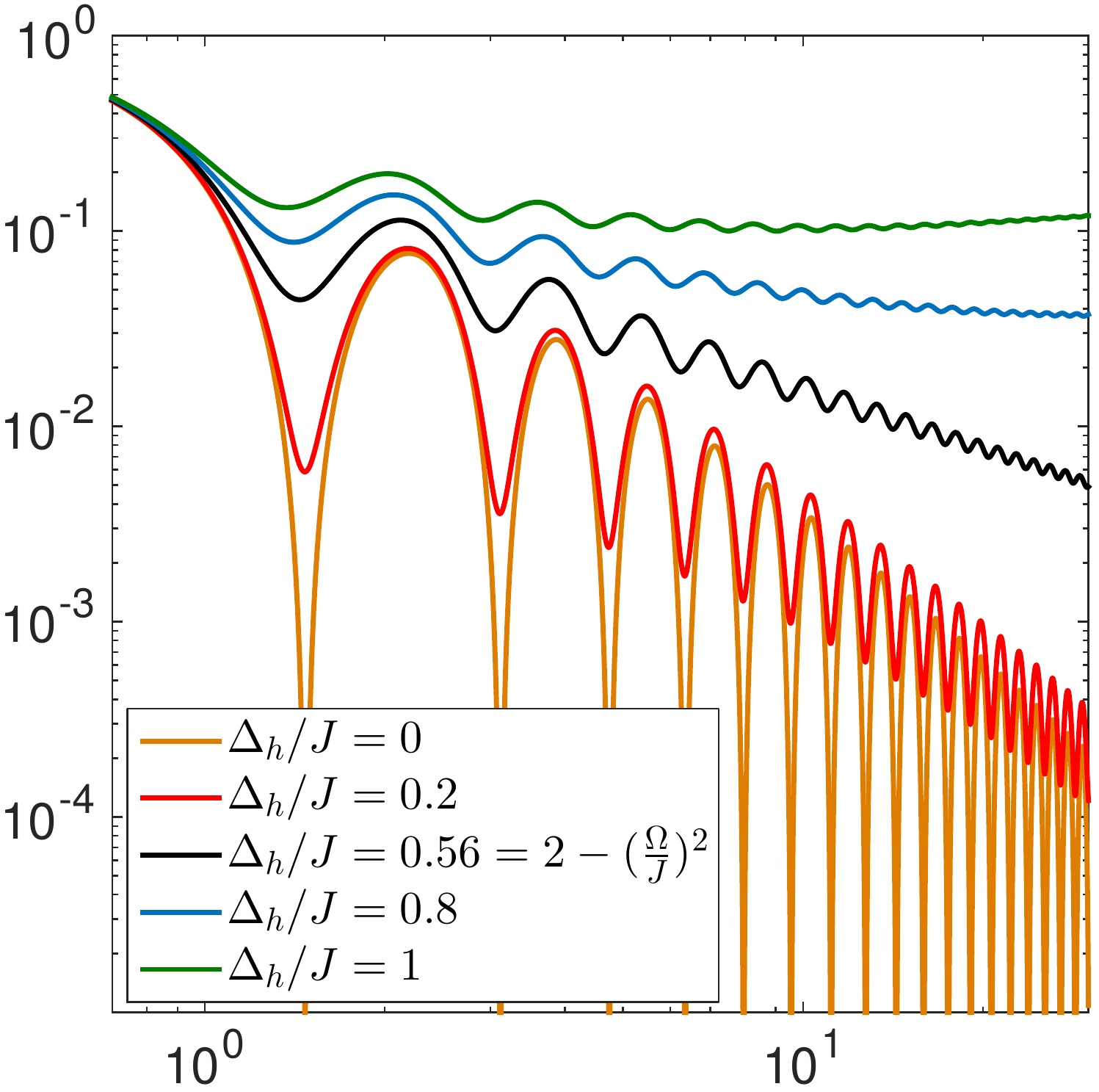}}
\hfill
\subfloat[\label{subfig:MPexplanation2Rates}]{\includegraphics[width=.24\textwidth]{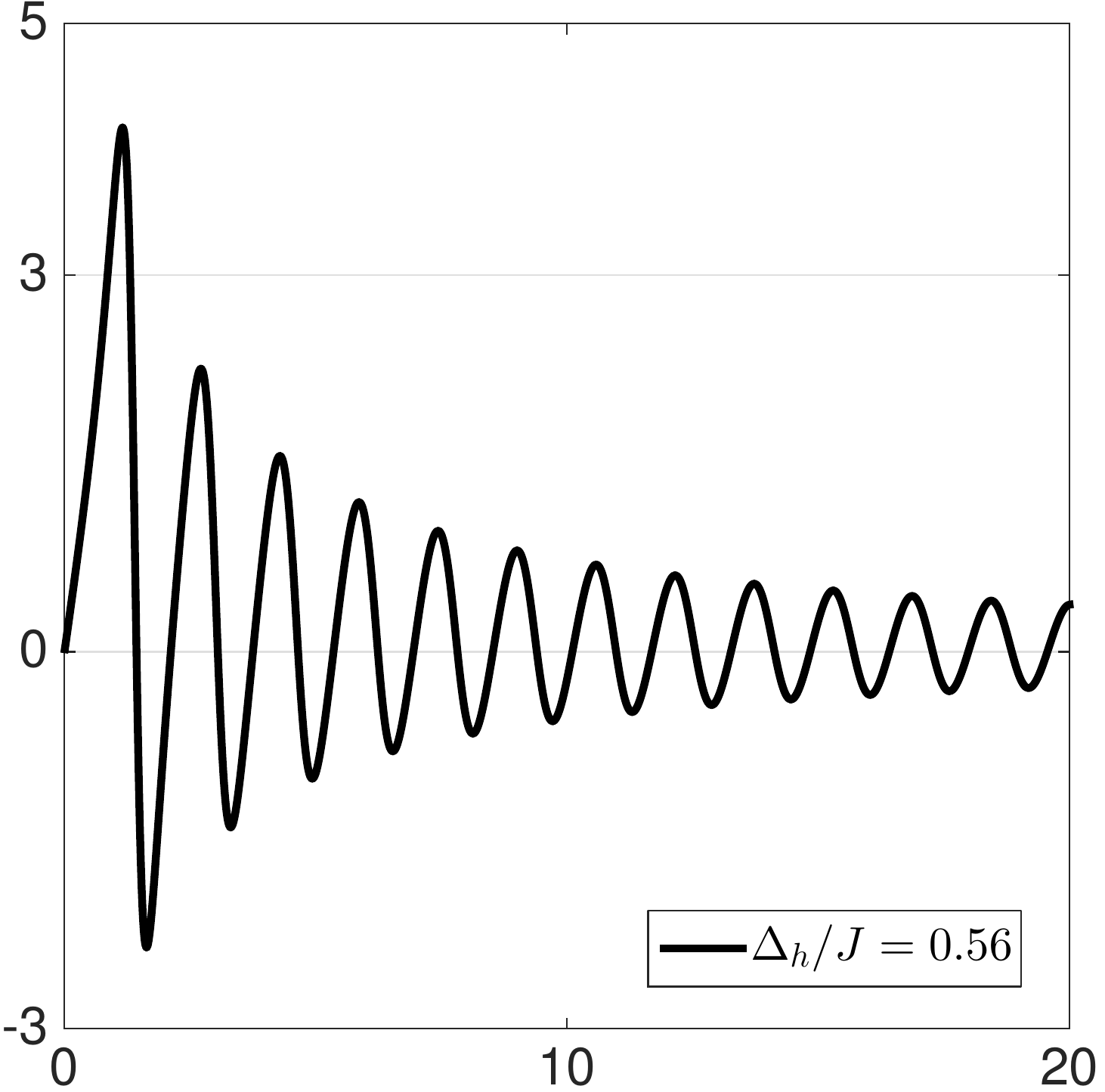}}
\caption{Time dependence of channel element and rate at different $\Delta_h$ in the $m_0=1$ model and for environment initially in the vacuum state ( $\Omega/J=1.2$). \protect\subref{subfig:MPexplanation2} $\mya^{\text{vac}} $ in log-log scale \protect\subref{subfig:MPexplanation2Rates} $\gamma^{\text{vac}}_3$  
}
\label{fig:MPexplanation2}
\end{figure}

As the detuning is shifted closer to the edge of the band, we observe that $\mya^{\text{vac}} $ decays as a power law modulated by damped oscillations, until, at the BLP-Markovian point, they do not break monotonicity anymore (Fig.~\ref{fig:MPexplanation}). This suggests a competition between the cross term involving both \emph{edge} contributions on the one hand and the monotonic power law decay of the absolute square of the relevant \emph{edge} contribution on the other hand (for detunings $\Delta_h\le2J-\frac{\Omega^2}{J}$). Generalising from the previous setup, the decreasing relevance of the oscillations towards the (upper) band edge is what we expect because the magnitudes of the upper/lower \emph{edge} contributions increase/decrease\cite{AlexPaper}. For detunings beyond the BLP-Markovian point ($\Delta_h>2J-\frac{\Omega^2}{J}$) we have a bound state, and cross terms will always break monotonicity of $\mya^{\text{vac}} $ at some times (see e.g. Fig.~\ref{fig:MPexplanation}, green line). Finally, for $\Omega/J>1$, even at $\Delta_h=2J-\frac{\Omega^2}{J}$ the cross term involving both \emph{edge} contributions is still strong enough to break monotonicity of $\mya^{\text{vac}} $ (see Fig.~\ref{fig:MPexplanation2}). Thus, at a fixed coupling strength, a region (in $\Delta_h$) without bound states is necessary (but not sufficient) for the existence of BLP Markovian points, and such a region can only exist if the spectral density does not diverge at the band edges (Appendix~\ref{app:Alex}). 

In the previous section and in Appendix~\ref{app:Alex} we argued that signatures of single excitation physics 
survive in the characteristic frequencies that modulate the channel elements, also in the setups with few excitations. The $m_0=1$ model provides an extreme example of this where these frequencies are present in setups with an arbitrary number of initially populated fermionic modes since $\mya -\myc $ does not depend on the initial state of the environment, as far as it commutes with the total number of excitations.

Notice that the BLP-Markovian points are located quite close to the band edge, where the standard derivation of the 'Markovian' master equation is not valid as discussed in Appendix~\ref{app:kernels}.\footnote{In the $m_0=1$ model the correlation functions of the vacuum and infinite temperature cases are proportional to each other.} Still, in the vacuum case, they are captured by a 'time-dependent Markovian' master equation at all times.

\section{Conclusion}
\label{sec:conclu}
In this work we have explored the dynamics of a single spin coupled to a quantum spin chain, when considered as an open quantum system. We have used a simulation of the real time evolution of the whole system to compute the dynamical map that governs the evolution of the spin, and to characterize and measure its non-Markovianity. We have identified situations, determined by the parameters of the system (coupling and detuning) and the initial state of the chain, in which the dynamics of the spin, at least until intermediate times, admits a description in terms of a time-dependent Markovian master equation, i.e. the map is divisible. Some of these scenarios occur in regimes that do not allow a standard derivation of the master equation.

We studied two scenarios. In the first one, we couple the spin to the center, in the second, to one edge of the environment chain.

In the first case, and when the chain is initialized in the vacuum, we find a Markovian parameter region when the detuning of the spin is deep within the band of the single-particle spectrum of the environment and the coupling is small to intermediate. This is in line with a characterisation based on the validity of the standard derivation of the master equation.\cite{AlexPaper} Setting the detuning close to the band edges produces strong non-Markovian effects, as does a strong coupling.

If the initial state of the chain contains a few excitations close to the lower band edge, the scenario changes: the Markovian regions disappear and the non-Markovianity close to the edge increases dramatically. The latter effect persists also beyond few excitations if one initializes the environment in a filled Fermi sea and sets the detuning close to the Fermi level. On the other hand, a high temperature initial state of the chain, which introduces a large number of excitations evenly spread across the spectrum, results in Markovian behavior, even for detunings close to the band edges (where the spectral density diverges). 

In the exactly solvable case of the chain initialized in the vacuum, the Markovian or non-Markovian character of the map can be completely explained in terms of the eigenstates of the full Hamiltonian and how they contribute to the scattering amplitude. In particular, non-Markovianity obeys to the presence of sufficiently strong cross terms between different contributions. A qualitatively similar picture holds in the case of few initial excitations. 

In the second case, when the spin is coupled to the edge of the chain, the problem is exactly solvable. A remarkable difference to the first case is that, if the chain is initialized in the vacuum, there are points in parameter space that are Markovian at all times. We explained this phenomenon with the cross term argument above and found that a non-diverging spectral density at the band edges is paramount to the existence of such points. Another difference to the first case is that while any Markovianity still disappears on introducing few excitations, a dramatic increase of non-Markovianity does not occur. Finally, high temperature does not impose Markovian dynamics. Instead, any non-Markovianity of the vacuum case survives at all temperatures. At high temperature this is in stark contrast to the Markovianity of the first case, but is consistent with the behavior of the environment correlation functions that we have computed, which, whilst showing a superexponential behavior inducing Markovianity in the first case, are characterized by a power law behavior for the second case, ruling out the standard derivation for detunings close to the band edges. The decay of the correlation functions becomes steeper as the position of the coupling is moved away from the edge of the chain.

We define Markovianity as divisibility of the evolution, but we can also compute other non-equivalent non-Markovianity measures. In particular, we have compared the results to the widely used BLP measure, less restrictive, which does neither detect the non-Markovianity deep within the band, nor the early time onset of the dramatic non-Markovianity close to the lower band edge, when few excitations are present in the environment.

\section{Acknowledgements}
\label{sec:acknow}
We gratefully acknowledge the insights of and discussions with T. Shi, A. Gonz\'alez-Tudela, I. de Vega and R. Verresen. JR is particularly grateful to T. Shi for pointing the way to the method in Appendix~\ref{app:Tao}, and to A. Gonz\'alez-Tudela for in-depth discussions on the vacuum case. JR acknowledges ExQM and IMPRS for support.
This work was supported by the Deutsche Forschungsgemeinschaft (DFG, German Research Foundation) under Germany's Excellence Strategy -- EXC-2111 -- 390814868, and by the European Union through the ERC grant
QUENOCOBA, ERC-2016-ADG (Grant no. 742102).

\appendix

\section{}
\label{app:Alex}
In reference\cite{AlexPaper} the amplitude $C_e(t)$ was explicitly computed for the case of an emitter coupled to a bosonic tight-binding chain, which, in the thermodynamic limit, is equivalent to our model with $m_0=N/2$ in the vacuum case. This is done by expressing the amplitudes as
\beq
C_\alpha(t)=-\frac{1}{2\pi i} \int_{-\infty}^\infty dE G_\alpha \left( E+i0^+\right) e^{-iEt},
\label{eq:contour}
\eeq
where $\alpha\in\{e,k\}$, and using the structure of singularities in the complex energy plane of the retarded Green functions
\begin{align}
G_e(z)=&\frac{1}{z-\Delta-\Sigma_e(z)} \label{eq:Green_e},\\
G_k(z)=&\frac{\Omega}{\left( z-E_k\right)\left(z-\Delta-\Sigma_e(z)\right)},\label{eq:Green_k}
\end{align}
with the self energy
\beq
\Sigma_e(z)=\pm\frac{\Omega^2}{\sqrt{(z-2h)^2-4J^2}}
\label{eq:selfenergy}.
\eeq
The $\pm$-sign depends on whether $\text{Re}{(z-2h)}\gtrless 0$. 

In particular it was found that $C_e(t)$ can be decomposed into a sum of contributions (roughly) due to different parts of the interacting spectrum
\beq
C_e(t)=\sum_{\alpha=\text{UE,LE}} A_\alpha(t) + \sum_{\beta=\text{UBS,LBS,RS}} R_\beta e^{-iz_\beta t}.
\label{eq:ce}
\eeq
where we chose labels representing the parts close to the (upper,lower) band edge (UE,LE), close to resonance $\Delta$ (RS) and the upper and lower bound states~\cite{AlexTaoBoundState} (UBS,LBS) of the interacting spectrum.

If we consider that the environment in the initial state contains a single excitation, the channel element $\myc$ from Eq.~\eqref{eq:chanT} can be computed as\footnote{because $H$ is time reversal invariant}
 \beq
\myc =|\langle e,0 | e^{-iHt} d^\dagger_k |g,0 \rangle |^2 =|C_k(t)|^2,
\eeq
such that, following reference\cite{AlexPaper}, its behavior will be determined by the singularities of Eq.~\eqref{eq:Green_k}. 

In the (large) finite case the \emph{resonance} and \emph{edge} contributions in Eq.~\eqref{eq:ce} correspond to terms that are dominated by a sum over terms $|\langle \tilde{n} |e,0 \rangle |^2 e^{-i \tilde{E}_n t}$ running over (scattering\cite{AlexTaoBoundState}) states $|\tilde{n} \rangle$ close to resonance $\Delta$ and close to the band edges respectively ($H |\tilde{n} \rangle = \tilde{E}_n |\tilde{n} \rangle$). The \emph{bound state} contributions are simply $|\langle \text{BS}^\pm |e,0 \rangle |^2 e^{-i \tilde{E}_{\text{BS}^\pm} t}$. With this intuitive picture one can anticipate that these contributions oscillate with frequencies $\nu$ approximately given by $\Delta$ ($\nu_{\mathrm{r}}\approx\Delta$), the band edges ($\nu_{\mathrm{e}\pm} \approx 2h \pm 2J$) and the bound state energies ($\nu_{\mathrm{b}\pm}= \tilde{E}_{\text{BS}^\pm}$) respectively, which we confirm by explicit computation. The magnitudes of the contributions decay exponentially (RS), with a power law (UE,LE) or are constant (UBS,LBS).\cite{AlexPaper}

Strictly, the \emph{resonance} and \emph{bound state} contributions stem from singularities in $G_e(z)$. In\cite{AlexTaoBoundState} it was proven that if $\operatorname{Im}\big( \Sigma_e(z)\big)$ diverges at the band edge, where it is proportional to the spectral density $D(E)$\cite{AlexPaper}, there always exists a pole associated to a bound state. This is the case when $m_0=\frac{N}{2}$, where the spectral density is proportional to the (diverging) density of states; otherwise it is not necessarily the case.\footnote{For $m_0=\frac{N}{2}$ (and in the thermodynamic limit) we may replace the OB expression $W_{m_0,k}=\sqrt{\frac{2}{N+1}}\sin(\frac{\pi k m_0}{N+1})$ by its PB form $W_{m_0,k}^{\text{PB}}=\frac{1}{\sqrt{N}}e^{-i k m_0}$.}

These singularities are still present\footnote{We find by explicit computation that the factor outside the bracket is finite at the singularities of the terms inside the bracket.} 
in $G_k(z)$:
 \beq
G_k(z)=\Omega \left( \frac{1}{ z-E_k}-G_e(z) \right)\frac{1}{E_k-\Delta-\Sigma_e(z)},
\label{eq:Green_k2}
\eeq
which has an additional pole corresponding to the free propagator of the mode $k$, leading to a constant magnitude \emph{free propagator} contribution (oscillating with frequency $\nu_{\mathrm{f}}=E_k$) to $C_k(t)$. If the initial environment excitation is close to the lower band edge ($E_k\approx 2h-2J$), we thus expect to have terms in $C_k(t)$ that oscillate with similar frequencies as those in $C_e(t)$, but since $|C_k(0)|^2=0$, in this case the \emph{resonant} contribution cannot be dominating all the other contributions.\footnote{Due to the minus sign in Eq.~\eqref{eq:Green_k2}, at time zero the \emph{free propagator} contribution is equal to the sum of the other contributions.}  
This is the mechanism behind the early time non-Markovianity (at $\Delta_h=0$) in \ref{subsec:newRates}.

\begin{figure}
\subfloat[\label{subfig:1exc}]{\includegraphics[width=1\columnwidth]{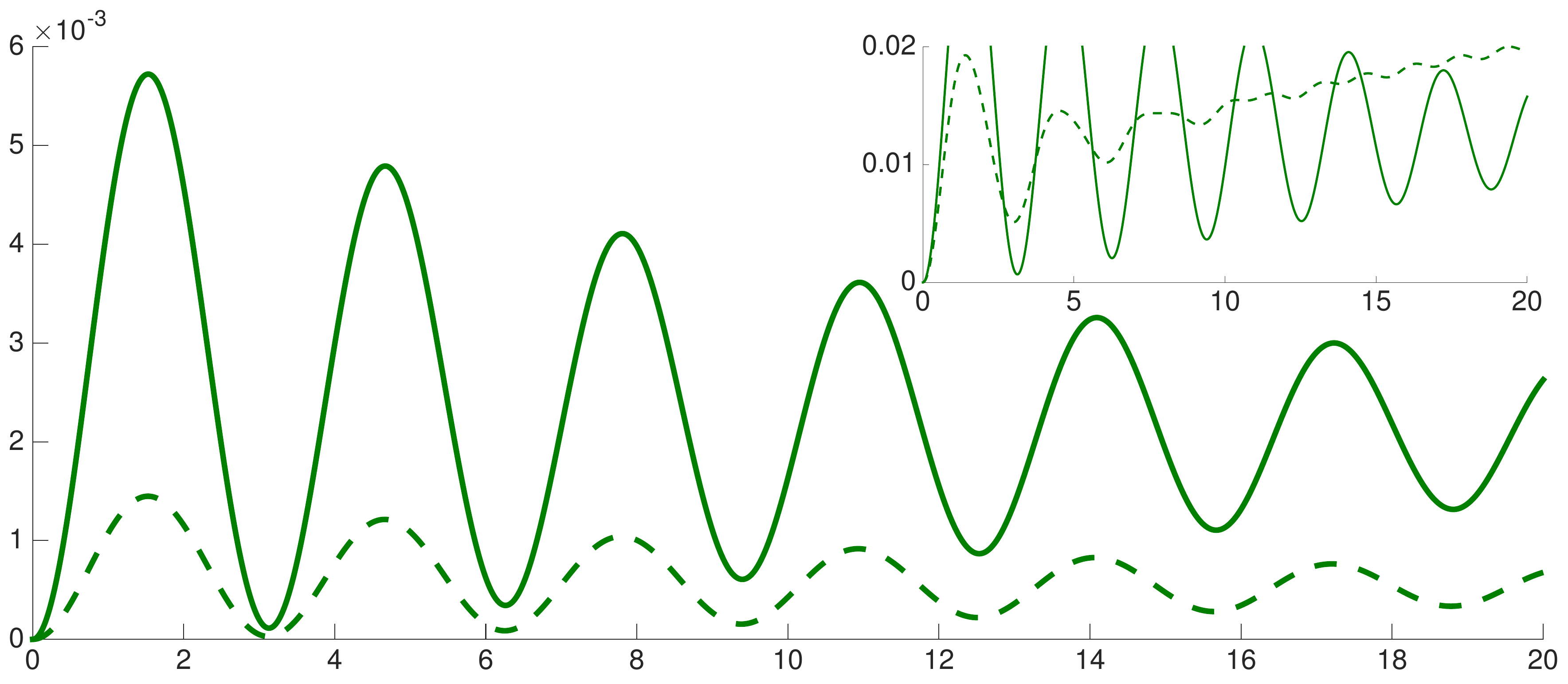}}\\
\subfloat[\label{subfig:1andmoreExc}]{\includegraphics[width=1\columnwidth]{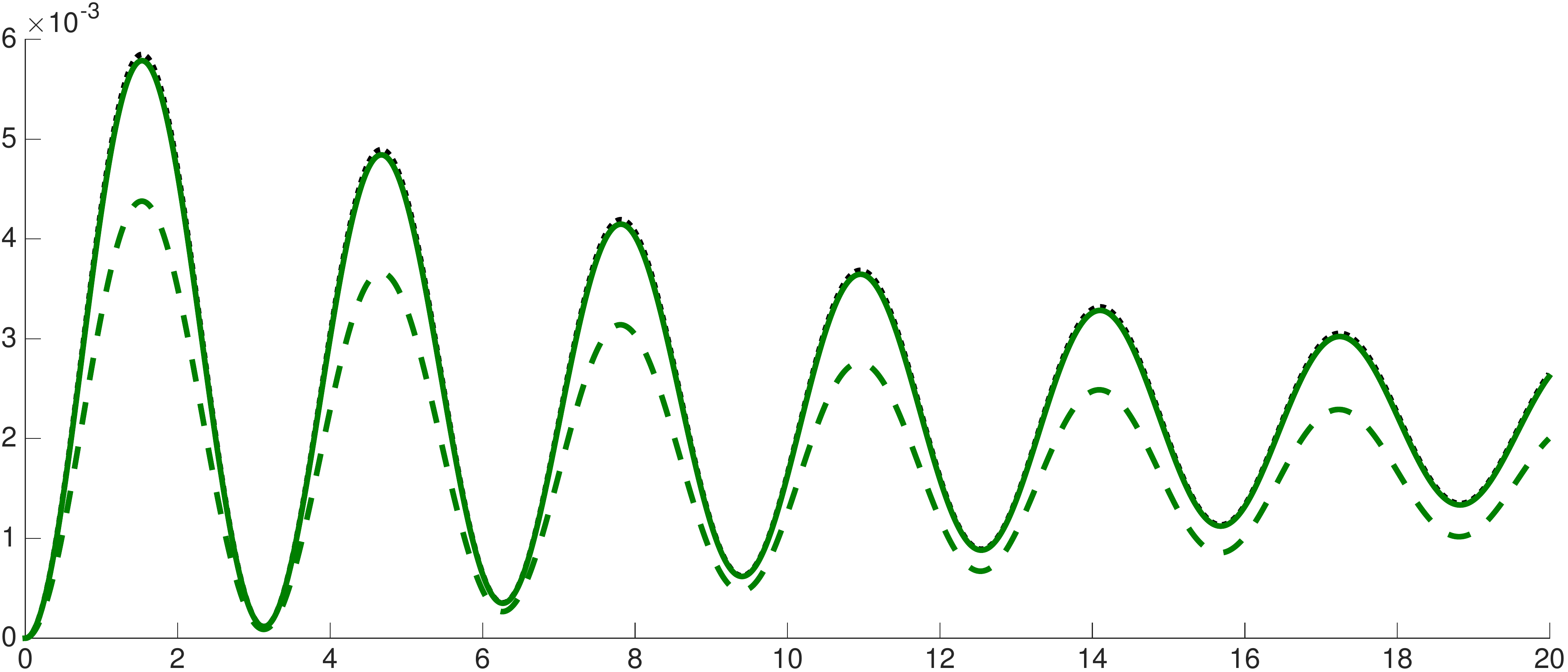}}
\caption{Time dependence of $\myc $ at the band center ($\Delta_h/J=0$). \protect\subref{subfig:1exc} Environment initially in state $|\Psi(0)\rangle=d^\dagger_N |g,0\rangle$ for $\Omega/J=0.4$ and $N=50$ (solid line), $N=200$ (dashed line). Inset: $N=50$ and $\Omega/J=0.85$ (dashed line), $\Omega/J=0.4$ (solid line and scaled by factor 6). \protect\subref{subfig:1andmoreExc} Environment initially in ground state ($h/J=0.995$) for $\Omega/J=0.4$ and $N=50$ (solid line), $N=100$ (dotted line), $N=200$ (dashed line).}
\label{fig:figapp1}
\end{figure}

Fig.~\ref{fig:figapp1} illustrates this for a finite chain as studied in our paper. We set the initial state to $d^\dagger_N |g,0\rangle$, i.e. the environment contains a single excitation in the lowest energy mode, and compute $c$ with exact diagonalisation, which, in this case, can be done efficiently since the dynamics is restricted to the one excitation sector. At the band center ($\Delta_h=0$), we find that the frequency of the oscillations is approximately equal to $2J$ ($|\nu_{\mathrm{r}}-\nu_{\mathrm{e}\pm}|$, $|\nu_{\mathrm{r}}-\nu_{\mathrm{b}\pm}|$, $|\nu_{\mathrm{r}}-\nu_{\mathrm{f}}|$) and that this feature is stable upon increasing the chain length (upper panel). Also, in the inset, we observe that there is a transition after which the frequency is approximately equal to $4J$ ($|\nu_{\mathrm{e}+}-\nu_{\mathrm{e}-}|$, $|\nu_{\mathrm{e}+}-\nu_{\mathrm{b}-}|$, $|\nu_{\mathrm{e}+}-\nu_{\mathrm{f}}|$, $\dots$). These two frequencies are characteristic for the interplay between these contributions. The frequency survives in the few excitations case (bottom panel), where an increasing number of excitations is obtained by initialising longer chains in their ground state at fixed chemical potential. 

\begin{figure}
\begin{center}
\includegraphics[width=1\columnwidth]{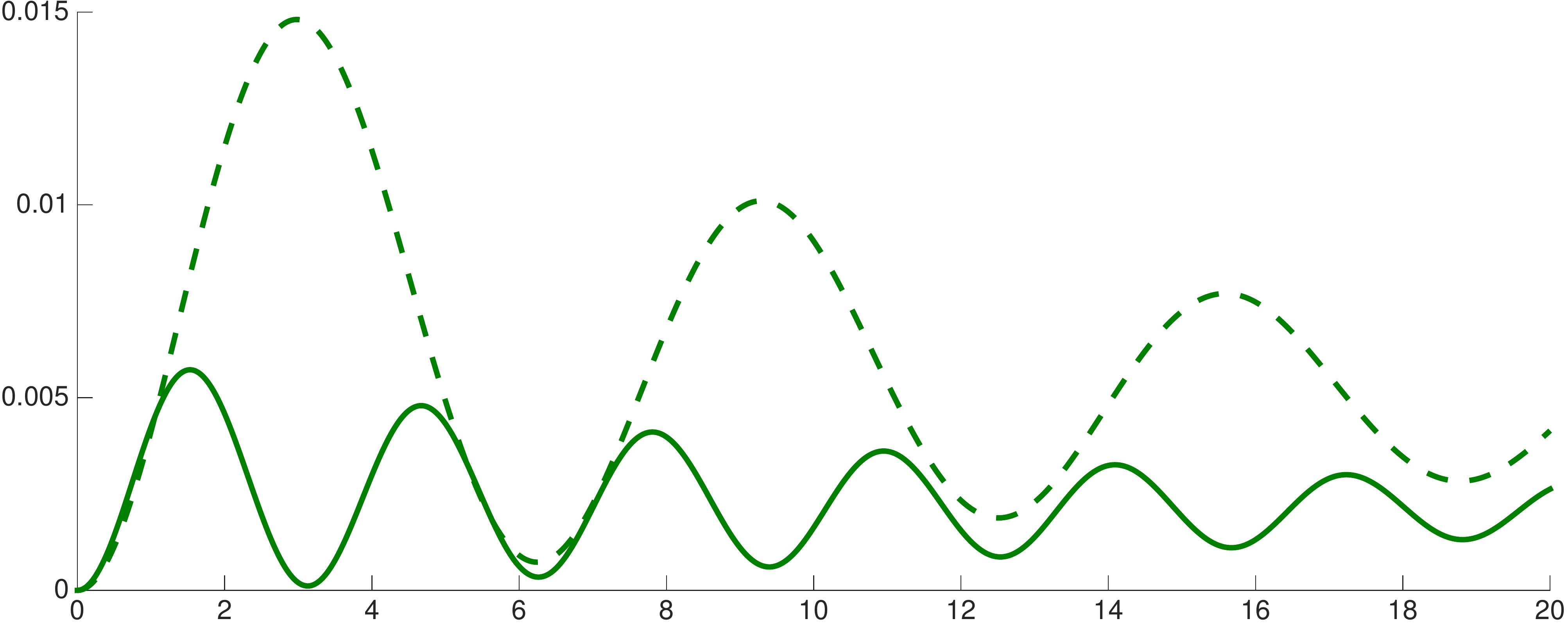}
\caption{Time dependence of $\myc $ in the band center ($\Delta_h/J=0$) for environment initially in the state $|\Psi(0)\rangle=d^\dagger_k |g,0\rangle$, for $\Omega/J=0.4$. We have $N=50$ and $E_k=2h-2J$ (solid line), $E_k=2h-J$ (dashed line).}
\label{fig:1excInsideBand}
\end{center}
\end{figure}

Fig.~\ref{fig:1excInsideBand} illustrates the importance of the \emph{free propagator} contribution, since for the case that the environment in the initial state contains a single excitation somewhere within the band, the frequency of the oscillations of the component $c$ is approximately given by $|\nu_{\mathrm{r}}-\nu_{\mathrm{f}}|$. Looking more carefully at the oscillations in Fig.~\ref{subfig:GSosc040}, we find that their frequency lies somewhere between the highest occupied mode (at the Fermi level $E_k=2h-1.9J$) and the lowest occupied mode (at the lower band edge $E_k=2h-2J$).

In the $m_0=1$ model of section~\ref{sec:Apollaro} we find that $\mya -\myc =\mya^{\text{vac}} $ is independent of the environment initial state and can thus show no signature of the \emph{free propagator} contributions. This means that beyond the vacuum case, \emph{free propagator} contributions must also contribute to $\mya $ such that their effect cancels out in $\mya -\myc $.
 
We remark that the presence of oscillations due to \emph{free propagator} contributions is independent of the existence of bound states and band edges, and should induce non-Markovianity also in unstructured environments (constant spectral density) if few initial excitations are present.

\section{}
\label{app:blp}
We prove here two statements. 

\emph{$\bm{1.}$ There is no information backflow as characterized by BLP during time intervals where the conditions $\mya -\myc \ge0$, $\frac{d}{dt}(\mya -\myc )\le0$ and $\frac{d}{dt}|\myb|^2\le0$ are all satisfied.} 

Proof: The density matrix of a spin can be written as $\rho=\frac{1}{2}\left(1 + v_1 \sigma^x + v_2 \sigma^y+ v_3 \sigma^z \right)$, where $v_i$ are real numbers. We then have for the trace distance between two spin states
\beq
\left| \rho-\tilde{\rho} \right|_1 =  \frac{1}{2} \sqrt{\sum_{i=1}^3 (v_i-\tilde{v}_i)^2}
\eeq
Information back flow $\frac{d}{dt}|\rho-\tilde{\rho}|_1>0$ at a given time implies
\beq
\sum_{i=1}^3 (v_i-\tilde{v}_i)(\dot v_i-\dot{\tilde{v}}_i)>0,
\label{eq:appc2}
\eeq
where the dot denotes the time derivative. The state resulting after the application of the channel Eq.~\eqref{eq:chanT} on an initial state $\rho(0)$ (with $\rho_{ij}(0) \equiv \langle ij | \rho(0) \rangle$) has parameters $v_1=2\operatorname{Re} \left( \myb  \rho_{10}(0) \right) $, $v_2=2\operatorname{Im} \left( \myb  \rho_{10}(0) \right) $ and $v_3=2(\mya -\myc )\rho_{00}(0)+2\myc -1$. Condition~\eqref{eq:appc2} then reads
\begin{align}
0<&4\operatorname{Re} \left[ \myb \left(\rho_{10}(0)-\tilde{\rho}_{10}(0) \right)\right] \operatorname{Re} \left[ \dot \myb \left(\rho_{10}(0)-\tilde{\rho}_{10}(0) \right)\right] \label{eq:inequalAppB}\\
&+4\operatorname{Im} \left[ \myb \left(\rho_{10}(0)-\tilde{\rho}_{10}(0) \right)\right] \operatorname{Im} \left[ \dot \myb \left(\rho_{10}(0)-\tilde{\rho}_{10}(0) \right)\right] \nn \\
&+(\dot \mya -\dot \myc  ) \cdot 4 (\mya -\myc ) \left( \rho_{00}(0) - \tilde{\rho}_{00}(0)\right)^2 \nn.
\end{align}
This reduces to
\begin{align}
0&<\frac{d|\myb |^2}{dt}\cdot \left[  2\left| \rho_{10}(0)-\tilde{\rho}_{10}(0) \right|^2 \right] \label{eq:blpInEqual}\\
  &+(\dot \mya -\dot \myc  ) \cdot\left[4(\mya -\myc )\left( \rho_{00}(0) - \tilde{\rho}_{00}(0)\right)^2 \right],\nn
\end{align}
and since the expressions in square brackets are positive if $\mya -\myc \ge0$, at least one of the expressions $\frac{d|\myb |^2}{dt} $ and $\dot \mya -\dot \myc  $ must be positive and the statement is proved. 

\emph{$\bm{2.}$ For cases where $\mya -\myc =|\myb|^2$ the BLP measure is nonzero iff $\frac{d}{dt}(\mya -\myc )>0$ at some time.}
 
Proof: We note that if $\mya -\myc =|\myb |^2$, condition~\eqref{eq:blpInEqual} reduces to 
\begin{align}
0&<(\dot \mya -\dot \myc  )\cdot \Big(  2\left| \rho_{10}(0)-\tilde{\rho}_{10}(0) \right|^2 \\
  & \quad \quad \quad \quad \quad + 4|b|^2\left( \rho_{00}(0) - \tilde{\rho}_{00}(0)\right)^2 \Big).\nn  \\
  &\iff \dot \mya -\dot \myc  > 0,
\end{align}
such that the direction $\text{"BLP measure nonzero} \implies \frac{d}{dt}(\mya -\myc )>0 \text{ at some time"}$ is proved.

To prove the other direction we choose the initial pair $\rho(0)=|e\rangle \langle e |$, $\tilde{\rho}(0)=|g\rangle \langle g|$. Taking into account the general structure of the channel Eq.~\eqref{eq:chanT}, we have $v_1=v_2=\tilde{v}_1=\tilde{v}_2=0$ at all times and thus:
\begin{align}
\frac{d}{dt}\left| \rho-\tilde{\rho} \right|_1 &= \frac{1}{2}\frac{d}{dt}| v_3-\tilde{v}_3| \\
&= \frac{1}{2}\frac{d}{dt} \Big| \mya \left( \rho_{00}(0)-\tilde{\rho}_{00}(0) \right) + \myc  \left( \rho_{11}(0)-\tilde{\rho}_{11}(0) \right)\Big| \nn\\
&= \frac{1}{2}\frac{d}{dt}| \mya - \myc | \nn.
\end{align}
Hence, since $\mya -\myc \ge 0$, we find that $\dot \mya -\dot \myc  >0$ at some time implies that the trace distance is increasing (BLP measure nonzero) and the statement is proved.

\section{}
\label{app:ApollaroFreeFermions}
Here we give the concrete expressions of the channel elements in the case of the spin coupled to the edge of the chain (section~\ref{sec:Apollaro}).

The total system can be mapped to the free fermion Hamiltonian:
\begin{align}
H=&\Omega(\tilde{c}_1^\dagger \tilde{c}_0+\text{h.c.})+\Delta \tilde{c}_0^\dagger \tilde{c}_0 \nn\\
&+J\sum_{i=1}^{N-1}(\tilde{c}_{i+1}^\dagger \tilde{c}_i +\text{h.c.})+2h\sum_{i=1}^N \tilde{c}_i^\dagger \tilde{c}_i,
\label{eq:freeHE}
\end{align}
with real space fermionic operators $\tilde{c}_i=e^{i\pi \sum_{j=0}^{i-1} \sigma^+_j \sigma^-_j } \sigma^-_i $, where $\sigma^\alpha_0\equiv \tau^\alpha_0$ are the spin operators on the subsystem.

The channel elements of Eq.~\eqref{eq:chanT} can be written in terms of the Heisenberg picture fermionic operators as follows:
\begin{align}
\mya (t)=&1-\langle  \tilde{c}_0(t)\tilde{c}^\dagger_0(t)  \rangle_e \label{eq:AppTeeee}\\
\myc (t)=&1-\langle  \tilde{c}_0(t)\tilde{c}^\dagger_0(t)  \rangle_g \label{eq:AppTeegg}\\
\myb (t)=&\langle  \tilde{c}_0(t) \rangle_{x+} + i\langle  \tilde{c}_0(t) \rangle_{y+}, \label{eq:AppTegeg}
\end{align}
where $I\in\{e,g,x+,y+\}$ denotes the global initial state $|I\rangle \langle I | \otimes \rho_E$ and we have defined $|x+\rangle=\frac{1}{\sqrt{2}}\left(|e\rangle+|g\rangle \right) $, $|y+\rangle=\frac{1}{\sqrt{2}}\left(|e\rangle+i |g\rangle \right)$. Note that in contrast to fundamental fermion models we can have $\langle \tilde{c}_0 \rangle \neq 0$ since the $0$-th mode corresponds to the spin subsystem. The parity (of $N_{\text{exc}}$) is nevertheless conserved, such that if one starts with a superposition in the spin, one can solve each sector independently and then add them up, where in each of the calculations $\langle \tilde{c}_0 \rangle = 0$.
These expectation values can be computed exactly. 

We write down the Heisenberg equation of motion
\begin{align}
i \frac{d}{dt} (\tilde{c}_i(t) \tilde{c}^\dagger_j(t)) =& \left[ \tilde{c}_i(t) \tilde{c}^\dagger_j(t),H \right] \\
=&\sum_l \left( \mathcal{H}_{il}\tilde{c}_l(t) \tilde{c}^\dagger_j(t) - \tilde{c}_i(t) \tilde{c}^\dagger_l(t)\mathcal{H}_{lj} \right) \nn,
\end{align}
where $\mathcal{H}$ is the real, symmetric, $N+1$ dimensional tridiagonal matrix in $H=\sum_{ij=0}^N \tilde{c}^\dagger_i \mathcal{H}_{ij} \tilde{c}_j$. Taking the expectation value with respect to global initial state $I$ on both sides and defining, for each initial state, a matrix $M^I_{ij}(t)\equiv \langle \tilde{c}_i(t) \tilde{c}^\dagger_j(t) \rangle_I$, we get the matrix equations:
\beq
i\partial_t M^I(t) = \left[ \mathcal{H},M^I(t) \right]
\eeq
Analogously we define $\xi^I_i(t)\equiv\langle \tilde{c}_i(t) \rangle_I$ and get the vector equations:
\beq
i\partial_t \xi^I(t) = \mathcal{H} \xi^I(t)
\eeq

Thus, to find $M^I(t)=e^{-i\mathcal{H}t} M^I(0) e^{i \mathcal{H}t}$ and $\xi^I(t)=e^{-i\mathcal{H}t} \xi^I(0)$ and solve Eqs.~\eqref{eq:AppTeeee} to~\eqref{eq:AppTegeg} we compute the initial conditions $ M^I(0)$ and $\xi^I(0)$. Since our initial states are product states, the initial matrices $M^I(0)$ can be written as a direct sum $M^I(0)= M^I_S \oplus  M_E$ with $M^e_S=0$, $M^g_S=1$. Correspondingly, we write the initial vectors as a direct sum $\xi^I(0)=\xi^I_S \oplus \xi_E$  with $\xi^{x+}_S=\frac{1}{2}$, $\xi^{y+}_S=\frac{-i}{2}$. We have $M^E=\frac{1}{e^{-\beta \mathcal{H}^E}+1}$ where $\mathcal{H}^E$ is the real, symmetric, $N$ dimensional tridiagonal matrix corresponding to the environment Hamiltonian $H_E=\sum_{ij=1}^N \tilde{c}^\dagger_i \mathcal{H}^E_{ij} \tilde{c}_j$. Also, $\xi^E=0$.

The difference $\mya -\myc $ is independent of the environment initial state:
\begin{align}
\mya -\myc =&M_{00}^g(t)-M_{00}^e(t)=
\left|\left(e^{-i \mathcal{H}t}\right)_{00}\right|^2,
\end{align}
where in our convention $X_{00}$ denotes the first matrix element of a $N+1$ dimensional matrix X with matrix indices running from $0$ to $N$. Similarly, 
\begin{align}
\myb =&\left(e^{-i \mathcal{H}t}\right)_{00}, 
\end{align}
and thus $|\myb |^2=\mya -\myc $. This immediately implies that $\gamma_1=0$ (see Eq.~\eqref{eq:gamma1}).

\section{}
\label{app:Tao}
In this appendix we provide an explicit scheme for evaluating the environment correlation functions of our model (defined in Eqs.~\eqref{eq:alphap} and~\eqref{eq:alpham}). We express them as a sum over Gaussian operator terms that can be computed efficiently.

The chain Hamiltonian can be written as $H_E=\sum_{ij=1}^N c^\dagger_i \mathcal{H}_{ij}^E c_j$ with the real, symmetric, $N$ dimensional tridiagonal matrix $\mathcal{H}^E=W\Lambda W$, where $W_{jk}=\sqrt{\frac{2}{N+1}}\sin(\frac{\pi k j}{N+1})$ and $\Lambda_{kq}=\delta_{kq}E_k$. With $\rho_E=\frac{e^{-\beta H_E}}{Z}$ we have
\begin{align}
\alpha^+(t)&=\operatorname{tr} \Big( \rho_E c^\dagger_{m_0} u_{m_0} e^{-i H_E t} u_{m_0}  c_{m_0} e^{i H_E t} \Big) \nn \\
&= \frac{\tilde{Z}}{Z}\operatorname{tr}\Big( G  c_{m_0}^\dagger  e^{-i H_E' t}  c_{m_0}\Big),
\label{eq:taostep1}
\end{align}
where we transformed the first exponential via a unitary conjugation with $u_{m0}$, and we introduced the Gaussian operator $G=\frac{e^{-(\beta-it)H_E}}{\tilde{Z}}$ with $\tilde{Z}=\operatorname{tr}\big( e^{-(\beta-it)H_E}\big)$. We have $H_E'=\sum_{ij=1}^N c^\dagger_i \mathcal{H}_{ij}^{E'} c_j$ with $\mathcal{H}^{E'}=V \mathcal{H}^E V$, where $V_{ij}=-\delta_{ij}$ if $i<m_0$ and $V_{ij}=\delta_{ij}$ otherwise. Defining $U=V W$, Eq.~\eqref{eq:taostep1} reduces to:
\begin{align}
\alpha^+(t)&=\frac{\tilde{Z}}{Z} \sum_{k,n} e^{i E_k t} U_{m_0,k}  U_{m_0, n}\operatorname{tr}\Big(G e^{-i t \sum_i E_i \fmnd{i} \fmn{i}}\fmnd{k}  \fmn{n}\Big),
\end{align}
with fermionic operators $\fmn{i}=\sum_{j}U_{ji}c_j$. Following analogous steps for $\alpha^-(t)$, we can finally write 
\beq
\alpha^\pm(t)=\sum_k e^{\pm i E_k t} \alpha_k^\pm(t),
\label{eq:taoresult}
\eeq
 with
\begin{align}
\alpha_k^+(t) & = \frac{\tilde{Z}}{Z} \sum_{n}  U_{m_0,k}  U_{m_0, n}\operatorname{tr}\Big(G e^{-i t \sum_i E_i \fmnd{i} \fmn{i}}\fmnd{k}  \fmn{n}\Big) \label{eq:alphakp} \\
\alpha_k^-(t) & = \frac{\tilde{Z}}{Z} \sum_{n}  U_{m_0,k}  U_{m_0, n} \Big[ \delta_{kn} \operatorname{tr}\Big(G e^{-i t \sum_i E_i \fmnd{i} \fmn{i}}\Big) \nn \\
& \quad \quad \quad  \quad \quad \quad \quad \quad \quad - \operatorname{tr}\Big(G e^{-i t \sum_i E_i \fmnd{i} \fmn{i}}\fmnd{n}  \fmn{k}\Big) \Big] \label{eq:alphakm}.
\end{align}

The traces involving the Gaussian operator in Eqs.~\eqref{eq:alphakp} and~\eqref{eq:alphakm} can be calculated exactly as shown in Appendix D of~\cite{Tao}, the only required ingredient being the $N^2$ dimensional (complex) matrix $\Gamma_f=\operatorname{tr}\Big(G C'  {C'}^{\dagger} \Big)$, where we introduced the vector $C'=(\fmn{1},\dots,\fmn{N},\fmnd{1},\dots,\fmnd{N})^T$. In our case it has the block diagonal form
\begin{align}
\Gamma_f=&W V W  \frac{1}{1+e^{-(\beta-i\tau)\Lambda}}  WVW
\nn \\
  &  \oplus WVW\frac{1}{1+e^{+(\beta-i\tau)\Lambda}}  WV W . 
\end{align}
We compute Eq.~\eqref{eq:taoresult} numerically for a (finite) $N$ chosen such that the result is converged in system size.

\section{}
\label{app:kernels}
Here we illustrate a complementary perspective to the Markovianity discussion of section \ref{subsec:kessler} for the two scenarios studied in this paper, (i) $m_0=N/2$ and (ii) $m_0=1$, at infinite temperature in the thermodynamic limit, where we have exact (closed) expressions for the correlation functions available: 

\begin{align}
\alpha^{+\text{(i)}}(t)&=\frac{1}{2}e^{+ i2ht}e^{-J^2t^2} \label{eq:spininfiniteT}\\
\alpha^{+\text{(ii)}}(t)&=e^{+ i 2ht}\frac{J_1(2Jt)}{2Jt}, \label{eq:ApoCorrFun}
\end{align} 
where Eq.~\eqref{eq:spininfiniteT} was taken from reference\cite{spinspin}, and Eq.~\eqref{eq:ApoCorrFun} was derived using $W_{m_0,k}=\sqrt{\frac{2}{N+1}}\sin(\frac{\pi k m_0}{N+1})$. $J_1(x)$ is a Bessel function of the first kind. Limiting the discussion to the first dissipative term of Eq.~\eqref{eq:integrodiff2} (at infinite temperature the other term follows analogously), for (i) we find:
 \begin{align}
\lim_{\Gamma / J \to 0} \int_0^t & ds \alpha^{+\text{(i)}}(s)e^{-i\Delta s}  g(t-s) \nn \\
& = \frac{1}{2} g(t) \int_0^\infty  ds e^{-J^2s^2} e^{- i\Delta_h s},
\end{align} 
where $g(x)$ is a function changing on a time scale characterised by $\frac{1}{\Gamma}$, and $\Gamma$ is the characteristic frequency of the spin (e.g. decay rate), which in our model is set by the coupling strength $\Omega$. Thus for small enough coupling, $\frac{\Gamma}{J}\ll1$, the Markovian master equation is valid at all detunings $\Delta_h$. 

For (ii) we can write, using the asymptotic form of the Bessel functions $ \lim_{x\to \infty} J_\alpha(x)=\sqrt{\frac{2}{\pi x}} \cos \big( x - \frac{\alpha \pi}{2} - \frac{\pi}{4} \big) $, 
 \begin{align}
\lim_{\Gamma / J \to 0} & \int_0^t  ds \alpha^{+\text{(ii)}}(s)e^{-i\Delta s}  g(t-s) \nn \\
= g(t) & \int_0^{t_1}  ds  \frac{J_1(2Js)}{2Js} e^{-i\Delta_h s} \nn\\
+ \frac{1}{4\sqrt{\pi}} & \int_{t_1}^t  ds g(t-s) (Js)^{-\frac{3}{2}} \Big( e^{+ i (2J-\Delta_h)s} e^{-\frac{3}{4}\pi i} \nn\\
 & \quad \quad \quad \quad \quad \quad \quad \quad \quad +  e^{- i (2J+\Delta_h)s} e^{\frac{3}{4}\pi i} \Big),
\end{align} 
where $\frac{1}{J}\ll t_1\ll \frac{1}{\Gamma}$. For detunings far enough away from the band edges $|\pm2-\frac{\Delta_h}{J}|\gg \frac{\Gamma}{J}$ the second integral can be neglected, and the 'Markovian' master equation is valid. At the band edges, where the oscillations are too slow to kill the integral, the slow power law decay violates the Markovian approximation at any $\Gamma>0$. 

We thus find that whilst, deep within the band and at small enough coupling, the dynamics is captured by the 'Markovian' master equation in both cases, the underlying mechanism is completely different: superexponentially decaying correlation functions in (i); rapid oscillations of the correlation functions in (ii). The self consistency condition~\eqref{eq:kessler} is blind to what is the fundamental origin of its validity. Too close to the band edges ($|\pm2-\frac{\Delta_h}{J}| \not \gg \frac{\Gamma}{J}$) the Markovian master equation is only valid for (i) (at small enough coupling).

\bibliography{nonMarkPaper}
\bibliographystyle{apsrev4-1}
\end{document}